\documentclass[aps,prc,twocolumn,superscriptaddress,showpacs,floatfix,nofootinbib,altaffilletter]{revtex4}
\usepackage{xspace}
\usepackage{longtable}
\usepackage{graphicx}
\usepackage{afterpage}
\usepackage{placeins}
\newcommand{\GeV}{\ensuremath{\mbox{GeV}}\xspace}

\newcommand{\GeVc}{\ensuremath{\mbox{GeV}/c}\xspace}

\newcommand{\mm}{\ensuremath{\mbox{mm}}\xspace}

\newcommand{\mrad}{\ensuremath{\mbox{mrad}}\xspace}
\newcommand{\rad}{\ensuremath{\mbox{rad}}\xspace}

\newcommand{\ps}{\ensuremath{\mbox{ps}}\xspace}

\newcommand{\pip}{\ensuremath{\pi^+}\xspace}
\newcommand{\pim}{\ensuremath{\pi^-}\xspace}

\def\be{\begin{equation}}
\def\ee{\end{equation}}
\def\bea{\begin{eqnarray}}
\def\eea{\end{eqnarray}}

\newcommand{\bfGeVc}{\ensuremath{\mathbf {\mbox{\bf GeV}/c}}\xspace}

\begin{document}
\title{{\Large EUROPEAN ORGANIZATION FOR NUCLEAR RESEARCH} \\
\vskip 2cm
\hspace*{10cm}
{\rm CERN-PH-EP/2010-016} \\
\hspace*{10cm}
{\rm 12 May 2010} \\
\vskip 2cm
\centerline {\large  HARP Collaboration}
\noaffiliation
\vskip 3 cm 
{\large  Measurements of forward proton production
with incident protons and charged pions on nuclear targets at the
CERN Proton Synchroton.}

\vskip 3cm

{\rm Measurements of the  double-differential proton production
  cross-section
$
  {{\mathrm{d}^2 \sigma}}/{{\mathrm{d}p\mathrm{d}\Omega }}
  $
in the range of momentum $0.5~\GeVc \leq p < 8.0~\GeVc$
  and angle $0.05~\rad \leq \theta  < 0.25~\rad$
  in collisions of charged pions and protons on
  beryllium, carbon, aluminium, copper,
  tin, tantalum and lead are presented.
  The data were taken with the large acceptance HARP detector in the T9 beam
  line of the CERN Proton Synchrotron.
  %
  Incident particles were identified by an elaborate system of beam
  detectors and impinged on a target of 5\% of a nuclear
  interaction length.
  The tracking and identification of the
  produced particles was performed using the forward spectrometer of the
  HARP experiment.
  Results are obtained for the double-differential cross-sections
  mainly at four incident beam momenta (3~\GeVc, 5~\GeVc, 8~\GeVc and
  12~\GeVc).
  Measurements are compared with predictions of the GEANT4 and MARS Monte Carlo
  generators. }

\vskip 5cm
\centerline{ {\bf (submitted to Physical Review C)}}
\clearpage
 }
\author{M.~Apollonio} 
\altaffiliation{Now at Imperial College, University of London, UK.}
\affiliation{Universit\`{a} degli Studi e Sezione INFN, Trieste, Italy}
\author{A.~Artamonov}   
\altaffiliation{ITEP, Moscow, Russian Federation.}
\affiliation{ CERN, Geneva, Switzerland}
\author{A. Bagulya} 
\affiliation{P. N. Lebedev Institute of Physics (FIAN), \\
Russian Academy of Sciences, Moscow, Russia}
\author{G.~Barr}
\affiliation{Nuclear and Astrophysics Laboratory, University of Oxford, UK} 
\author{A.~Blondel}
\affiliation{Section de Physique, Universit\'{e} de Gen\`{e}ve, Switzerland} 
\author{F.~Bobisut$^{(**)}$} 
\affiliation{Sezione INFN$^{(*)}$ and Universit\'a degli Studi$^{(**)}$, 
Padova, Italy}
\author{M.~Bogomilov}
\affiliation{ Faculty of Physics, St. Kliment Ohridski University, \\
Sofia,
  Bulgaria}
 \author{M.~Bonesini}
\thanks{Corresponding author (M.~Bonesini).\\
E-mail: maurizio.bonesini@mib.infn.it}
\affiliation{Sezione INFN Milano Bicocca, Milano, Italy} 
\author{C.~Booth} 
\affiliation{ Dept. of Physics, University of Sheffield, UK}
\author{S.~Borghi}  
\altaffiliation{Now at CERN}
\affiliation{Section de Physique, Universit\'{e} de Gen\`{e}ve, Switzerland}
\author{S.~Bunyatov}
\affiliation{Joint Institute for Nuclear Research, JINR Dubna, Russia} 
\author{J.~Burguet--Castell}
\affiliation{Instituto de F\'{i}sica Corpuscular, IFIC, CSIC and Universidad de Valencia, Spain}
\author{M.G.~Catanesi}
\affiliation{Sezione INFN, Bari, Italy} 
\author{A.~Cervera--Villanueva}
\affiliation{Instituto de F\'{i}sica Corpuscular, IFIC, CSIC and Universidad de Valencia, Spain}
\author{P.~Chimenti}  
\affiliation{Universit\`{a} degli Studi e Sezione INFN, Trieste, Italy}
\author{L.~Coney}
\altaffiliation{Columbia University, New York, USA (MiniBooNE Coll).
Now at University of California, Riverside, USA.}
\noaffiliation 
\author{E.~Di~Capua}
\affiliation{Universit\`{a} degli Studi e Sezione INFN, Ferrara, Italy} 
\author{U.~Dore}
\affiliation{ Universit\`{a} ``La Sapienza'' e Sezione INFN Roma I, Roma,
  Italy}
\author{J.~Dumarchez}
\affiliation{ LPNHE, Universit\'{e}s de Paris VI et VII, Paris, France}
\author{R.~Edgecock}
\affiliation{Rutherford Appleton Laboratory, Chilton, Didcot, UK} 
\author{M.~Ellis}          
\altaffiliation{Now at FNAL, Batavia, Illinois, USA.} 
\affiliation{Rutherford Appleton Laboratory, Chilton, Didcot, UK}
\author{F.~Ferri}           
\affiliation{Sezione INFN Milano Bicocca, Milano, Italy}
\author{U.~Gastaldi}
\affiliation{Laboratori Nazionali di Legnaro dell' INFN, Legnaro, Italy}
\author{S.~Giani} 
\affiliation{ CERN, Geneva, Switzerland}
\author{G.~Giannini} 
\affiliation{Universit\`{a} degli Studi e Sezione INFN, Trieste, Italy}
\author{D.~Gibin$^{(**)}$}
\affiliation{Sezione INFN$^{(*)}$ and Universit\'a degli Studi$^{(**)}$, 
Padova, Italy}
\author{S.~Gilardoni}       
\affiliation{ CERN, Geneva, Switzerland} 
\author{P.~Gorbunov}  
\altaffiliation{ITEP, Moscow, Russian Federation.}
\affiliation{ CERN, Geneva, Switzerland}
\author{C.~G\"{o}\ss ling}
\affiliation{ Institut f\"{u}r Physik, Universit\"{a}t Dortmund, Germany}
\author{J.J.~G\'{o}mez--Cadenas} 
\affiliation{Instituto de F\'{i}sica Corpuscular, IFIC, CSIC and Universidad de Valencia, Spain}
\author{A.~Grant}  
\affiliation{ CERN, Geneva, Switzerland}
\author{J.S.~Graulich}
\altaffiliation{Now at Section de Physique, Universit\'{e} de Gen\`{e}ve, Switzerland.}
\affiliation{Institut de Physique Nucl\'{e}aire, UCL, Louvain-la-Neuve,
  Belgium} 
\author{G.~Gr\'{e}goire}
\affiliation{Institut de Physique Nucl\'{e}aire, UCL, Louvain-la-Neuve,
  Belgium} 
\author{V.~Grichine}  
\affiliation{P. N. Lebedev Institute of Physics (FIAN), \\
Russian Academy of Sciences, Moscow, Russia}
\author{A.~Grossheim} 
\altaffiliation{Now at TRIUMF, Vancouver, Canada.}
\affiliation{ CERN, Geneva, Switzerland} 
\author{A.~Guglielmi$^{(*)}$}
\affiliation{Sezione INFN$^{(*)}$ and Universit\'a degli Studi$^{(**)}$, 
Padova, Italy}
\author{L.~Howlett}
\affiliation{ Dept. of Physics, University of Sheffield, UK}
\author{A.~Ivanchenko}
\altaffiliation{Now at CNRS, CENBG Bordeaux, France.} 
\affiliation{ CERN, Geneva, Switzerland}
\author{V.~Ivanchenko}  
\altaffiliation{On leave  from Ecoanalitica, Moscow State University,
Moscow, Russia}
\affiliation{ CERN, Geneva, Switzerland}
\author{A.~Kayis-Topaksu}
\thanks{Now at \c{C}ukurova University, Adana, Turkey.}
\affiliation{ CERN, Geneva, Switzerland}
\author{M.~Kirsanov}
\affiliation{Institute for Nuclear Research, Moscow, Russia}
\author{D.~Kolev} 
\affiliation{ Faculty of Physics, St. Kliment Ohridski University, \\
Sofia,
  Bulgaria}
\author{A.~Krasnoperov} 
\affiliation{Joint Institute for Nuclear Research, JINR Dubna, Russia}
\author{J. Mart\'{i}n--Albo}
\affiliation{Instituto de F\'{i}sica Corpuscular, IFIC, CSIC and Universidad de Valencia, Spain}
\author{C.~Meurer}
\affiliation{Institut f\"{u}r Physik, Forschungszentrum Karlsruhe, Germany}
\noaffiliation{}
\author{M.~Mezzetto$^{(*)}$}
\affiliation{Sezione INFN$^{(*)}$ and Universit\'a degli Studi$^{(**)}$, 
Padova, Italy}
\author{G.~B.~Mills}
\altaffiliation{Los Alamos National Laboratory, Los Alamos, USA (MiniBooNE Coll.)}
\noaffiliation
\author{M.C.~Morone}
\altaffiliation{Now at University of Rome Tor Vergata, Italy.}   
\affiliation{Section de Physique, Universit\'{e} de Gen\`{e}ve, Switzerland}
\author{P.~Novella} 
\affiliation{Instituto de F\'{i}sica Corpuscular, IFIC, CSIC and Universidad de Valencia, Spain}
\author{D.~Orestano$^{(**)}$}
\affiliation{Sezione INFN$^{(*)}$ and Universit\'a$^{(**)}$  Roma Tre, 
Roma, Italy}
\author{V.~Palladino}
\affiliation{Universit\`{a} ``Federico II'' e Sezione INFN, Napoli, Italy}
\author{J.~Panman}
\affiliation{ CERN, Geneva, Switzerland}
 \author{I.~Papadopoulos}  
\affiliation{ CERN, Geneva, Switzerland}
\author{F.~Pastore$^{(**)}$} 
\affiliation{Sezione INFN$^{(*)}$ and Universit\'a$^{(**)}$  Roma Tre, 
Roma, Italy}
\author{S.~Piperov}
\affiliation{ Institute for Nuclear Research and Nuclear Energy,
Academy of Sciences, \\
Sofia, Bulgaria}
\author{N.~Polukhina}
\affiliation{P. N. Lebedev Institute of Physics (FIAN), \\
Russian Academy of Sciences, Moscow, Russia}
\author{B.~Popov} 
\altaffiliation{Also supported by LPNHE, Paris, France.}
\affiliation{Joint Institute for Nuclear Research, JINR Dubna, Russia}
\author{G.~Prior}   
\altaffiliation{Now at CERN}
\affiliation{Section de Physique, Universit\'{e} de Gen\`{e}ve, Switzerland}
\author{E.~Radicioni}
\affiliation{Sezione INFN, Bari, Italy}
\author{D.~Schmitz}
\altaffiliation{Columbia University, New York, USA (MiniBooNE Coll.)}
\noaffiliation
\author{R.~Schroeter}
\affiliation{Section de Physique, Universit\'{e} de Gen\`{e}ve, Switzerland}
\author{G~Skoro}
\affiliation{ Dept. of Physics, University of Sheffield, UK}
\author{M.~Sorel}
\affiliation{Instituto de F\'{i}sica Corpuscular, IFIC, CSIC and Universidad de Valencia, Spain}
\author{E.~Tcherniaev}
\affiliation{ CERN, Geneva, Switzerland}
 \author{P.~Temnikov}
\affiliation{ Institute for Nuclear Research and Nuclear Energy,
Academy of Sciences, \\
Sofia, Bulgaria}
\author{V.~Tereschenko}  
\affiliation{Joint Institute for Nuclear Research, JINR Dubna, Russia}
\author{A.~Tonazzo$^{(**)}$}
\affiliation{Sezione INFN$^{(*)}$ and Universit\'a$^{(**)}$  Roma Tre, 
Roma, Italy}
\author{L.~Tortora$^{(*)}$}
\affiliation{Sezione INFN$^{(*)}$ and Universit\'a$^{(**)}$  Roma Tre, 
Roma, Italy}
\author{R.~Tsenov}
\affiliation{ Faculty of Physics, St. Kliment Ohridski University, \\
Sofia,
  Bulgaria}
\author{I.~Tsukerman}   
\altaffiliation{ITEP, Moscow, Russian Federation.}
\affiliation{ CERN, Geneva, Switzerland}
\author{G.~Vidal--Sitjes}  
\altaffiliation{Now at Imperial College, University of London, UK.}
\affiliation{Universit\`{a} degli Studi e Sezione INFN, Ferrara, Italy}
\author{C.~Wiebusch}    
\altaffiliation{Now at III Phys. Inst. B, RWTH Aachen, Germany.}
\affiliation{ CERN, Geneva, Switzerland}
\author{P.~Zucchelli} 
\altaffiliation{Now at SpinX Technologies, Geneva, Switzerland;\\
on leave  from INFN, Sezione di Ferrara, Italy.}
\affiliation{ CERN, Geneva, Switzerland}
\vskip 5cm
\maketitle

\section{Introduction}

In many particle and astroparticle physics experiments the 
knowledge of hadron production is required as an external input
to make optimal use of
the recorded data and to help design the experimental facilities.
The HARP experiment~\cite{harp-prop} is motivated by this need for
precise hadron production measurements.
It has taken data with beams of pions and protons
with momenta from 1.5~\GeVc to 15~\GeVc hitting nuclear targets made of a large
range of materials.
To provide a large angular and momentum coverage of the produced charged
particles the experiment comprises two spectrometers, a forward
spectrometer built around a dipole magnet covering the angular range up
to 250~mrad and a large-angle spectrometer constructed in a solenoidal
magnet with an angular acceptance of 
$0.35~\rad \leq \theta  < 2.15~\rad$, based on a time-projection-chamber
(TPC).

The main HARP objectives are to measure pion yields for a quantitative
design of the proton driver of future superbeams (high-intensity
conventional beams) and a neutrino factory~\cite{ref:nufact}, 
to provide measurements to improve calculations of the atmospheric neutrino
flux~\cite{Battistoni,Stanev,Gaisser,Engel}
and to measure particle yields as input for the flux
calculation of accelerator neutrino experiments~\cite{ref:physrep}, 
such as K2K~\cite{ref:k2k,ref:k2kfinal},
MiniBooNE~\cite{ref:miniboone} and SciBooNE~\cite{ref:sciboone}. 
The momentum range of the incoming particles presented here corresponds
to a momentum region of great interest for neutrino beams and are
in a region far from coverage by earlier dedicated hadroproduction 
experiments~\cite{ref:na56,ref:atherton}.
In addition to these specific aims, the data provided by HARP are
valuable for validating hadron production models used in simulation
programs. 
These simulations are playing an important role in the interpretation
and design of modern particle-physics experiments.
In particular, the simulation of calorimeter response and secondary
interactions in tracking systems needs to be supported by experimental
hadron production data.
Results on forward production of charged pions by incident protons are the
subject of previous HARP publications 
\cite{ref:harp:alPaper,ref:harp:bePaper,ref:harp:carbonfw,ref:harp:cnofw,ref:harp:protonforward}.  
The analysis of results on charged pion production with charged pion beams
on the full range of targets can be found in Ref.~\cite{ref:harp:pionforward} \footnote{
results on the large angle charged pion production are instead presented
in Refs.~\cite{ref:harp:la} and \cite{ref:harp:pionla}}. 

In this paper, measurements of the double-differential cross-section, 
$
{{\mathrm{d}^2 \sigma^{p}}}/{{\mathrm{d}p\mathrm{d}\Omega }}
$,
for forward production of protons by
incident charged pions or protons of 3~\GeVc, 5~\GeVc, 8~\GeVc,
8.9~\GeVc (Be only), 12~\GeVc and 12.9~\GeVc (Al only) momentum impinging
on a thin solid beryllium, carbon, aluminium, copper, tin, tantalum  
and lead targets of 5\% nuclear interaction length 
($\lambda_{\mathrm{I}}$) thickness are presented.

To our knowledge no high statistics proton production data 
at low momenta  ($\leq 15$ \GeVc)
in the forward direction ($\leq 250$~mrad) with incident protons or
charged pions on nuclear targets have been published.

Data were taken in the T9 beam of the CERN PS. The collected statistics,
for the different nuclear targets, are reported in Table \ref{tab:events}. 

\begin{table*}[hbpt]
\caption{Total number of events and tracks used in the various nuclear
  5\%~$\lambda_{\mathrm{I}}$ target data sets and the number of
  incident protons and charged pions on target as calculated from the pre-scaled incident beam triggers.
  First entries (total DAQ events) are for the positive and negative beam; then numbers are given for
  incident protons, $\pi^{+},\pi^{-}$ in units of $10^3$ events.
\label{tab:events}
}
{\small
\begin{center}
\begin{tabular}{ll|cc|cc|cc|c|cc|c} \hline
\bf{Data set (\bfGeVc)}          &         &\multicolumn{2}{c}{\bf{3}}&\multicolumn{2}{c}{\bf{5}}&\multicolumn{2}{c}{\bf{8}}&\multicolumn{1}{c}{\bf{8.9}}&\multicolumn{2}{c}{\bf{12}} & \multicolumn{1}{c}{\bf{12.9}}\\
beam polarity   &         & + & $-$ & + & $-$ & + & $-$ & + & + & $-$ & +  \\ \hline
    Total DAQ events     & (Be)     &  1113&2233 & 1296&1798  & 1935&1585    & 5868&  1207&1227& \\
                         &  (C)     &  1345&1831 & 2628&1279      & 1846&1399&      &  1062&646& \\
                         & (Al)     & 1159&1523 & 1789&920     & 1707&1059   &   & 619&741  & 4713 \\
                         & (Cu)     & 624&3325  & 2079&1805      & 2089&1615     & &   745&591  & \\
                         & (Sn)     & 1637&1972 & 2828&1625      & 2404&1408     & &  1803&937  & \\
                         & (Ta)     & 1783&994 & 2084&1435      & 1965&1505      & & 866&961   & \\
                         & (Pb)     & 1911&1282 & 2111&2074      & 2266&1496     & & 487&1706  & \\
\end{tabular}
\begin{tabular}{ll|ccc|ccc|ccc|cc|ccc|cc} \hline
\bf{Data set (\bfGeVc)}          &         &\multicolumn{3}{c}{\bf{3}}&\multicolumn{3}{c}{\bf{5}}&\multicolumn{3}{c}{\bf{8}}&\multicolumn{2}{c}{\bf{8.9}}&\multicolumn{3}{c}{\bf{12}} & \multicolumn{2}{c}{\bf{12.9}}\\
particle type   &         & p & $\pi^+$ & $\pi^-$ & p & $\pi^+$ & $\pi^-$ & p & $\pi^+$ & $\pi^-$ & p & $\pi^+$ & p & $\pi^+$ & $\pi^-$ & p & $\pi^+$  \\ \hline
  Acc.  beam part.      &  (Be)         & 99&246&731  &  289&384&914     &  761&341&826  & 2103&1278  &    580&76&693 & \\
  with forw. int.               & (C)           & 101&257&299      &  542&754&530         &  709&358&772 &  & & 470&41&352& \\
                         &  (Al)    & 86&213&486  &  376&523&308      & 637&335&611   &   &  & 306&27&435    & 2116&332 \\
                         &  (Cu)    & 73&168&1185  &  408&611&850     &  741&397&966 &    &  & 363&33&347  &  \\
                         &  (Sn)    & 217&467&778  &  528&819&732     &  818&481&804   &  &  & 856&79&584   &  \\
                         &  (Ta)    & 281&561&426  &  398&600&671     &  668&388&893   &  &  & 403&37&536  &   \\
                         &  (Pb)    & 310&611&473  &  387&594&997     &  758&444&896   &  &  & 221&20&839 &  \\
\hline
  \bf{Final state p } & (Be)  & 7.2&1.7&1.5   &   18.0&3.8&3.1      &  37.0&4.3&4.8 & 86.4&15.6   &  19.3&1.0&5.7& \\
  {selected with PID} & (C)   & 6.2&1.6&0.6   &   29.1&6.7&1.7     &  32.2&4.4&4.1   & &  &   14.8&0.5&2.5  & \\
                                  & (Al)  & 4.2&1.3&0.8    & 19.0&4.8&1.0      &  26.8&3.9&2.9 &  & &    10.6&0.4&2.8  &  73.3 & 5.0\\
                                  & (Cu)  & 2.5&0.8&1.6 & 18.1&5.6&2.9&  31.5&5.2&6.0 &  & &   12.9&0.6&3.1 & \\
                                  & (Sn)  & 4.9&1.5&0.9 & 20.8&6.9&2.5 &  33.6&6.4&3.9 & & &  30.2&1.3&5.2 & \\
                                  & (Ta)  & 4.8&1.4&0.4   & 12.9&4.6&2.1 &  25.6&5.0&4.2  &   &  & 14.2&0.6&4.8  & \\
                                  & (Pb)  & 4.4&1.4&0.3   & 12.0&4.0&2.8 &  26.6&5.2&4.4  & & &  7.5&0.3&7.4  & \\
\hline
\end{tabular}
\end{center}
}
\end{table*}

\subsection{Experimental apparatus}

\begin{figure*}[tb]]
\centering
\includegraphics[width=0.8\textwidth]{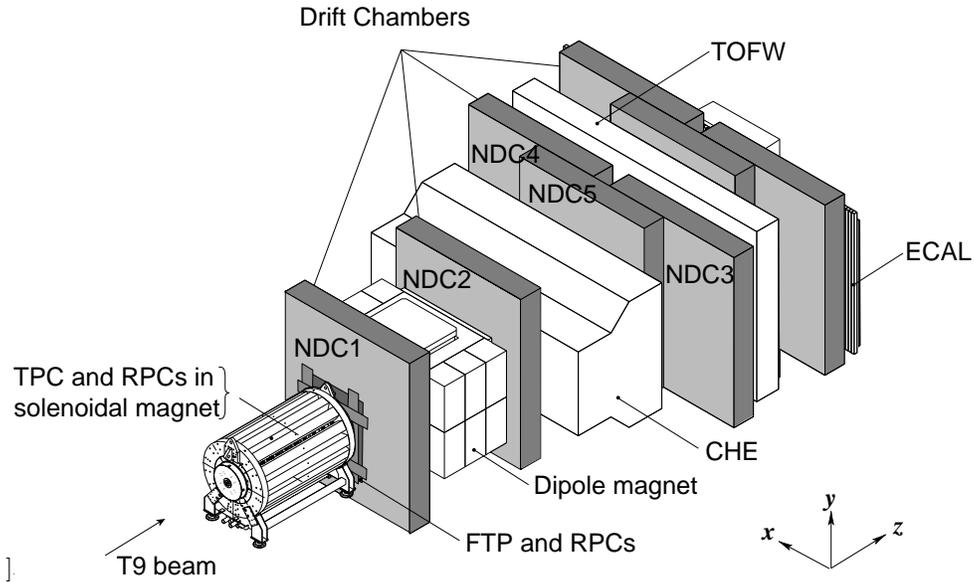} 
\caption{
\label{fig:harp}
Schematic layout of the HARP detector.
The convention for the coordinate system is shown in the lower-right
corner.
The three most downstream (unlabelled) drift chamber modules are only partly
equipped with electronics and are not used for tracking.
The detector covers a total length of 13.5 m along the beam axis and has 
a maximum width of 6.5 m perpendicular to the beam.
The beam muon identifier is visible as the most downstream detector
 (white box).
}
\end{figure*}

 The HARP experiment
 makes use of a large-acceptance spectrometer consisting of a
 forward and a large-angle detection system.
 The HARP detector is shown in Fig.~\ref{fig:harp} and is fully described
 in Ref.~\cite{ref:harpTech}.
 The forward spectrometer -- 
 based on five modules of large area drift chambers
 (NDC1-5)~\cite{ref:NOMAD_NIM_DC} and a dipole magnet
 complemented by a set of detectors for particle identification (PID): 
 a time-of-flight wall (TOFW)~\cite{ref:tofPaper}, a large Cherenkov detector (CHE) 
 and an electromagnetic calorimeter (ECAL) --
 covers polar angles up to 250~mrad, 
 which is well matched to the angular range of interest for the
 measurement of hadron production to calculate the properties of
 conventional neutrino beams.
The discrimination power of time-of-flight below 3 GeV/c and the Cherenkov
dectector above 3 GeV/c are combined to provide powerful separation
of forward pions and protons~\cite{ref:harp:pidPaper}. The calorimeter is used only 
for separating pions and electrons when characterizing the response
of the other detectors. 
The muon contamination of the beam is measured with a 
 1.4 m wide muon identifier (BMI).
At the downstream end of the spectrometer, after a 0.4 m long iron absorber,
 it is made of an iron-scintillator sandwich with five planes of six scintillator
 each, read out at both sides, giving a total thickness 
of $6.4 \ \lambda_{\mathrm{I}}$.

 The large-angle spectrometer consists of a Time Projection Chamber (TPC) 
 and Resistive Plate Chambers (RPCs), located inside a solenoidal magnet.
 It has a large acceptance in the momentum
 and angular range for the pions relevant to the production of the
 muons in a neutrino factory. 
 This system is not used in the present analysis. 

The HARP experiment, located in the T9 beam of the CERN PS, took data in 2001
and 2002.
The momentum definition of the T9 beam 
is known with a precision of the order of 1\%~\cite{ref:t9}. 

The target is placed inside the inner field cage (IFC) of the TPC,
in an assembly that can be moved in and out of the solenoid magnet.
The solid targets used for the measurements reported here
have a cylindrical shape with a nominal diameter of about 30~\mm.
Their thickness is equivalent to about 5\% 
of $\lambda_{\mathrm{I}}$.

A set of four multi-wire
proportional chambers (MWPCs) measures the position and direction of
the incoming beam particles with an accuracy of $\approx$1~\mm in
position and $\approx$0.2~\mrad in angle per projection.
A beam time-of-flight system (BTOF)
measures the time difference of particles over a $21.4$~m path-length. 
It is made of two
identical scintillation hodoscopes, TOFA and TOFB (originally built
for the NA52 experiment~\cite{ref:NA52}),
which, together with a small target-defining trigger counter (TDS,
also used for the trigger), provide particle
identification at low energies. This provides separation of pions, kaons
and protons up to 5~\GeVc and determines the initial time at the
interaction vertex ($t_0$). 
The timing resolution of the combined BTOF system is about 70~\ps.
A system of two N$_2$-filled Cherenkov detectors (BCA and BCB) is
used to tag electrons at low energies and pions at higher energies. 
The electron and pion tagging efficiency is found to be close to
100\%.
At the beam energy used for this analysis the Cherenkov counters select
all particles lighter than protons, while the BTOF is used to reject ions. 
A set of trigger detectors completes the beam instrumentation.

The beam of positive particles used for this measurement contains mainly 
positrons, muons, pions and protons, with small components of kaons,
deuterons and heavier ions.
Its composition depends on the selected beam momentum.
The proton fraction in the incoming positive-particle beam varies from
$\sim 35$\% at 3~\GeVc to $\sim 92$\% at 12~\GeVc.  
The negatively-charged particle beam is mainly composed of pions with
small background components of muons and electrons.

At the first stage of the analysis a favoured beam particle type is selected
using the beam time-of-flight system and the two Cherenkov
counters.
A value of the pulse height consistent with the absence of a signal in both beam
Cherenkov detectors distinguishes protons (and kaons) from electrons and pions.
We also ask for time measurements to be present which are needed for calculating 
the arrival time of the beam proton at the target. 
The beam TOF system is used to reject ions, such as deuterons, and to
separate protons from pions at low momenta.
At 3~\GeVc, the TOF measurement allows the selection of
pions from protons to be made at more than 5{$\sigma$}.
In most beam settings the nitrogen pressure in the beam Cherenkov
counters was too low for kaons to be above the threshold.
Kaons are thus contained in the proton sample.
However, the fraction of kaons has been measured in the 12.9~\GeVc beam
configuration and is found to contribute less than 0.5\%, and hence it is
negligible both in the pion and proton beam sample.
Electrons radiate in the Cherenkov counters and would be counted as
pions. 
In the 3~\GeVc beam, electrons are identified by both Cherenkov counters,
since the pressure was such that pions remained below threshold.
In the 5~\GeVc beam electrons could be tagged by one Cherenkov counter
only, while the other Cherenkov counter was used to tag pions.
The $e/\pi$ fraction was measured to be 1\% in the 3~\GeVc beam and 
$<10^{-3}$ in the  5~\GeVc beam.
By extrapolation from the lower-energy beam settings this electron
contamination can be estimated to be negligible ($<10^{-3}$) for the
beams where it cannot be measured directly.
More details on the beam particle selection can be found in 
\cite{ref:harp:alPaper} and Refs.~\cite{ref:harpTech}.

In addition to the momentum-selected beam of protons and pions originating
from the T9 production target one expects also the presence of muons
from pion decay both downstream and upstream of the beam momentum selection.  
Therefore, precise absolute knowledge of the muon rate incident on the
HARP targets is required when measurements 
of particle production with incident pions are performed. 
The particle identification detectors in the beam do not distinguish
muons from pions. 
A separate measurement of the muon component has been performed
using data-sets without target (``empty-target data-sets'').
Since the empty-target data were taken with the same beam parameter
settings as the data taken with targets, the beam composition can be
measured in the empty-target runs using the forward spectrometer  
and then used as an overall correction
for the counting of pions in the runs with targets. 
Muons are recognized by their longer range in the beam muon identifier
(BMI).
The punch-through background in the BMI is measured counting the
protons (identified with the beam detectors) thus mis-identified as muons
by the BMI. 
A comparison of the punch-through rate between simulated incoming pions
and protons was used to determine a correction for the difference
between pions and protons and to determine the systematic error.
This difference is the dominant systematic error in the beam composition
measurement. 
The aim was to determine the composition of the beam as it strikes the
target, thus muons produced in pion decays after the HARP target
should be considered as a background to the measurement of muons in the
beam. 
The rate of these latter background muons, which depends mainly on the total 
inelastic cross-section and pion decay,  was calculated by a Monte Carlo 
simulation using GEANT4~\cite{ref:geant4}.
The muon fraction in the beam (at the target) is obtained taking
into account the efficiency of the BMI selection criteria as well as the
punch-through and decay backgrounds.
The analyses for the various beam settings give results for
$R=\mu/(\mu+\pi)$ of (4.2$\pm$1)\% and (5.2$\pm$1)\% for the low-momentum beams (3~\GeVc
and 5~\GeVc) and between (4.1$\pm$1)\% and (2.8$\pm$1)\% for the highest momenta (from
8~\GeVc to 12.9~\GeVc).
The uncertainty in these fractions is dominated by the systematic
uncertainty in the punch-through background.
The fact that the background does not scale with the decay probability
for pions is due to the limited acceptance of the beam-line to transport
the decay muons. 
The muon contamination is taken into account in the normalization of the
pion beam and adds a systematic error of 1\% to the overall
normalization. 
Only events with a single reconstructed beam track  in the four MWPCs,
good timing measurements in BTOF and no signal in the beam halo counters
are accepted.

A downstream trigger in the forward scintillator trigger plane (FTP) 
was required to record an event,
accepting only tracks with a trajectory outside the central hole
(60~mm) which allows beam particles to pass. 

The length of the accelerator spill is 400~ms with a typical intensity
of 15~000 beam particles per spill.
The average number of events recorded by the data acquisition ranges
from 300 to 350 per spill.

The absolute normalization of the cross-section was
performed using ``incident-proton'' triggers. 
These are triggers where the same selection on the beam particle was
applied but no selection on the interaction was performed.
The rate of this trigger was down-scaled by a factor 64.
These unbiased events are used to determine $N_{\mathrm{pot}}$ in the 
cross-section formula (1), see later. 

\section{Data analysis and cross-section calculation.}

This  analysis is similar to the one reported in reference
\cite{ref:harp:pionforward}.
For the current analysis we have used identical reconstruction and PID
algorithms.
Secondary track selection criteria,
described in~\cite{ref:harp:carbonfw},  are optimized to ensure the quality
of momentum reconstruction and a clean time-of-flight measurement
while maintaining a high reconstruction efficiency. 

The background induced by
interactions of beam particles in the materials outside the target
is measured  by taking data without a
target in the target holder (``empty target data'').  
These data are  subject to the same  event and track
selection criteria as the standard  data sets and are subtracted
bin-by-bin. 

The collected event statistics on the different solid targets 
is summarised in Table \ref{tab:events}. 

\par
The double-differential cross-section for the production of a particle
of type $\alpha$ is calculated as follows:
\begin{eqnarray}
\frac{d^2 \sigma^{\alpha}}{dp d\Omega}(p_i,\theta_j)  =  
\frac{A}{N_A \rho t} \cdot \frac{1}{N_{\rm pot}} \cdot \frac{1}{\Delta p_i \Delta \Omega_j} \cdot \\ \nonumber
\sum_{p'_i,\theta'_j,\alpha'} \mathcal{M}^{\rm cor}_{p_i\theta_j\alpha p'_i\theta'_j\alpha'} \cdot 
N^{\alpha'}(p'_i,\theta'_j)\hspace{0.1cm},
\label{eq-1}
\end{eqnarray} 
where 
\begin{itemize}
\item $\frac{d^2 \sigma^{\alpha}}{dp d\Omega}(p_i,\theta_j)$ is the
  cross-section in mb/(\GeVc sr) for the particle type $\alpha$ ( a proton
  in our case) for each true momentum and angle bin ($p_i,\theta_j$)
  covered in this analysis;
\item $N^{\alpha'}(p'_i,\theta'_j)$  is the number of particles of
  type $\alpha'$ in bins of reconstructed momentum $p'_i$ and angle
  $\theta_j'$ in the raw data, after subtraction of empty target data 
  (due to beam protons interacting in material other than the nuclear
   target). These particles must satisfy the event, track 
  and PID selection criteria.
\item $\mathcal{M}^{\rm cor}_{p\theta\alpha p'\theta'\alpha'}$ is the
  correction matrix which accounts for finite efficiency and resolution of
  the detector. It unfolds the true variables $p_i, \theta_j, \alpha$ from
  the reconstructed variables $p'_i, \theta'_j, \alpha'$ and corrects the
  observed particle number to take into account effects such as reconstruction
  efficiency, acceptance, absorption, pion decay, tertiary production, PID
  efficiency and PID misidentification rate.
\item $\frac{A}{N_A \rho t}$, $\frac{1}{N_{\rm pot}}$ and
  $\frac{1}{\Delta p_i \Delta \Omega_j}$ are normalization factors,
  namely:
\subitem $\frac{N_A \rho t}{A}$ is the number of target nuclei per unit area 
\footnote{$A$ - atomic  mass, $N_A$ - Avogadro number, $\rho$ - target
  density and $t$ - target thickness};
\subitem $N_{\rm pot}$ is the number of protons on
  target (particles on target);
\subitem $\Delta p_i $ and $\Delta \Omega_j $ are the bin sizes in
  momentum and solid angle, respectively 
\footnote{$\Delta p_i = p^{\rm max}_i-p^{\rm min}_i$,\hspace{0.2cm}
  $\Delta \Omega_j = 2 \pi (\cos(\theta^{\rm min}_j)- 
  \cos(\theta^{\rm max}_j))$}.
\end{itemize}
We do not make a correction for the attenuation
of the  beam in the target, so that strictly speaking the
cross-sections are valid for $\lambda_{\mathrm{I}}=5\%$
targets.

The  calculation of the
correction matrix $M^{\rm cor}_{p_i\theta_j\alpha
  p'_i\theta'_j\alpha'}$ is 
a rather difficult task.
Various techniques are
described in the literature to obtain this matrix. 
The method applied here and called UFO in~\cite{ref:harp:alPaper}  
is the unfolding method introduced 
by D'Agostini~\cite{ref:DAgostini}~\footnote{
The  unfolding method tries to put in correspondence the
vector of measured observables (such as particle momentum, polar
angle and particle type) $x_{\rm meas}$ with the vector of true values
$x_{\rm true}$ using a migration matrix: $x_{\rm meas} = {\sl M}_{\rm
migr} \times x_{\rm true}$.
The goal of the method is to compute a transformation 
(correction matrix) to obtain the expected
values for $x_{\rm true}$ from the measured ones. The most simple
and obvious solution, based on simple matrix inversion 
${\sl M}^{-1}_{\rm migr}$, 
is usually unstable and is dominated by large variances and strong negative
correlations between neighbouring bins.
In the method of D' Agostini, 
the correction matrix ${\sl M}^{\rm UFO}$ tries to connect
the measurement space (effects) with the space of the true values (causes) 
using an iterative Bayesian approach, based on Monte Carlo simulations to
estimate the probability for a given effect to be produced by a certain 
cause.}.

The Monte Carlo simulation of the HARP setup is based on 
GEANT4~\cite{ref:geant4}. 
The detector
materials are accurately 
described
in this simulation as well as the
relevant features of the detector response and the digitization
process. All relevant physics processes are considered, including
multiple scattering, energy loss, absorption and
re-interactions. 
The track reconstruction used in this analysis and the simulation are
identical to the ones used for the $\pi^+$ production in p-Be
collisions~\cite{ref:harp:bePaper}. 
A detailed description of the corrections and their magnitude can be
found there. 

The reconstruction efficiency (inside the geometrical acceptance) is
larger than 95\% above 1.5~\GeVc and drops to 80\% at 0.5~\GeVc. 
The requirement of a match with a TOFW hit has an efficiency between
90\% and 95\% independent of momentum.
The electron veto rejects about 1\% of the pions and protons below
3~\GeVc with a remaining background of less than 0.5\%.
Below Cherenkov threshold the TOFW separates pions and protons with
negligible background and an efficiency of $\approx$94\% for protons at
low momentum increasing to $\approx$98\% at threshold.
Above Cherenkov threshold the efficiency for protons is greater than 98\%
with less than 1\% of pions mis-identified as protons.

The absorption and decay of particles is simulated by the Monte Carlo.
The absorption correction is on average 20\%, approximately independent
of momentum.
Uncertainties in the absorption of secondaries in the dipole
spectrometer material are taken into account by
a variation of 10\% of this effect in the simulation. 
The uncertainty in the production of background due to tertiary
particles is larger. 
The average correction is $\approx$10\% and up to 20\% at
1~\GeVc. 
The correction includes reinteractions in the detector material as well
as a small component coming from reinteractions in the target.
The validity of the generators used in the simulation was checked by an
analysis of HARP data with incoming protons and charged pions on
aluminium and carbon targets at lower momenta (3~\GeVc and 5~\GeVc).
A 30\% variation of the secondary production was applied.
The average empty-target subtraction amounts to $\approx$20\%.


Owing to the redundancy of the tracking system downstream of the
target the detection efficiency is very robust under the usual
variations of the detector performance during the long data taking
periods. 
Since the momentum is reconstructed without making use of the upstream
drift chamber module (which is more sensitive in its performance to the beam
intensity) the reconstruction efficiency is uniquely determined by the
downstream system.
No variation of the overall efficiency has been observed.
The performance of the TOFW and CHE system have been monitored to be
constant for the data taking periods used in this analysis.
The calibration of the detectors was performed on a day-by-day basis.

\subsection{Error estimation}
\label{errorest}

The total statistical error of the corrected data is composed of the
statistical error of the raw data and of the statistical error
of the unfolding procedure, as the unfolding matrix is obtained
from the data themselves, thus contributing also to the statistical
error. The statistical error provided by the unfolding program is
equivalent to the propagated statistical error of the raw data. In
order to calculate the statistical error of the unfolding procedure a
separate analysis is applied,
as described in~\cite{ref:harp:carbonfw},\cite{ref:grossheim}. 
Its conclusion is that the statistical error provided by the unfolding
procedure has to be multiplied globally by a factor of 2, which is done
for the analyses described here.
This factor is somewhat dependent on the shape of the distributions.

Different types of sources 
induce
systematic errors for the analysis
described here: 
track yield corrections ($\sim 5 \%$), particle identification ($\sim 0.1 \%$),
momentum and angular reconstruction ($\sim 1 \%$)~\footnote{
The quoted error in parenthesis refers to fractional error 
of the integrated cross-section ($\delta^{\pi}_{\rm int} (\%)$)
in the kinematic range covered by the HARP experiment}.
The strategy to calculate these systematic errors and the different
methods used for their evaluation are fully described in~\cite{ref:harp:carbonfw}.
As a result of these systematic error studies, each error source 
can be represented by a covariance matrix. The
sum of these matrices describes the total systematic error,
as explained in~\cite{ref:harp:carbonfw}. 

\begin{figure*}[htbp]
  \begin{center}
\includegraphics[width=0.49\textwidth]{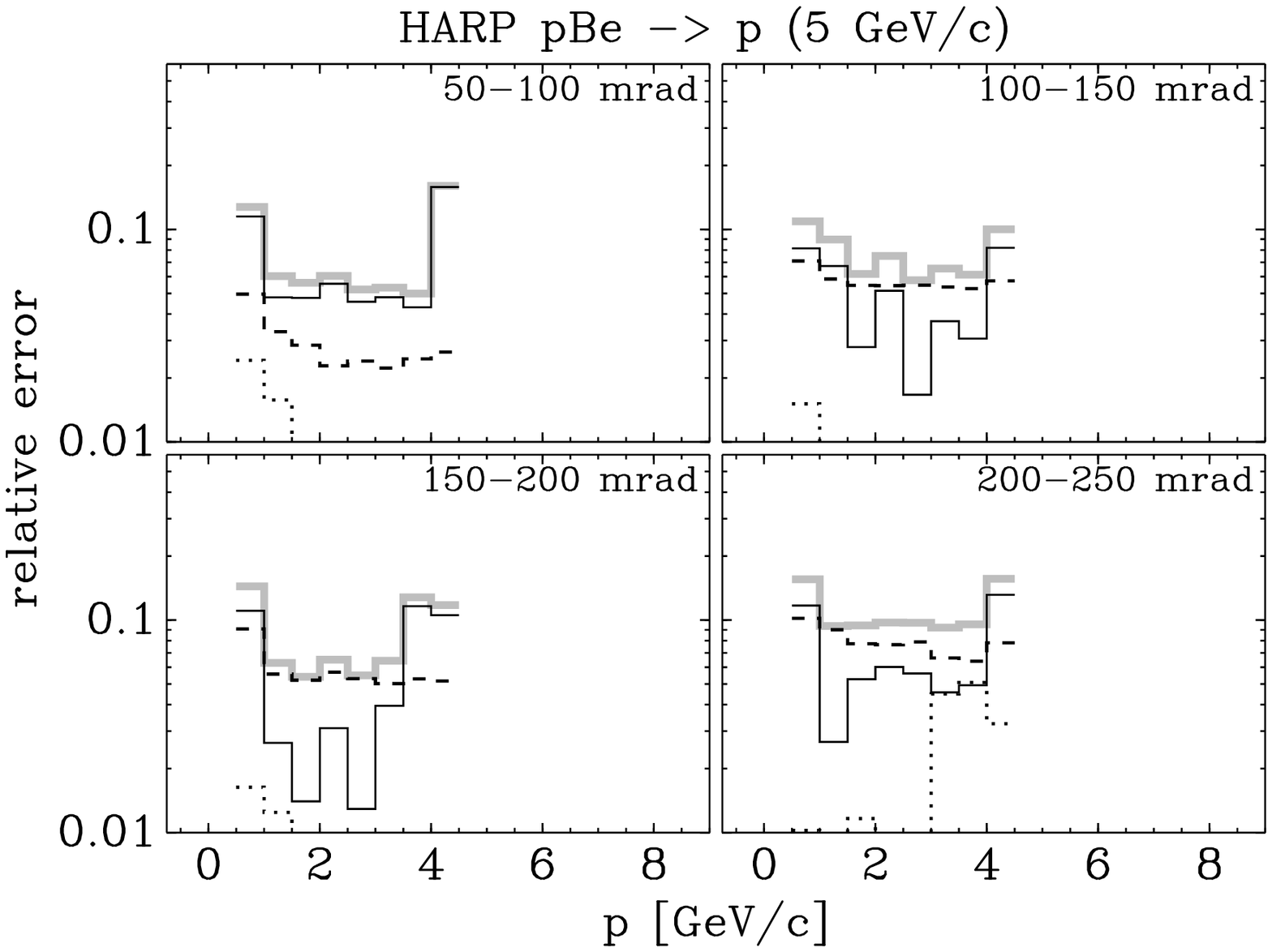}
\includegraphics[width=0.49\textwidth]{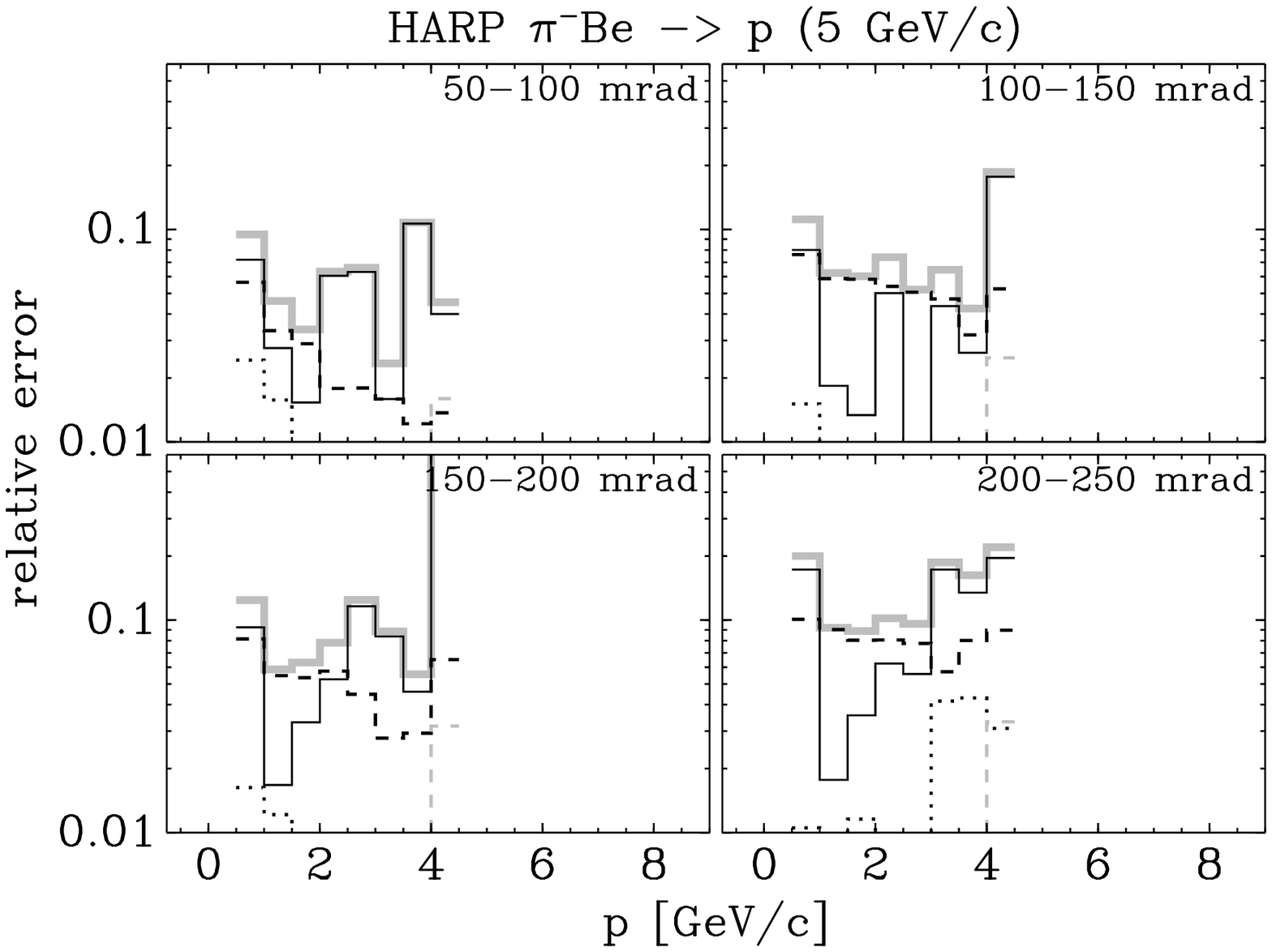}
\includegraphics[width=0.49\textwidth]{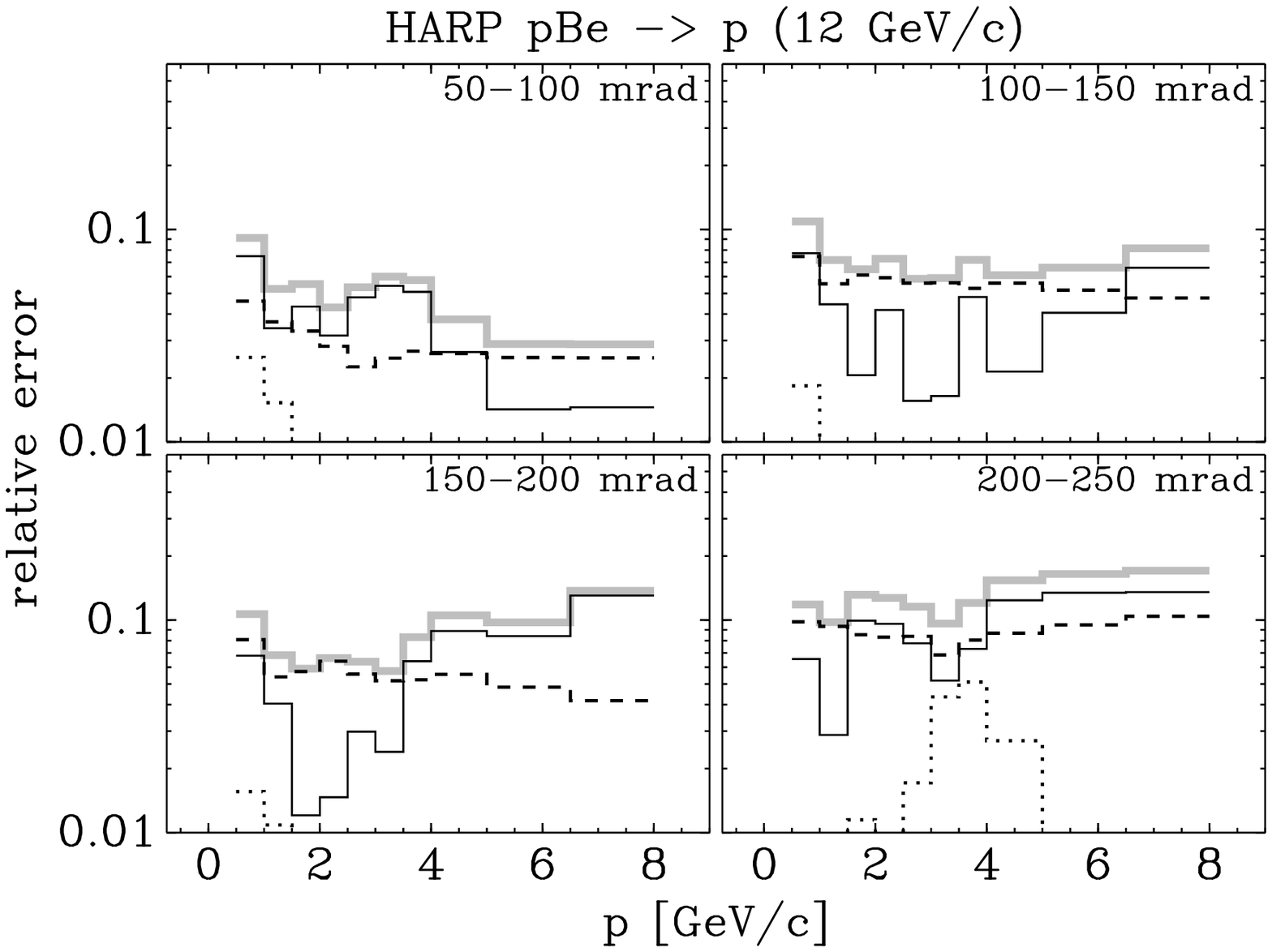}
\includegraphics[width=0.49\textwidth]{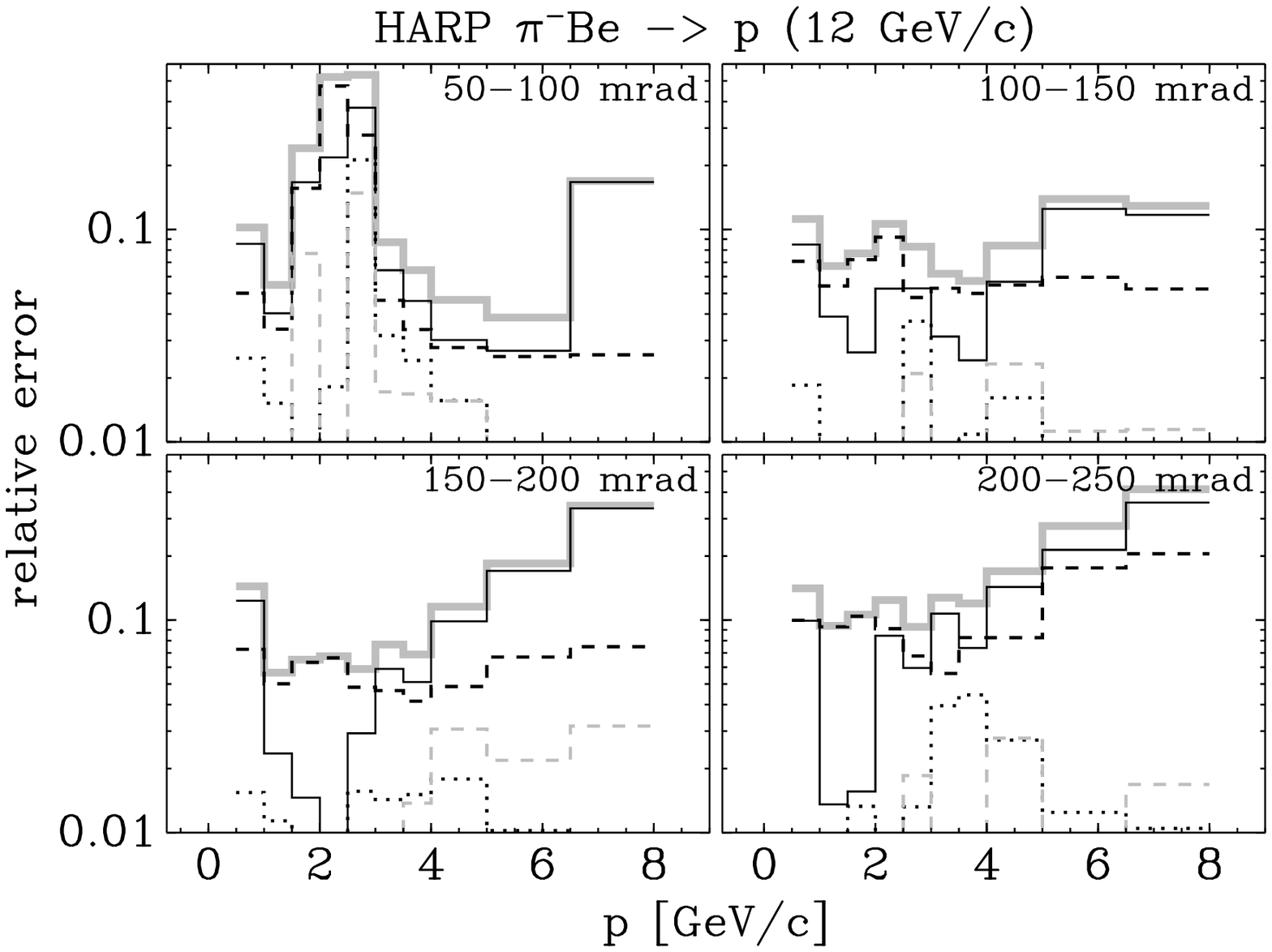}

\end{center}
\caption{Total systematic error (grey solid line) and main components
for a typical (Be) target with incident p and $\pi^{-}$
beams at 5 and 12~\GeVc
black short-dashed line for absorption+tertiares interactions,
black dotted line for track efficiency and target pointing efficiency,
black dot-dashed line for $\pi^{0}$ subtraction, 
black solid line for
momentum scale+resolution and angle scale, grey short-dashed line
for PID.
}
\label{fig:syst}
\end{figure*}

The experimental uncertainties are shown for a typical target (Be)
in Figure~\ref{fig:syst} at 5~\GeVc and 12~\GeVc 
incident beam momenta for incident protons and negative pions.
They are very similar for $\pi^{+}$ and at the other beam energies. 
Going from lighter (Be, C) to heavier targets (Ta, Pb)
the corrections for $\pi^{0}$ (conversion) and absorption/tertiares increase.

On average the total integrated systematic error is around $5-6\%$,
with a differential bin-to-bin systematic error of the order of
$10-11 \%$, to be compared with a statistical integrated (bin-to-bin
differential) error of $\sim 2-3 \%$ ($\sim 10-13 \%$).
Systematic and statistical errors are roughly of the same order. 

\section{Experimental results}
\label{sec:results}

\begin{figure*}[tb]
\centering
\includegraphics[width=.49\textwidth]{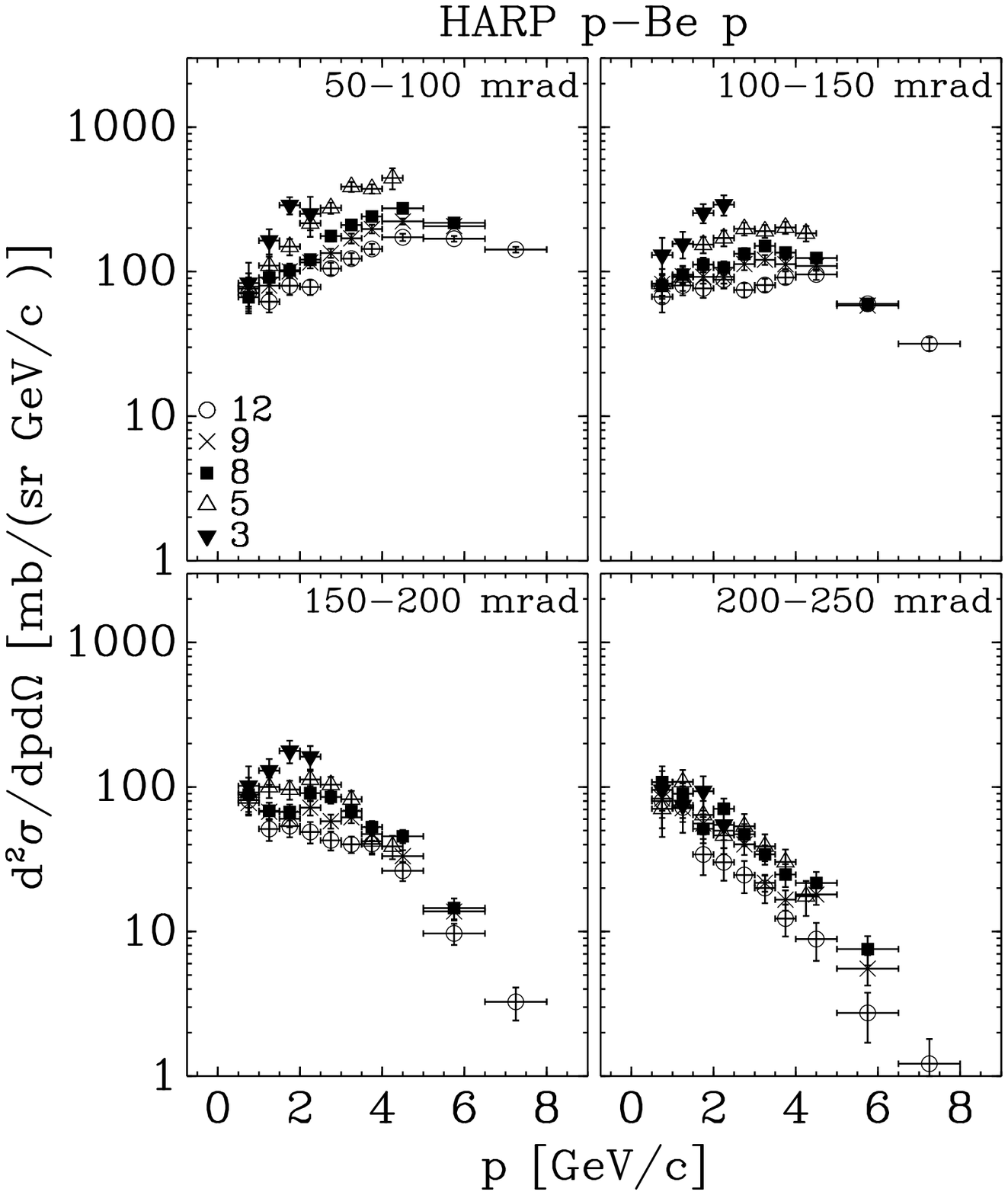}
\includegraphics[width=.49\textwidth]{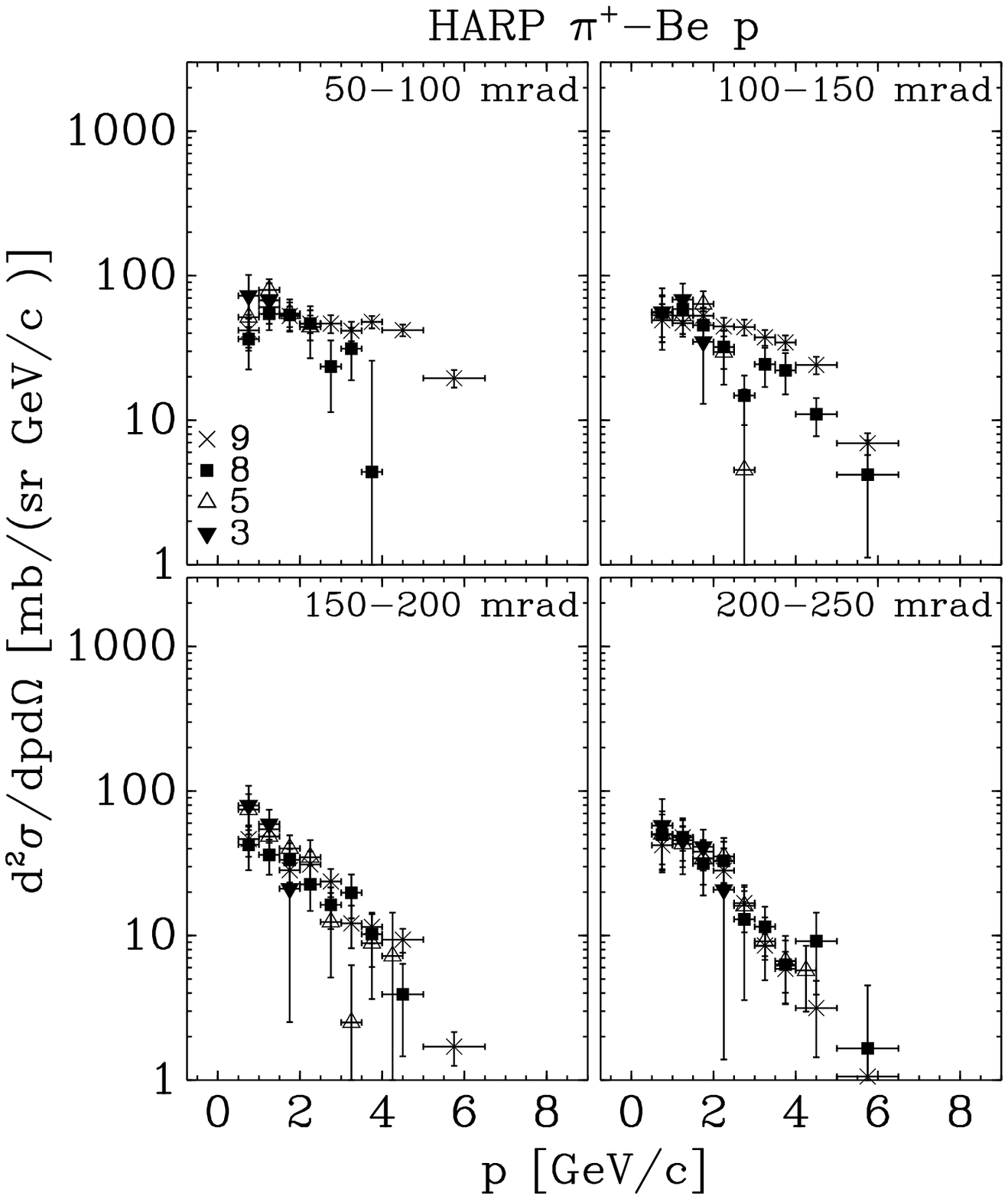}
\includegraphics[width=.49\textwidth]{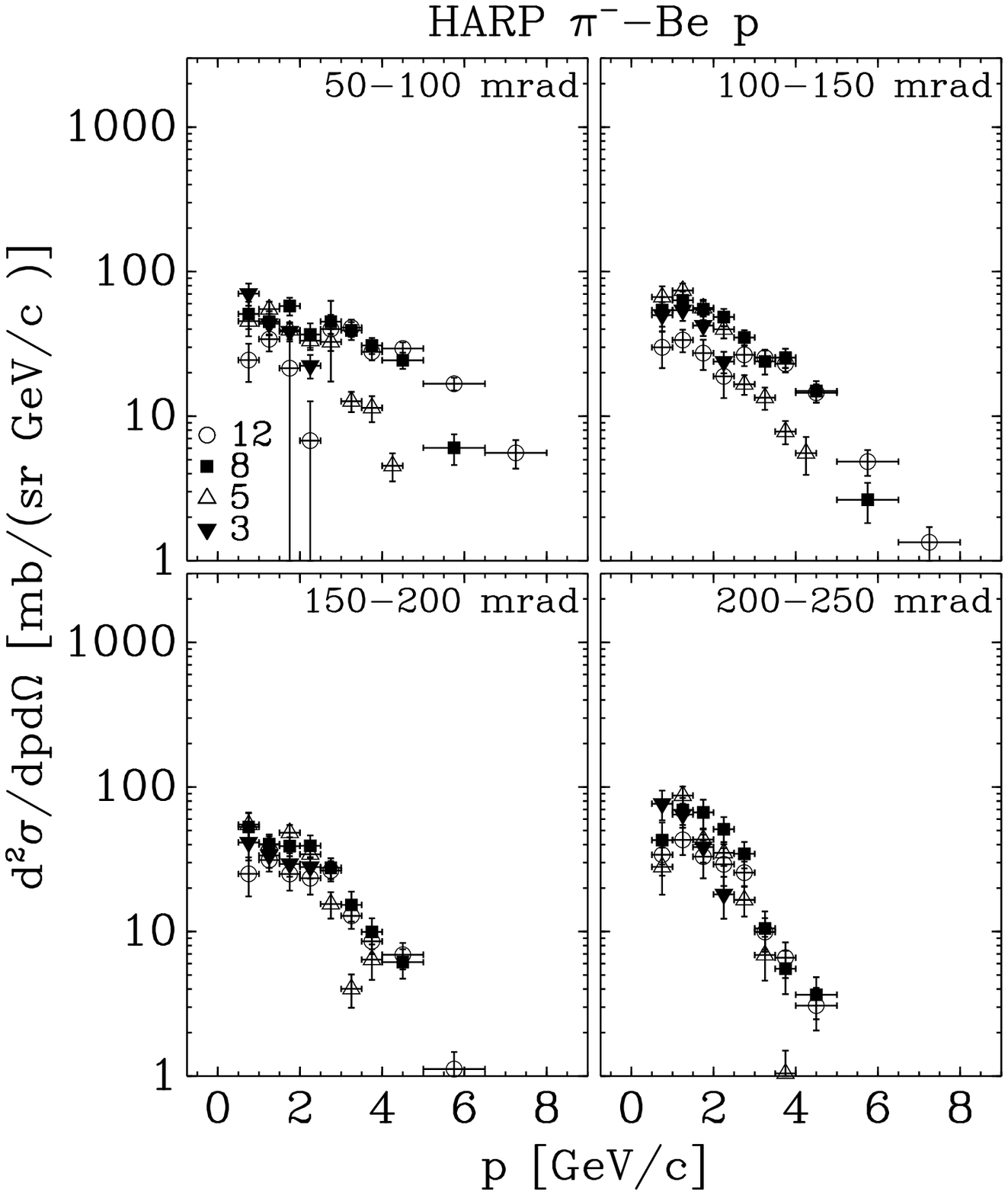}
\caption{Differential cross sections for proton forward production with incident p, $\pi^{\pm}$ on a 
thin Be target.   
In the top right corner of each plot the 
covered angular range is shown in mrad.}
\label{fig:Be}
\end{figure*}

\begin{figure*}[tb]
\centering
\includegraphics[width=.49\textwidth]{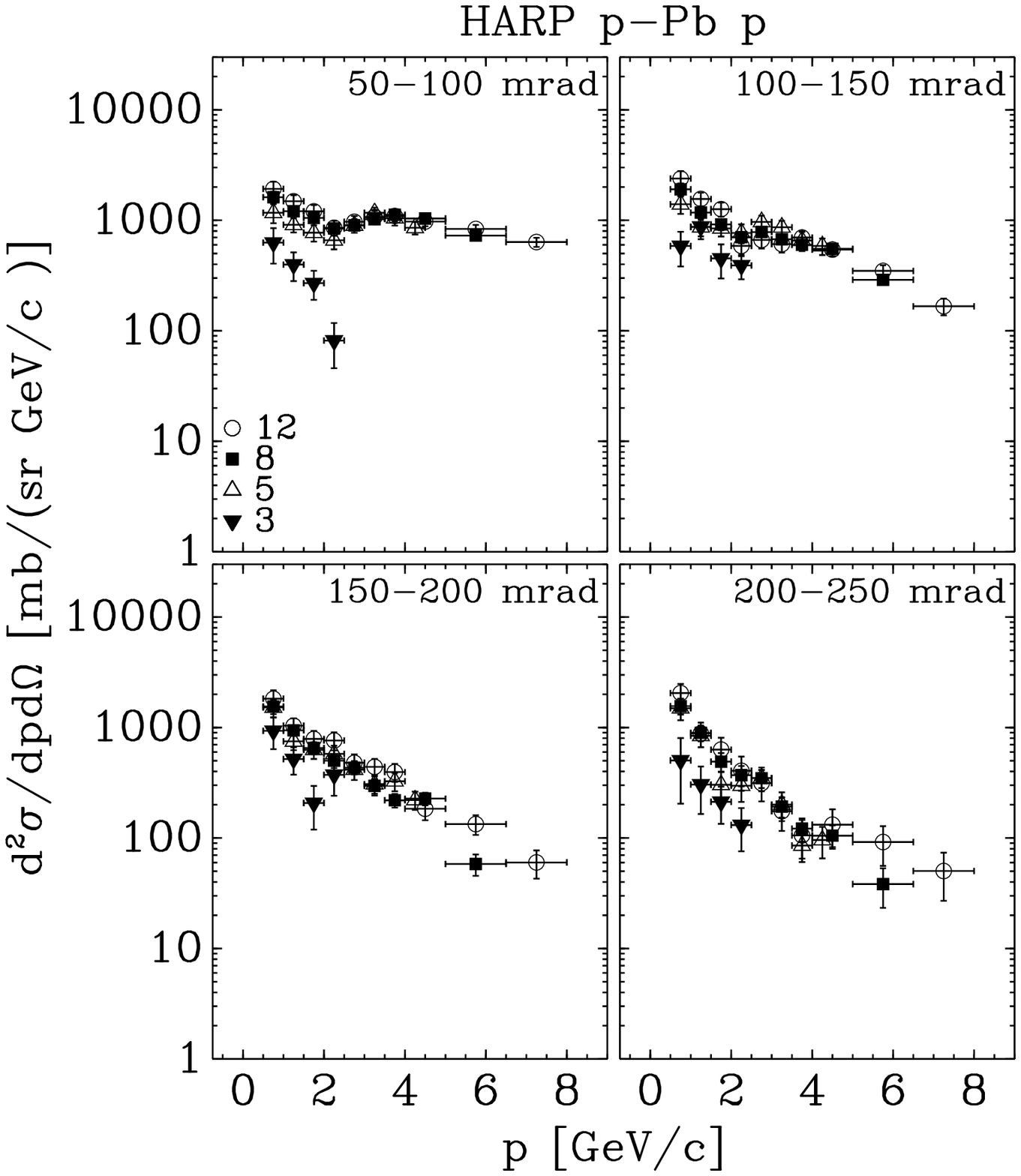}
\includegraphics[width=.49\textwidth]{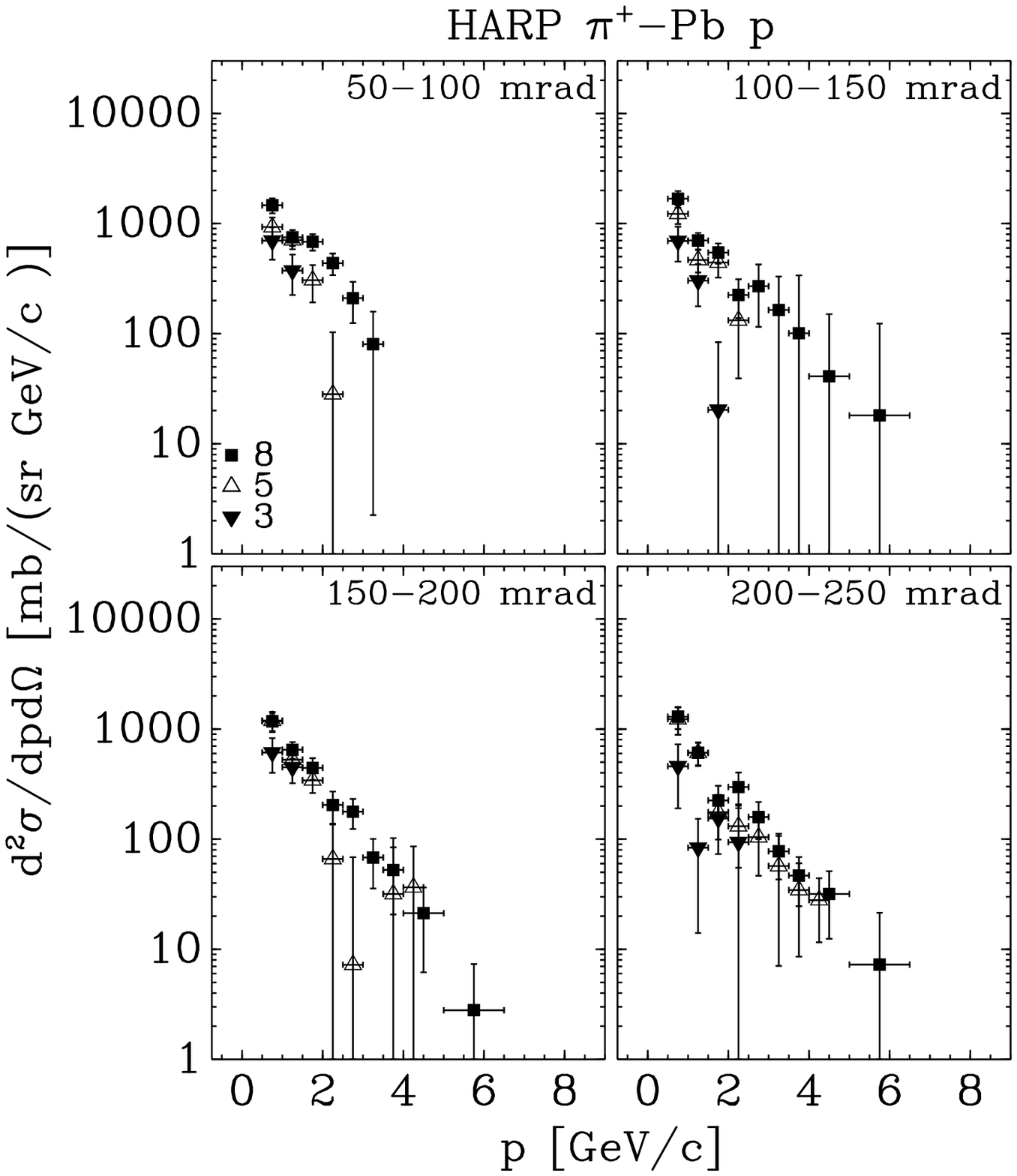}
\includegraphics[width=.49\textwidth]{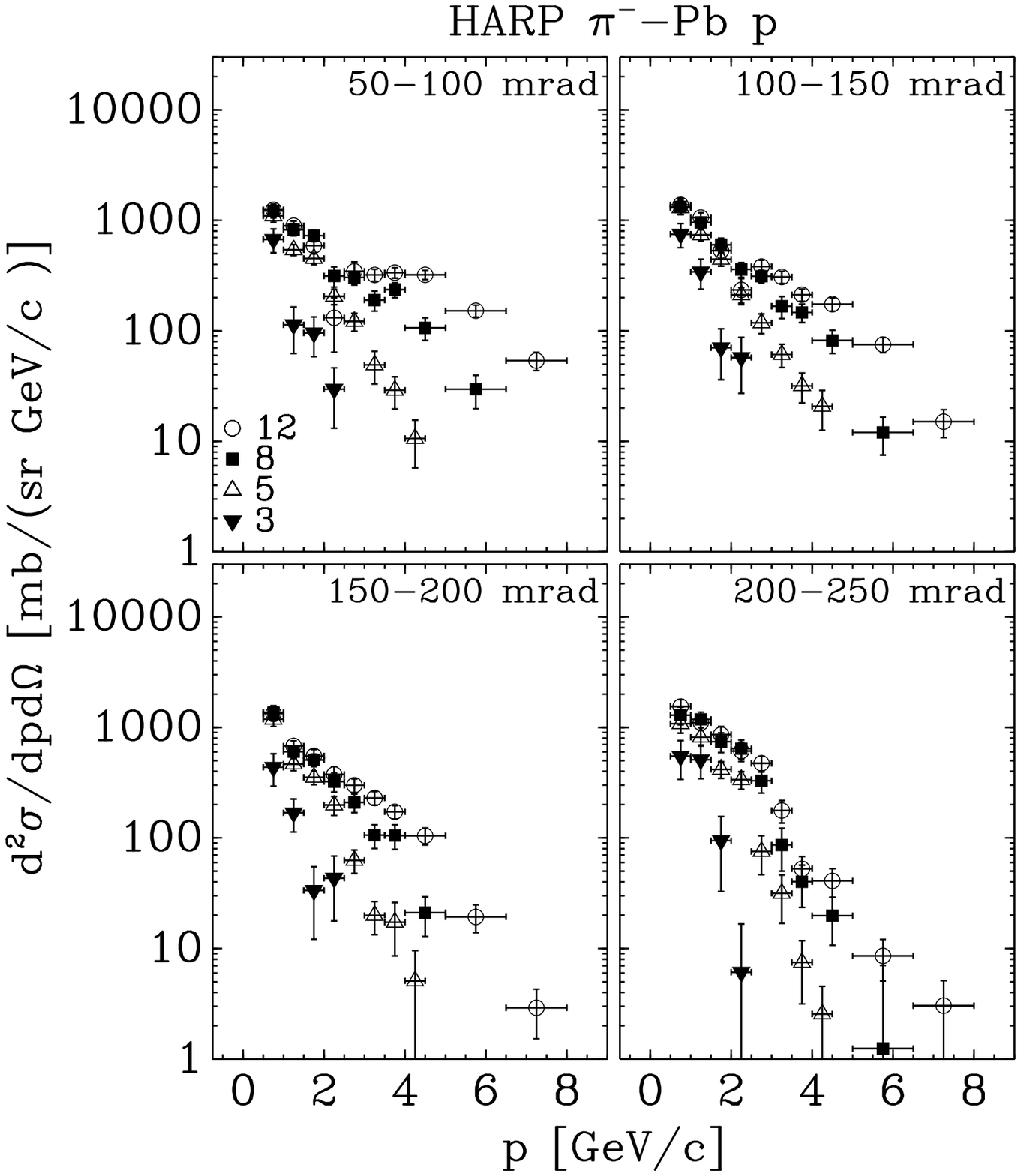}
\caption{Differential cross sections for proton forward production with incident p, $\pi^{\pm}$ on 
a thin Pb target.
In the top right corner of each plot the 
covered angular range is shown in mrad.}
\label{fig:Pb}
\end{figure*}

The measured double-differential cross-sections for the
production of forward protons in the laboratory system as a function of
the momentum and the polar angle for each incident beam momentum are
shown in Figures \ref{fig:Be} and \ref{fig:Pb} for two typical solid
targets: beryllium and lead, as an example of a light and a heavy
target. 
The error bars  shown are the
square-roots of the diagonal elements in the covariance matrix,
where statistical and systematic uncertainties are combined
in quadrature.
The correlation of the statistical errors (introduced by the unfolding
procedure) are typically smaller than 20\% for adjacent momentum bins and
even smaller for adjacent angular bins.
The correlations of the systematic errors are larger, typically 80\% for
adjacent bins.
The overall scale error ($<2\%$) is not shown.
The results of the analysis for all solid targets 
are fully tabulated in Appendix A~\footnote{
the scale error has not been included to make it possible to calculate e.g.
integrated particle ratios taking it into account only when applicable, 
i.e. when different beams are compared.}.

\begin{figure*}[tbp]
\begin{center}
  \includegraphics[width=0.42\textwidth]{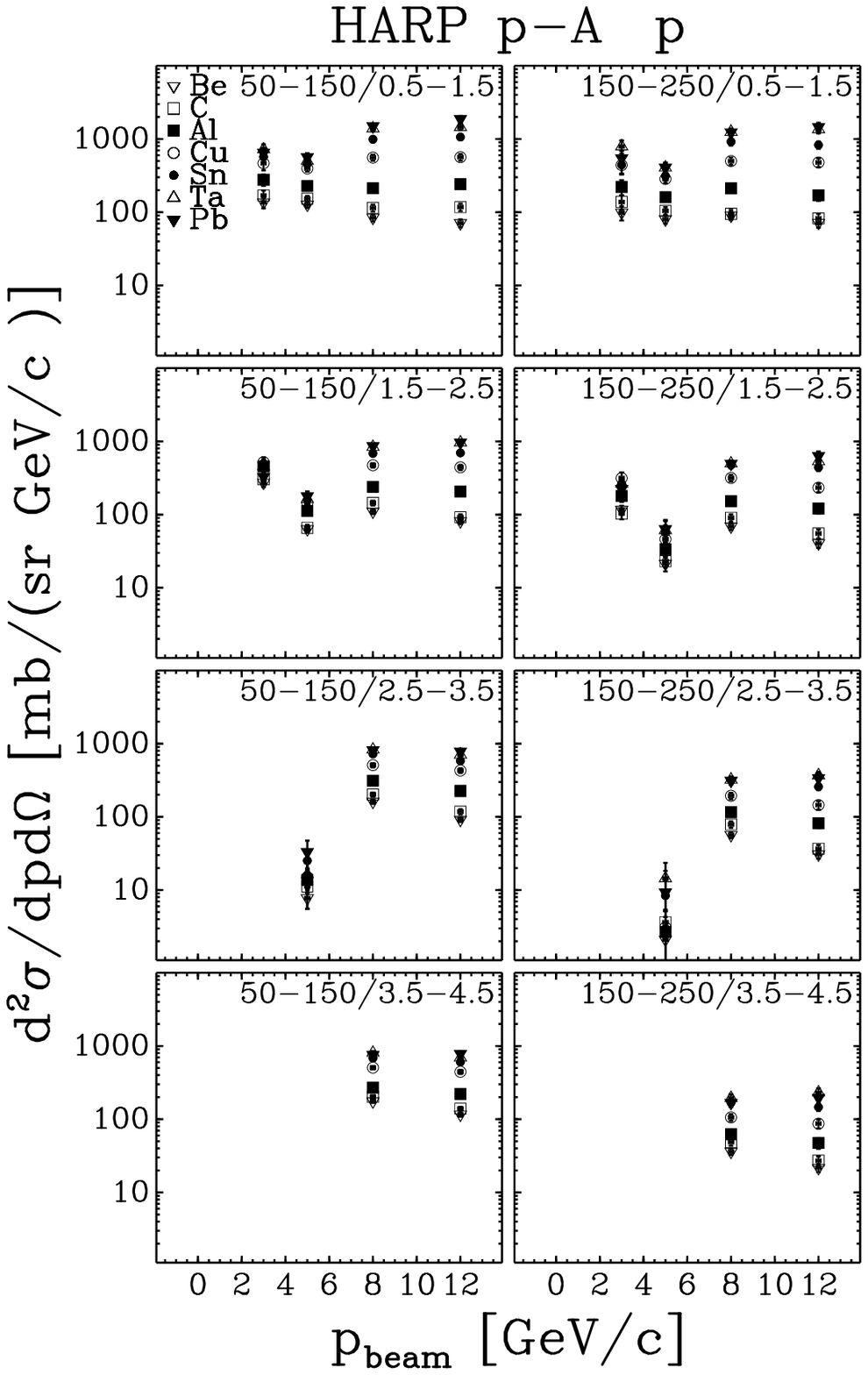} \\
  \includegraphics[width=0.42\textwidth]{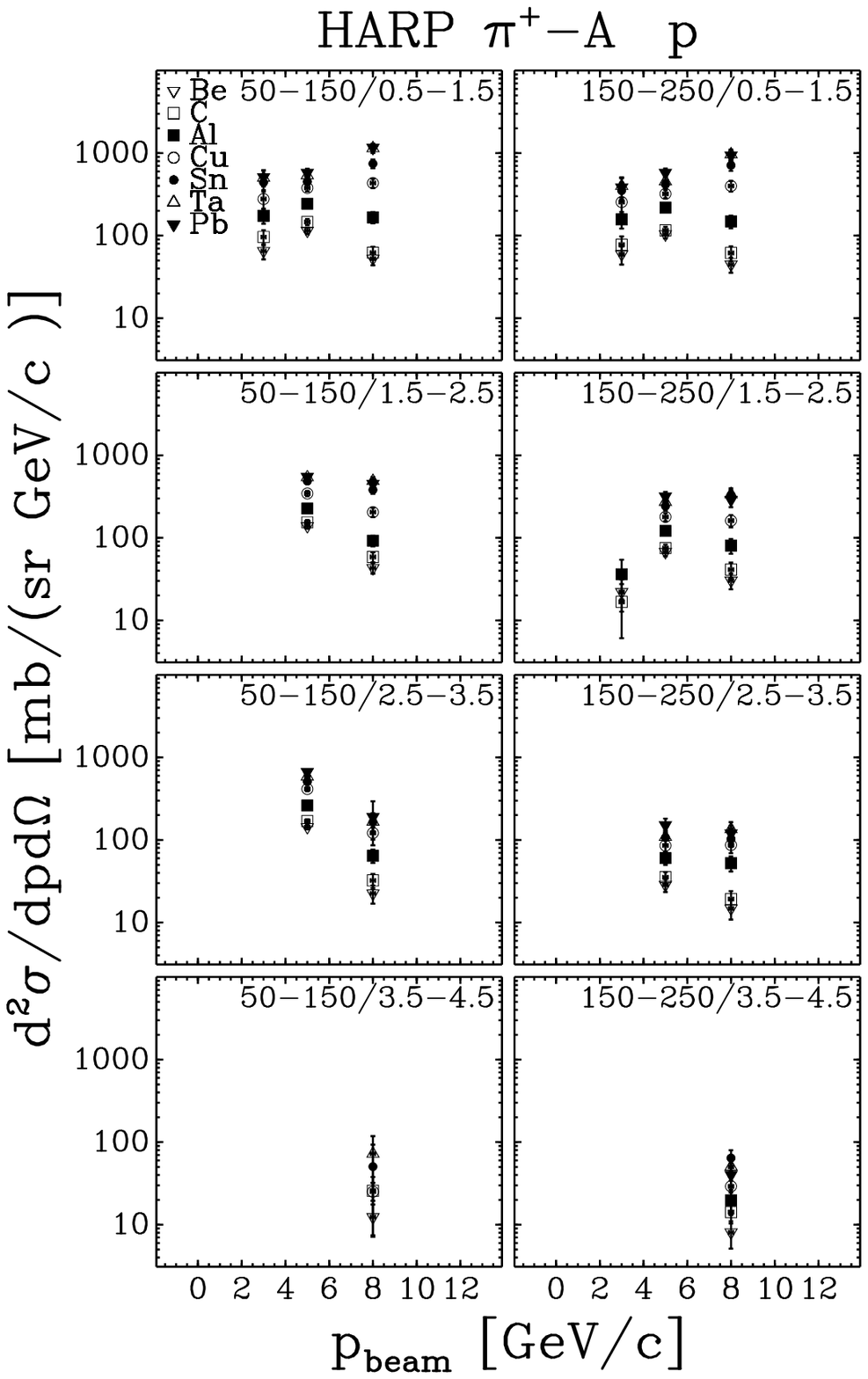}
  \includegraphics[width=0.42\textwidth]{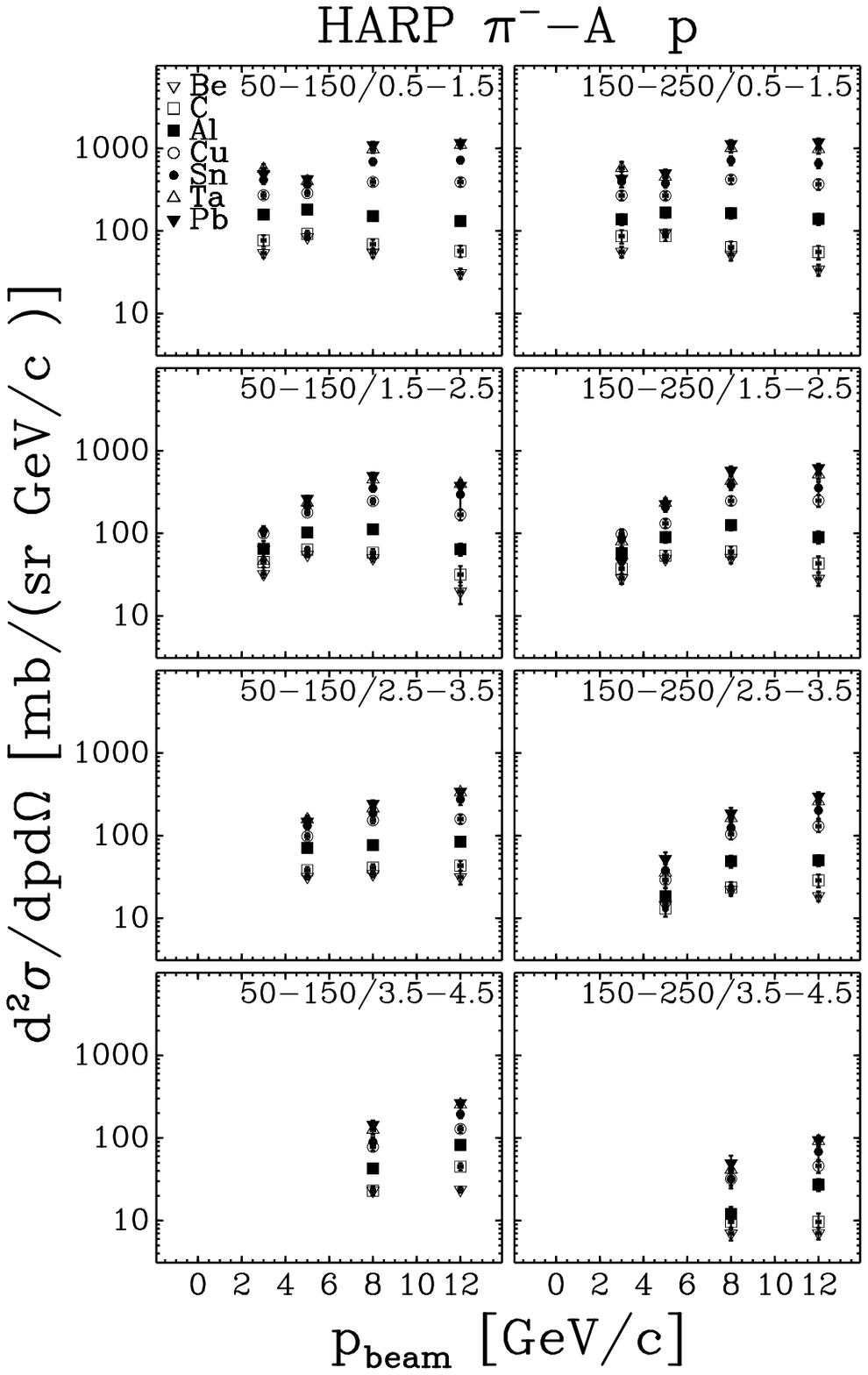}
\end{center}
\caption{
 The dependence on the beam momentum of the forward proton
  production yields
 in p--A and $\pi^{\pm}$--A (A = Be,C,Al,Cu,Sn,Ta, Pb)
 interactions averaged over two forward angular regions
 ($0.05~\rad \leq \theta < 0.15~\rad$ and
  $0.15~\rad \leq \theta < 0.25~\rad$)
 and four momentum regions
  ($0.5~\GeVc \leq p < 1.5~\GeV/c$,
   $1.5~\GeVc \leq p < 2.5~\GeV/c$,
   $2.5~\GeVc \leq p < 3.5~\GeV/c$ and
   $3.5~\GeVc \leq p < 4.5~\GeV/c$), for the four different
  incoming beam energies.
}
\label{fig:xs-trend}
\end{figure*}

The dependence of the averaged proton yields on the incident beam
momentum is shown in Fig.~\ref{fig:xs-trend}.
The proton yields, averaged over two angular regions
 ($0.05~\rad \leq \theta < 0.15~\rad$ and 
  $0.15~\rad \leq \theta < 0.25~\rad$)
 and four momentum regions 
  ($0.5~\GeVc \leq p < 1.5~\GeV/c$,
   $1.5~\GeVc \leq p < 2.5~\GeV/c$,
   $2.5~\GeVc \leq p < 3.5~\GeV/c$ and
   $3.5~\GeVc \leq p < 4.5~\GeV/c$),
are
shown.

\begin{figure*}[tbp]
\begin{center}
  \includegraphics[width=0.42\textwidth]{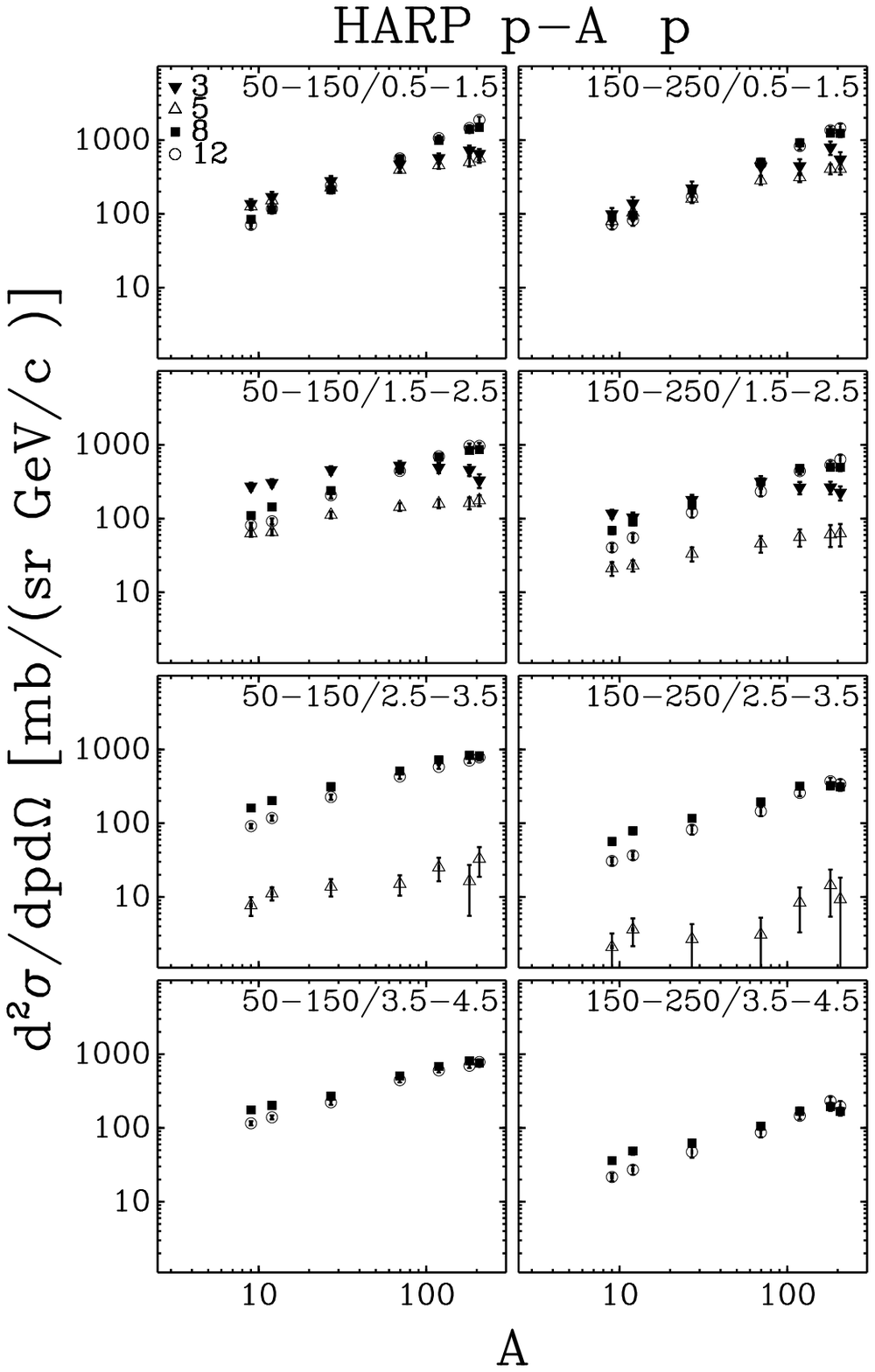} \\
  \includegraphics[width=0.42\textwidth]{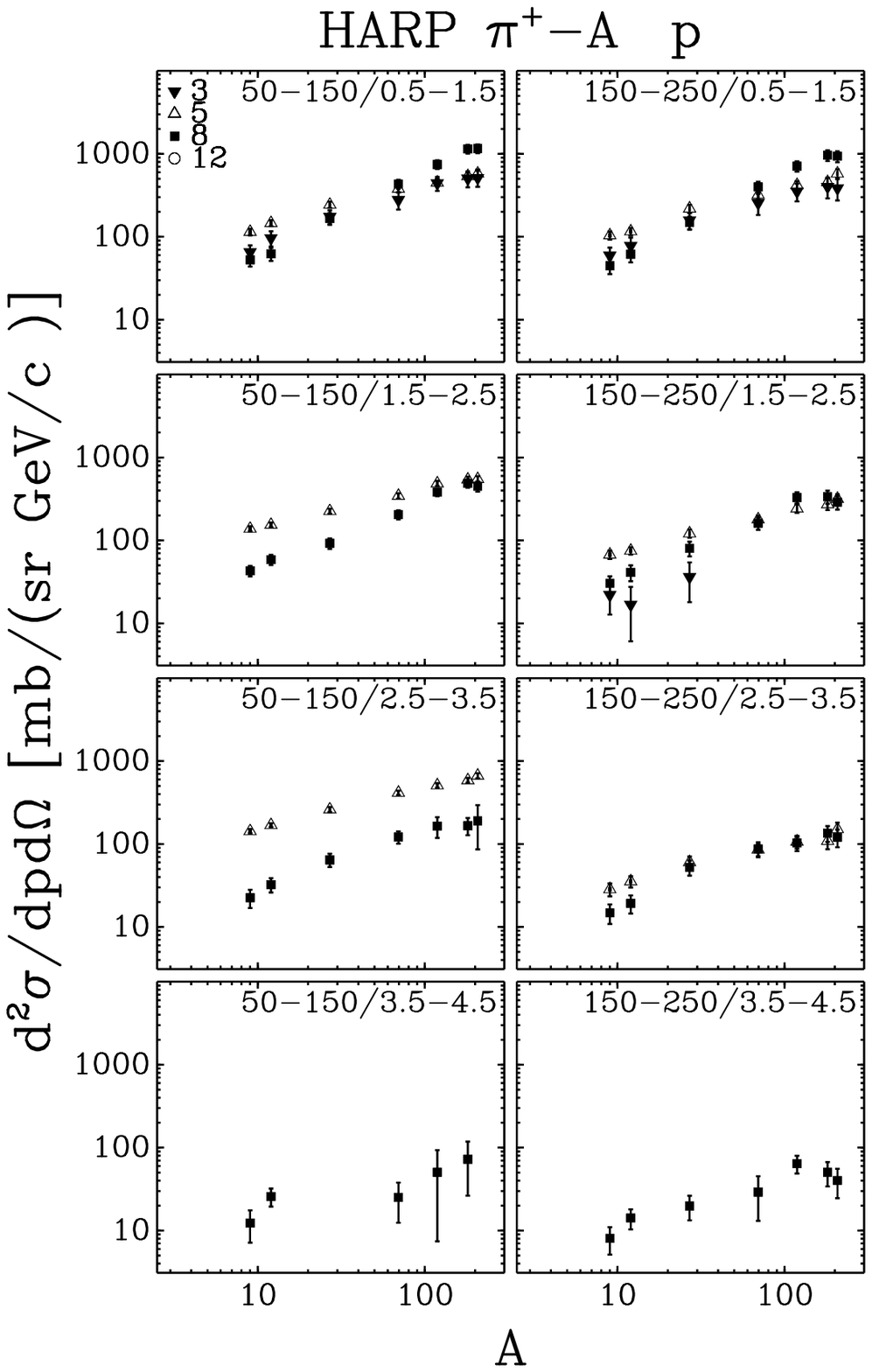}
  \includegraphics[width=0.42\textwidth]{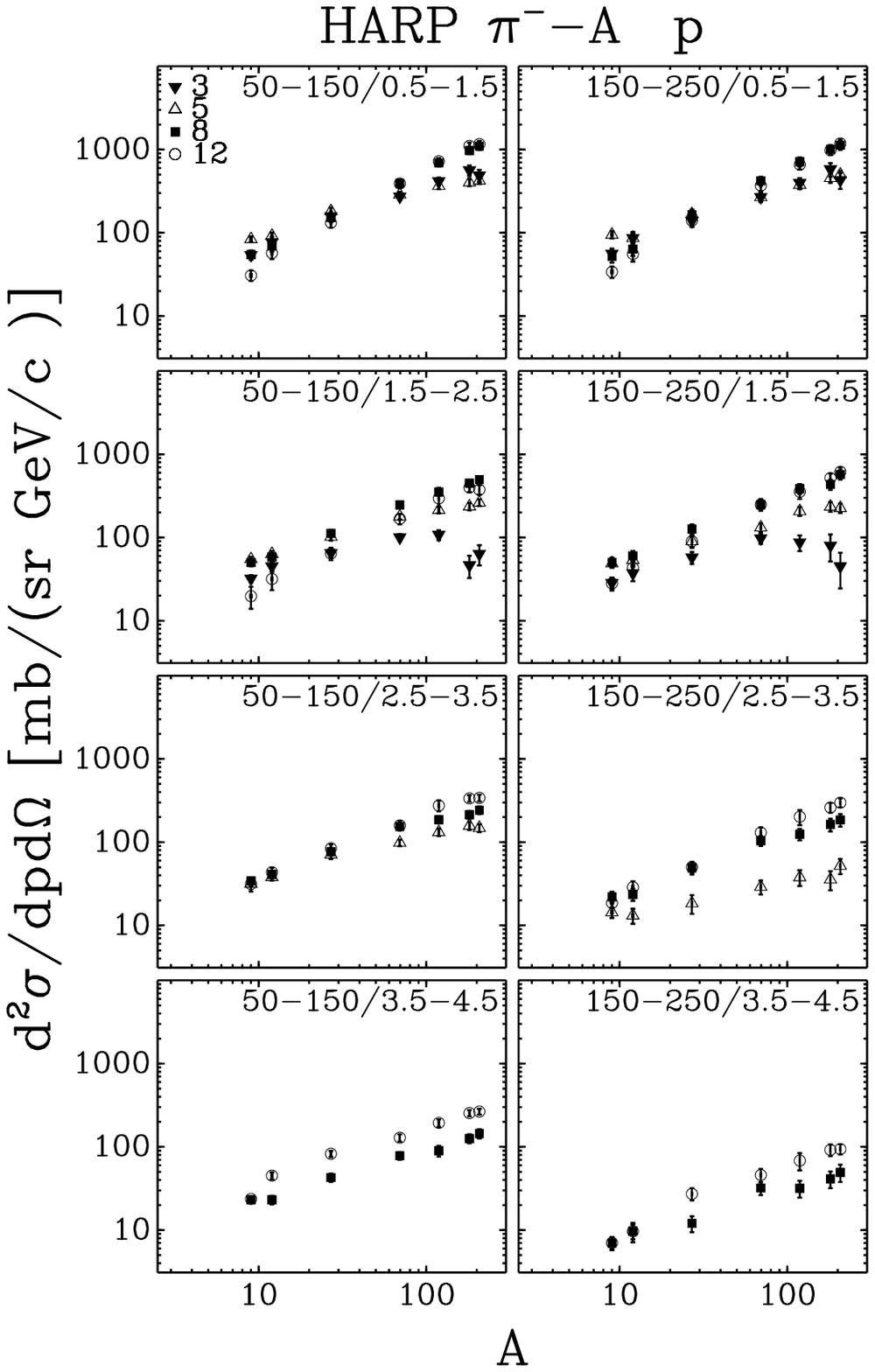}
\end{center}
\caption{
 The dependence on the atomic number $A$ of the forward proton production yields
 in $p$--A and $\pi^{\pm}$--A (A=Be,Al,C,Cu,Sn,Ta,Pb)
 interactions averaged over two forward angular regions
 ($0.05~\rad \leq \theta < 0.15~\rad$ and 
  $0.15~\rad \leq \theta < 0.25~\rad$)
 and four momentum regions 
  ($0.5~\GeVc \leq p < 1.5~\GeV/c$,
   $1.5~\GeVc \leq p < 2.5~\GeV/c$,
   $2.5~\GeVc \leq p < 3.5~\GeV/c$ and
   $3.5~\GeVc \leq p < 4.5~\GeV/c$), for the four different
  incoming beam energies.
}
\label{fig:xs-a-dep}
\end{figure*}

The dependence of the averaged 
proton yields on the atomic number $A$ is
shown in Fig.~\ref{fig:xs-a-dep}.
The yields are shown, averaged over two angular regions
 ($0.05~\rad \leq \theta < 0.15~\rad$ and 
  $0.15~\rad \leq \theta < 0.25~\rad$)
 and four momentum regions 
  ($0.5~\GeVc \leq p < 1.5~\GeV/c$,
   $1.5~\GeVc \leq p < 2.5~\GeV/c$,
   $2.5~\GeVc \leq p < 3.5~\GeV/c$ and
   $3.5~\GeVc \leq p < 4.5~\GeV/c$).

\begin{figure*}[tb]
\centering
\includegraphics[width=.26\textwidth]
{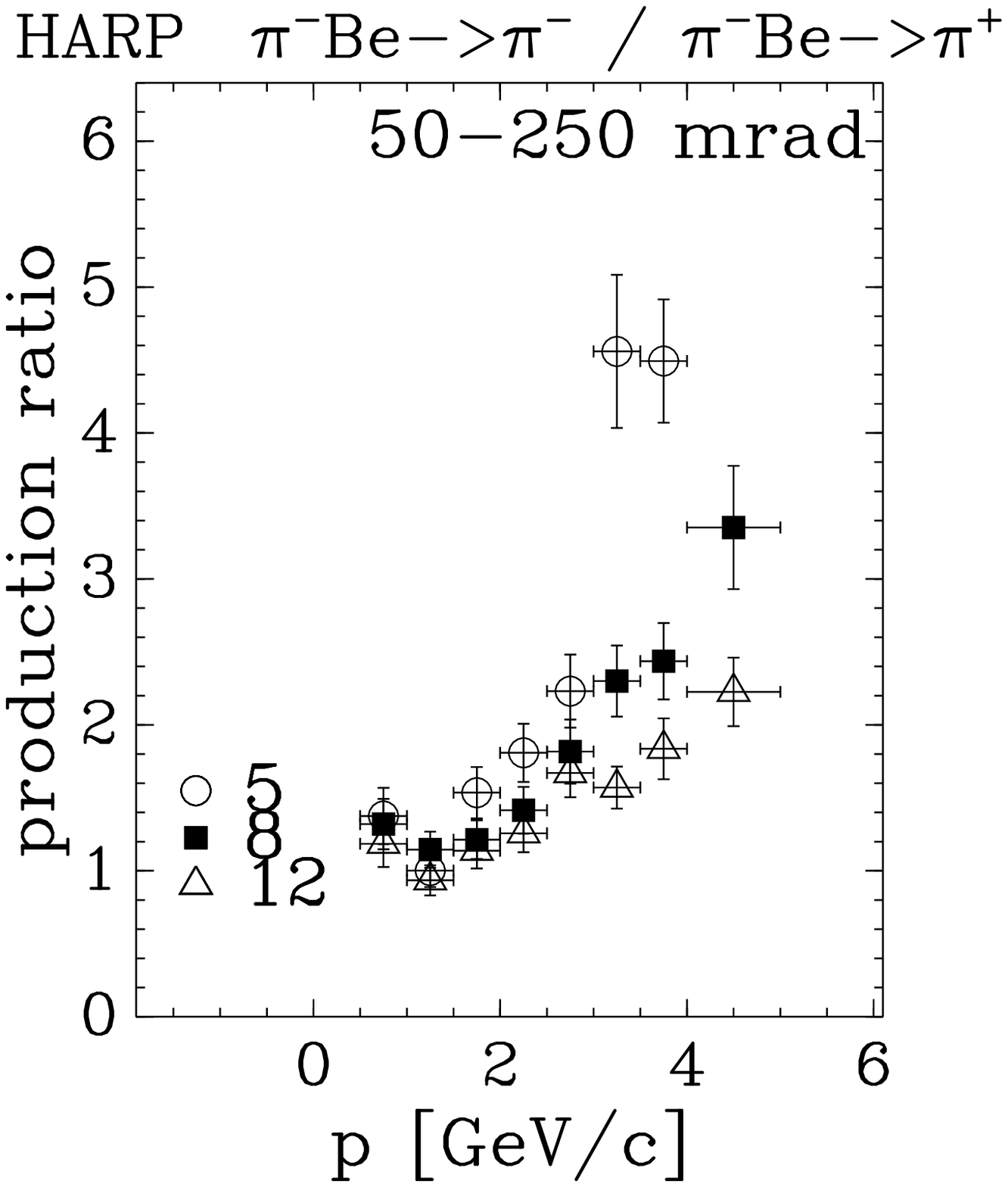}
\includegraphics[width=.28\textwidth]
{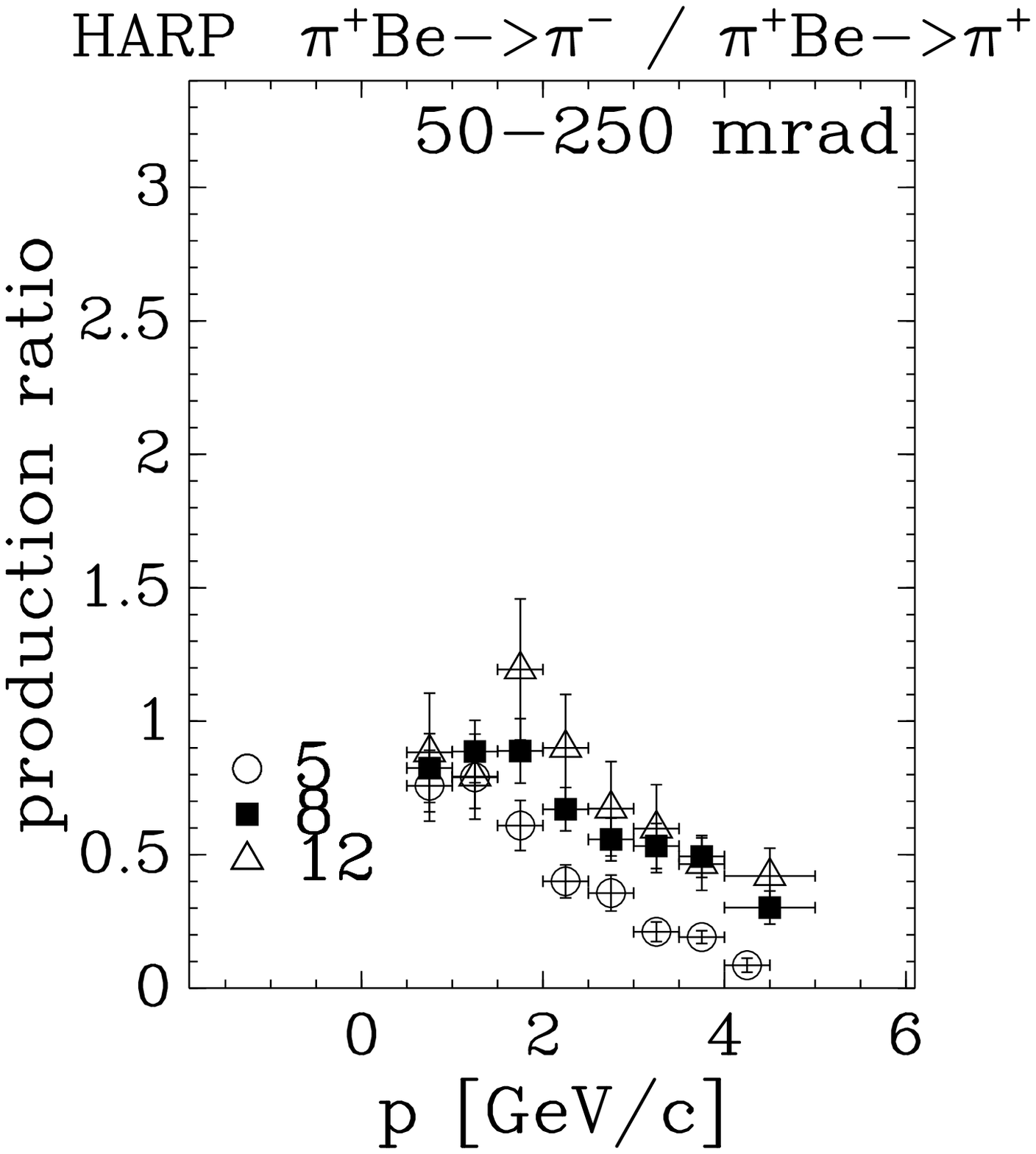}
\includegraphics[width=.28\textwidth]
{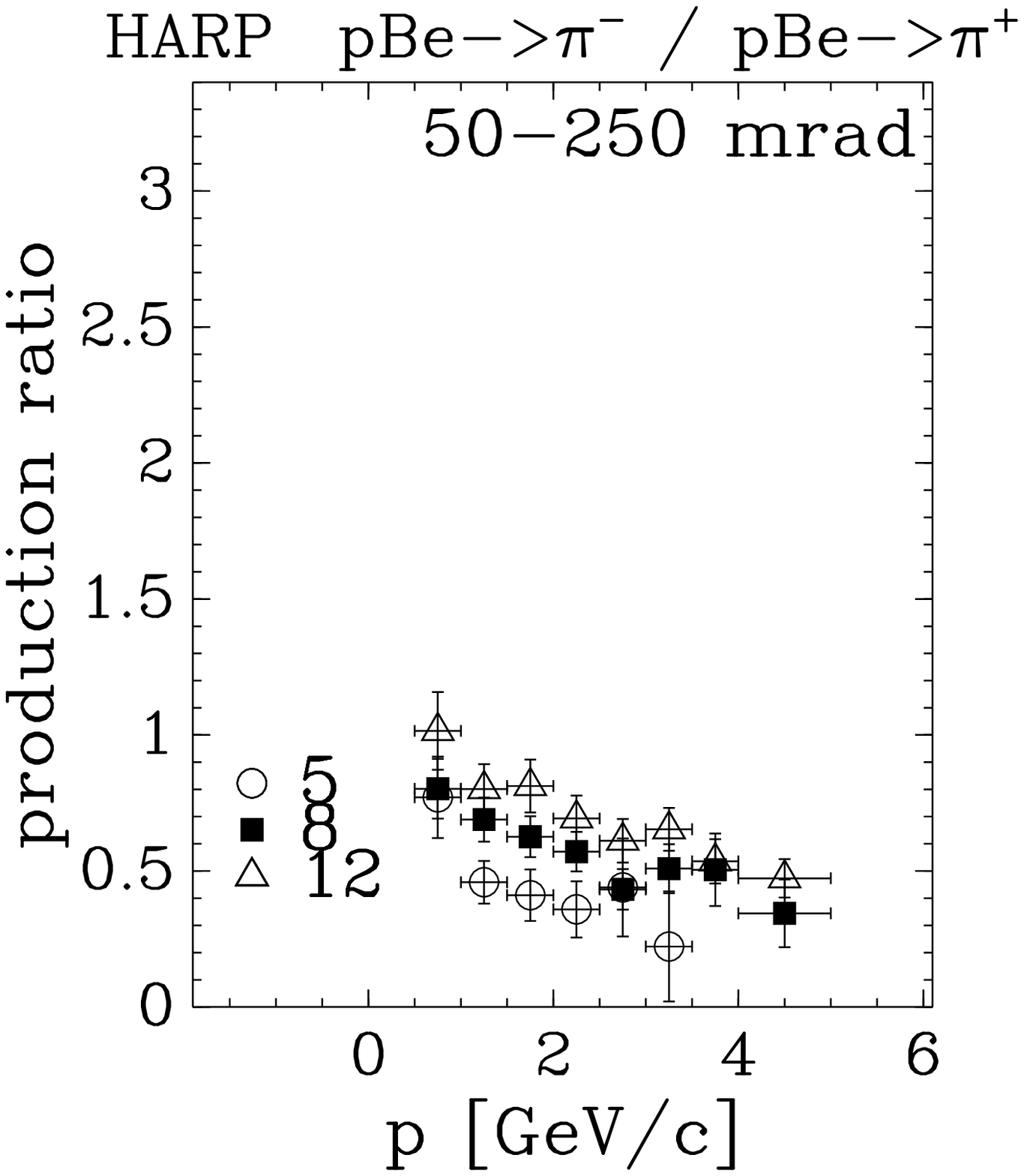}
\includegraphics[width=.26\textwidth]
{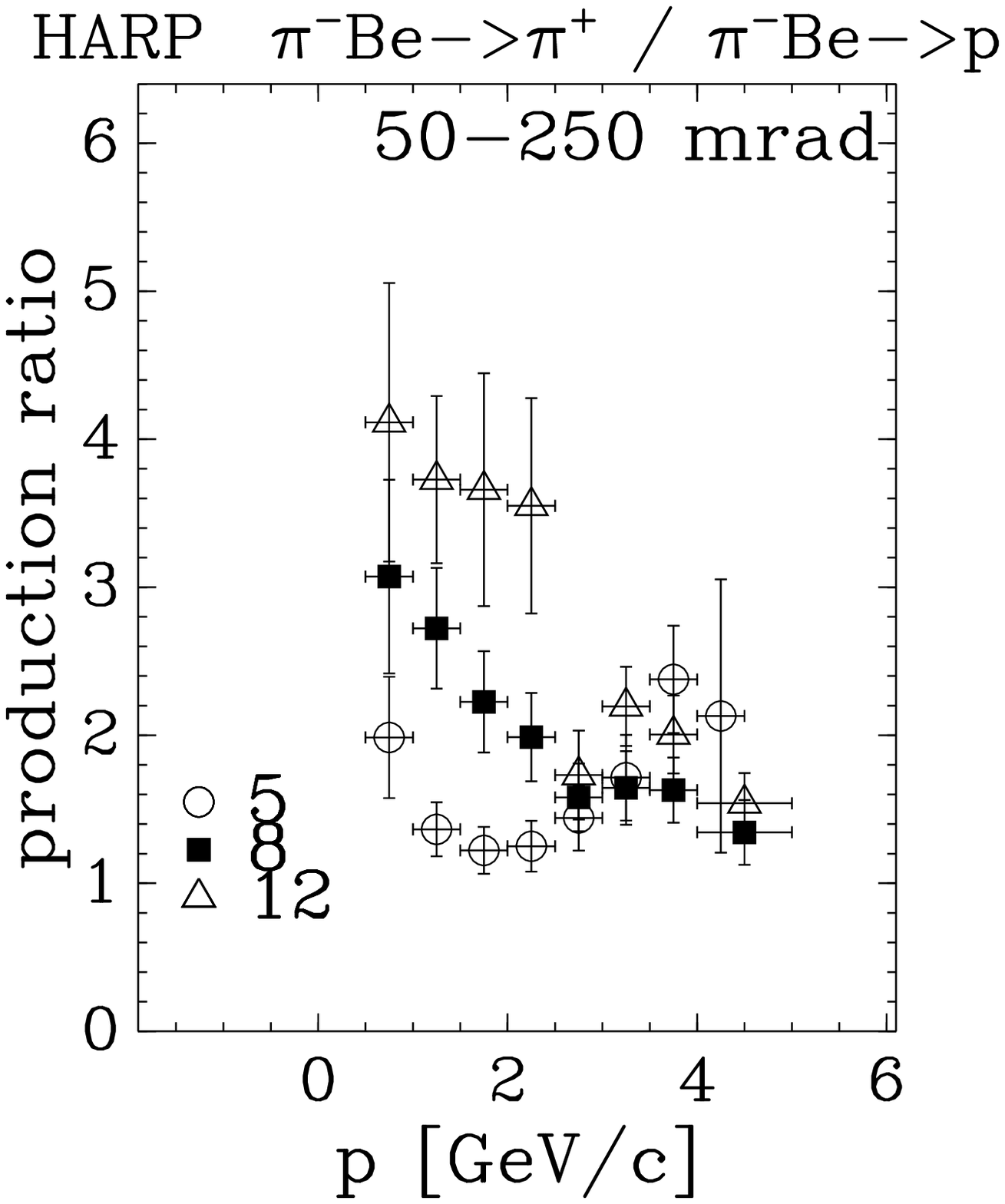}
\includegraphics[width=.28\textwidth]
{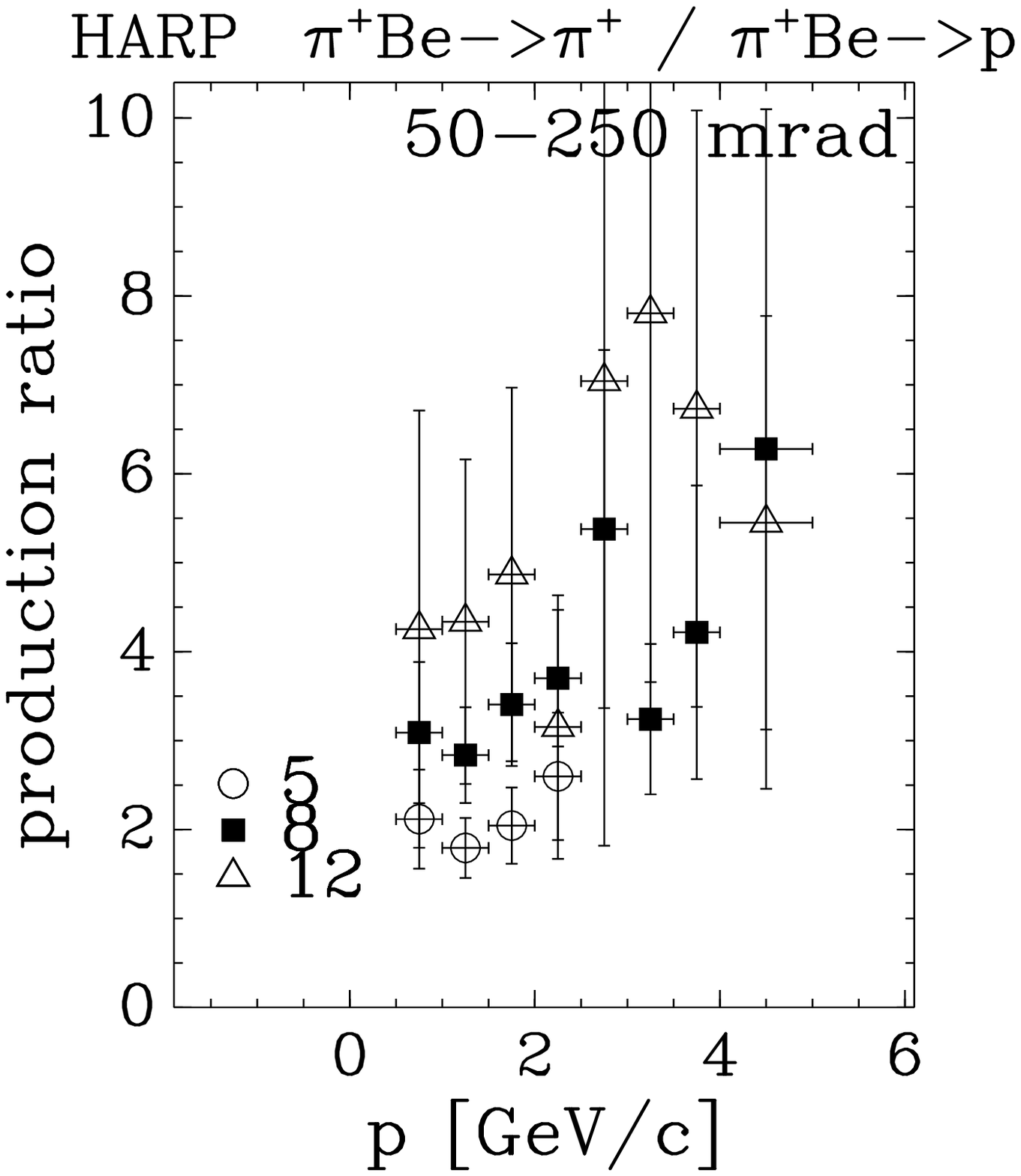}
\includegraphics[width=.28\textwidth]
{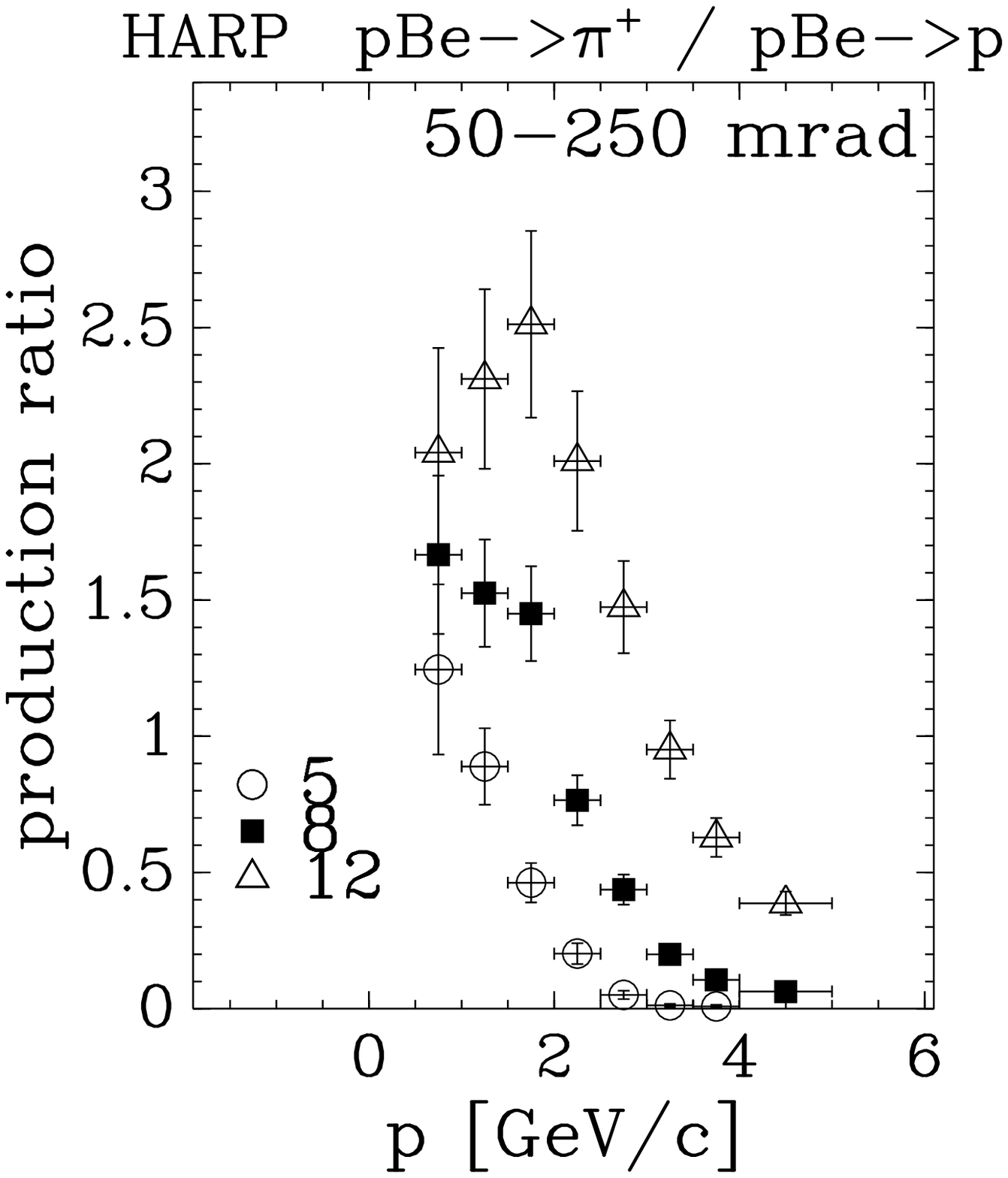}
\caption{ Particle production ratios on a Be target at 5, 8, 12~\GeVc.
Top panels: $\pi^{-}/\pi^{+}$ ratio with incident $\pi^{-}, \pi^{+}$ and  p;
 bottom panels: $\pi^{+}/p$ ratio with incident $\pi^{-}, \pi^{+}$ and p.}
\label{fig:rat_Be}
\end{figure*}

\begin{figure*}[tb]
\centering
\includegraphics[width=.26\textwidth]
{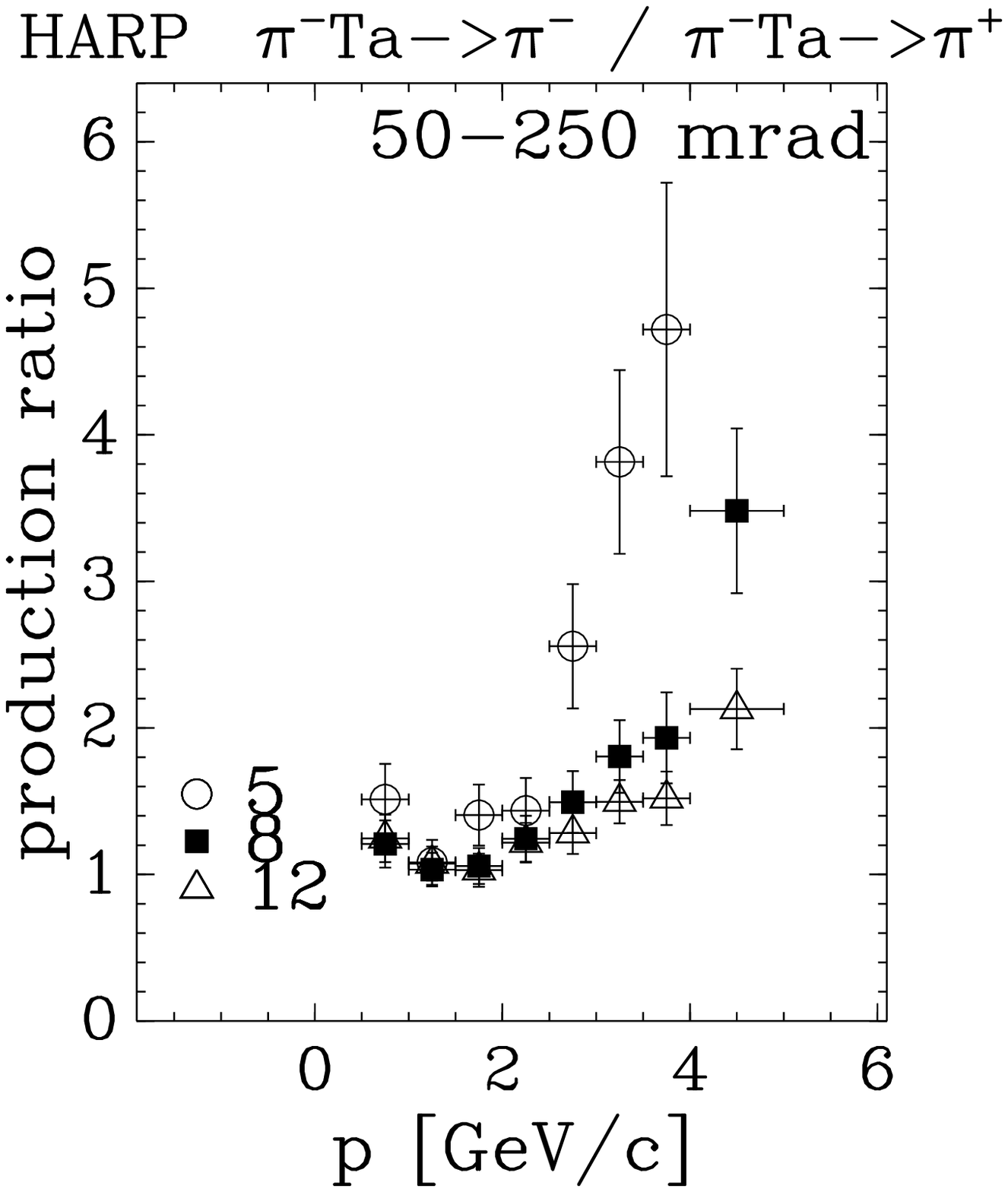}
\includegraphics[width=.28\textwidth]
{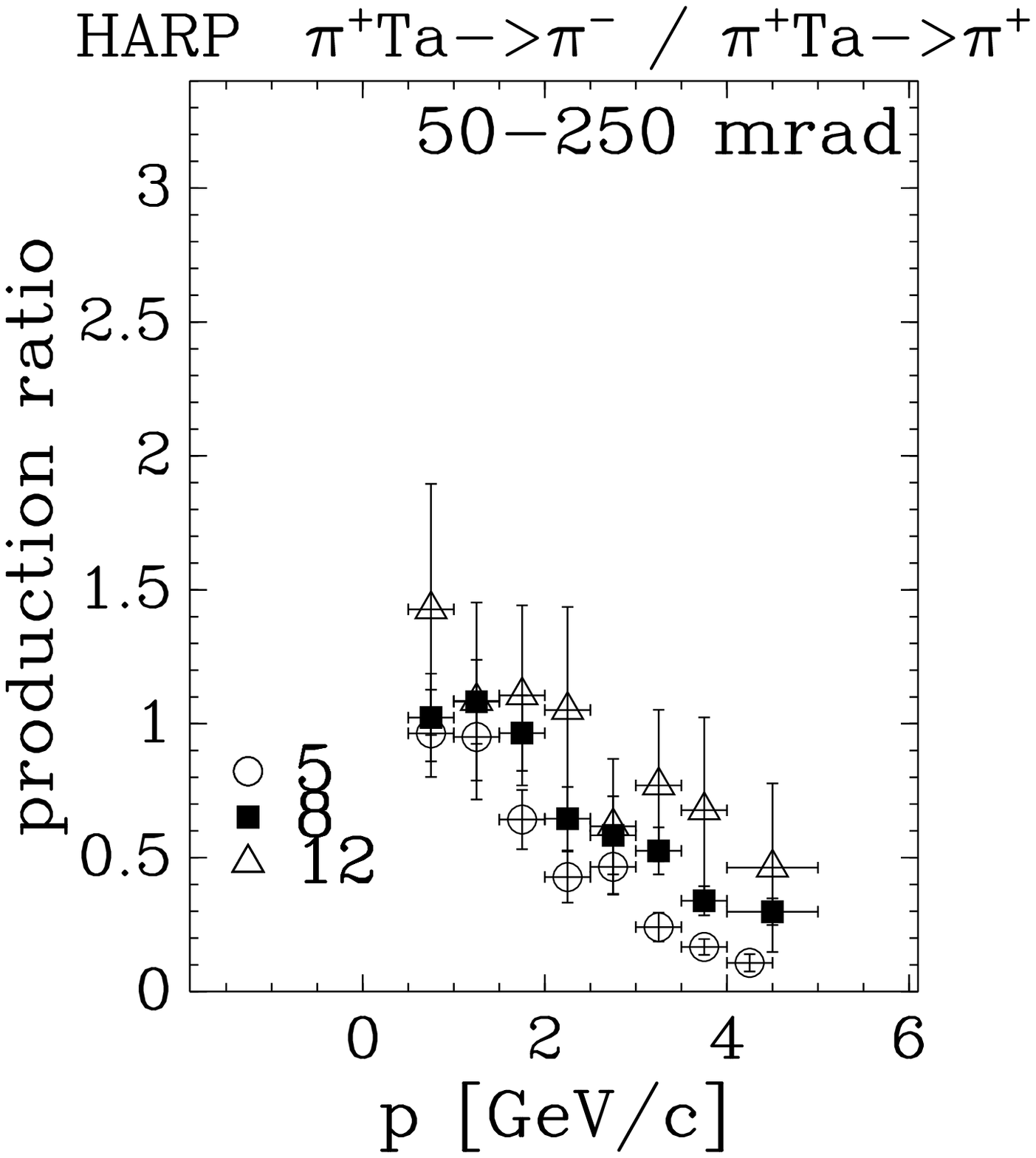}
\includegraphics[width=.28\textwidth]
{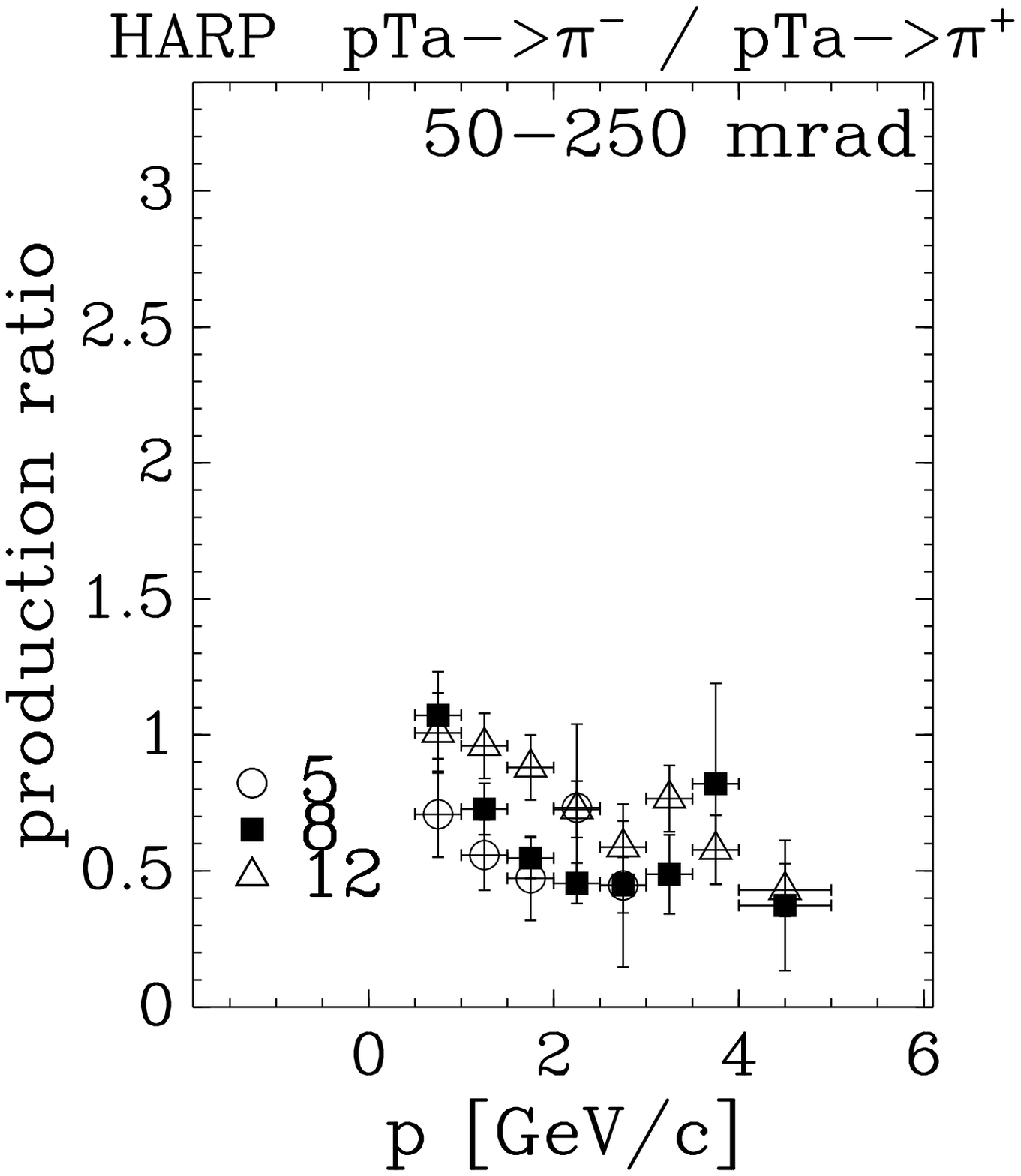}
\includegraphics[width=.26\textwidth]
{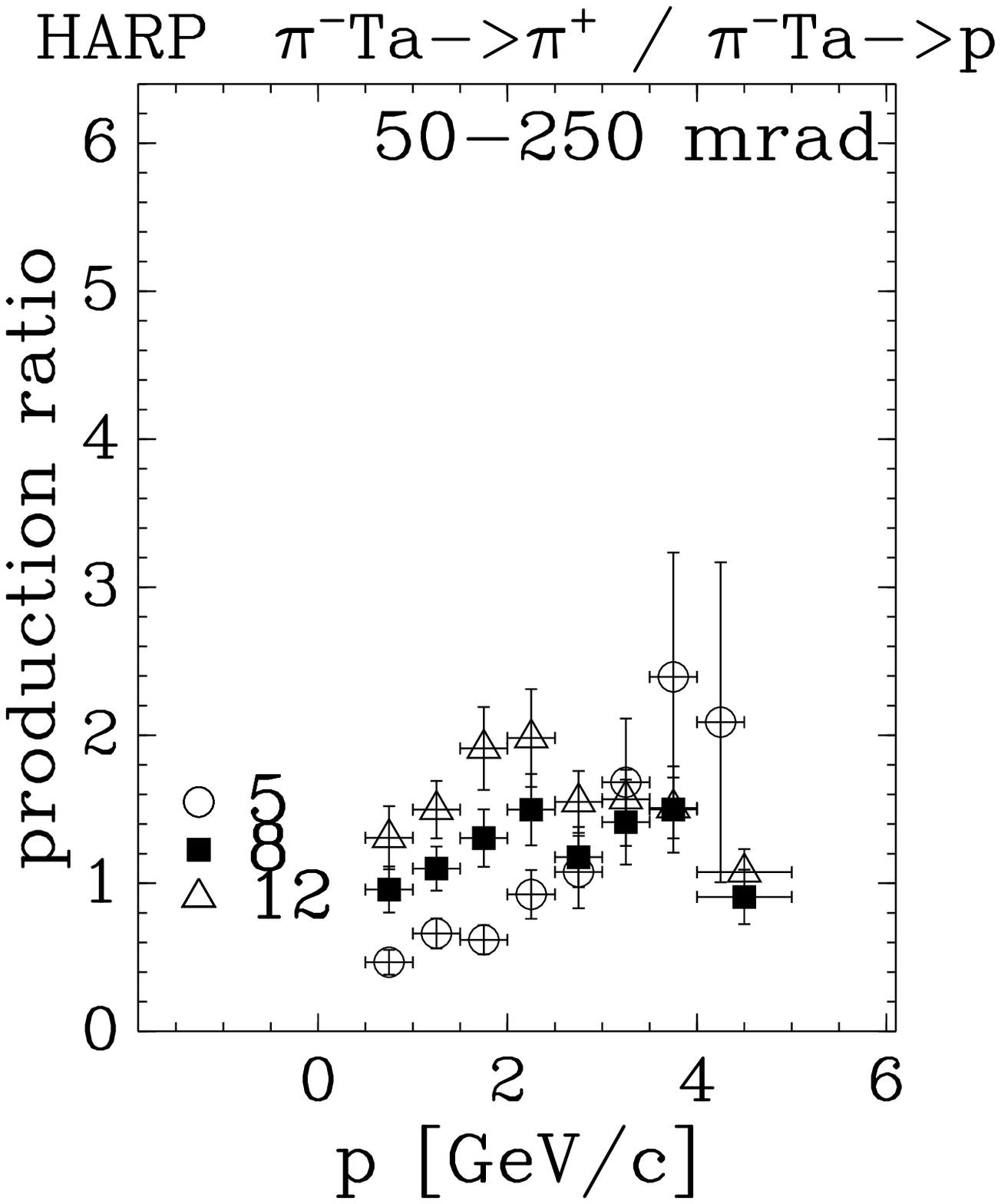}
\includegraphics[width=.28\textwidth]
{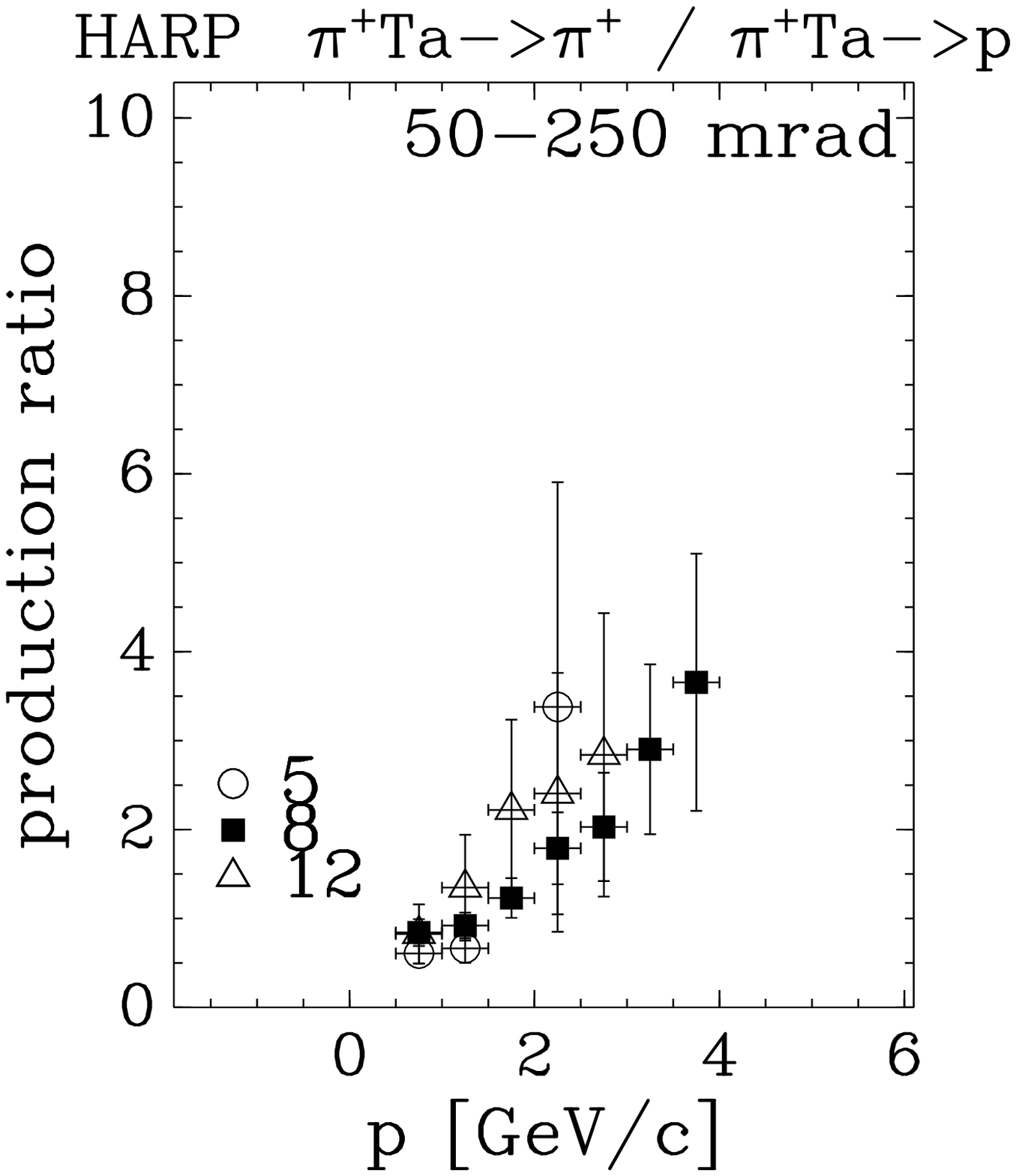}
\includegraphics[width=.28\textwidth]
{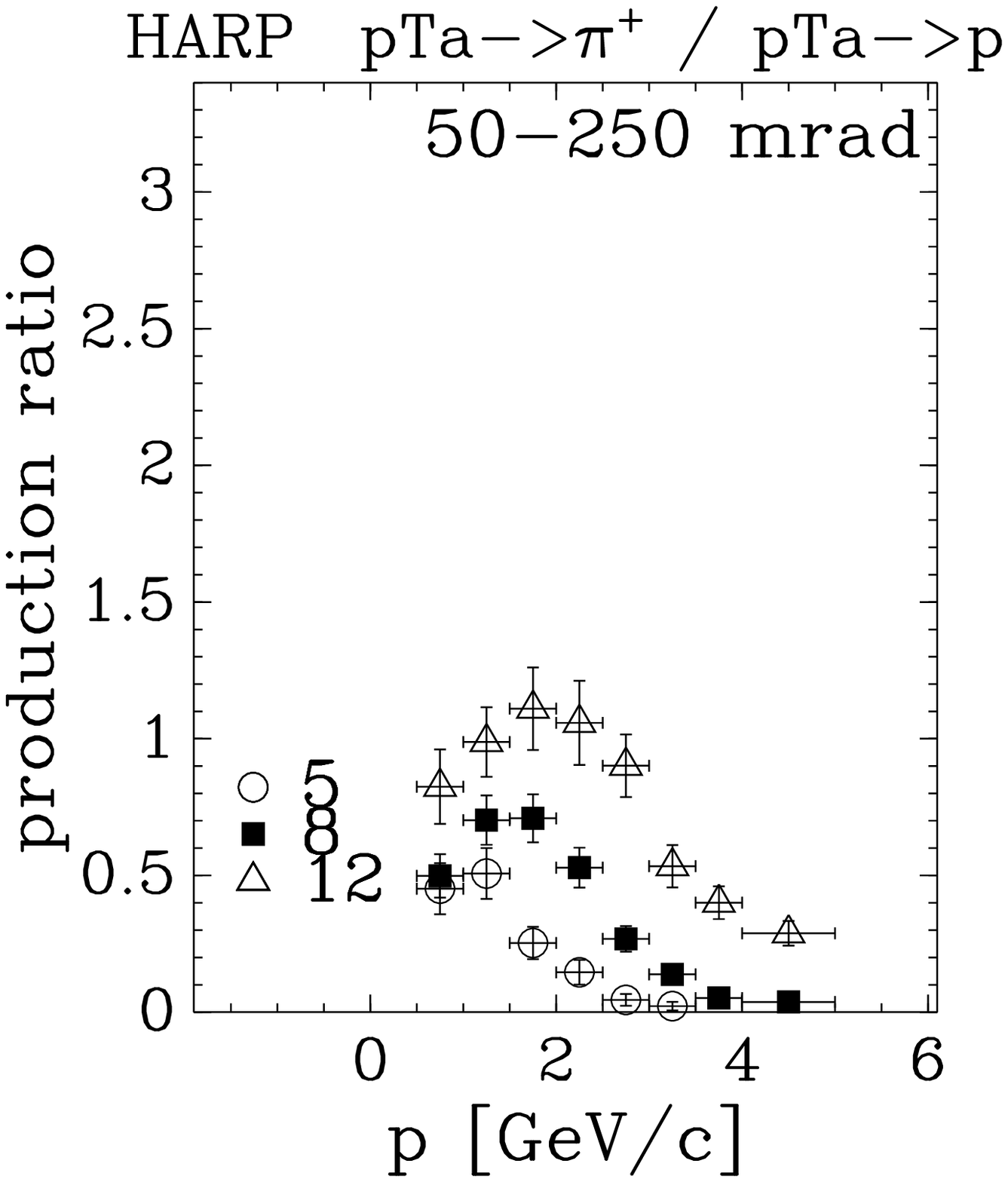}
\caption{ Particle production ratios on a Ta target at 5, 8, 12~\GeVc.
Top panels: $\pi^{-}/\pi^{+}$ ratio with incident $\pi^{-}, \pi^{+}$ and p; 
bottom panels: $\pi^{+}/p$ ratio with incident $\pi^{-}, \pi^{+}$ and p.}
\label{fig:rat_Ta}
\end{figure*}
The particle production ratios $\pi^{-}/\pi^{+}$ and $\pi^{+}/p$,
in the integrated angular range $0.05~\rad \leq \theta < 0.25~\rad$, 
are reported instead in Figures 
\ref{fig:rat_Be} and \ref{fig:rat_Ta}
for two typical targets: beryllium and tantalum and the different incident
beam particles. They clearly reflect the charge polarity of the incident 
beam particle, especially at high secondary momenta.

\FloatBarrier
 
\subsection{Comparison with Monte Carlo generators.}

In the following we will show only some comparisons with  two
widely used Monte Carlo simulation packages: MARS~\cite{ref:mars}
and GEANT4~\cite{ref:geant4}~\footnote{The GEANT4 version used is 9.2p01}, using different
generator models.
The comparison will be shown for a limited set of plots
and only for the Be and Ta targets, as examples of a light and a heavy target.
In both generators, no single model is applicable to all energies and a
transition between low energy models and high energy models, at about 5--10
GeV, is needed. 

The lack of hadron nuclei collisions data with small errors (both statistics
and systematics) on an extended set of thin targets has been, up to now,
an obstacle for a serious tuning of these models.
Dedicated simulations, such as GiBUU \cite{ref:gall}, may also
profit from the availability of our data.

At intermediate energies (up to 5--10 GeV),
GEANT4 uses two types of intra-nuclear cascade models: the Bertini
model~\cite{ref:bert,ref:bert1} (valid up to $\sim 10$ GeV) and the binary
model~\cite{ref:bin} (valid up to $\sim 3$ GeV). Both models treat the target
nucleus in detail, taking into account density variations and tracking in the
nuclear field.
The binary model is based on hadron collisions with nucleons, giving
resonances that decay according to their quantum numbers. The Bertini
model is based on the cascade code reported in \cite{ref:bert2}
and hadron collisions are assumed to proceed according to free-space partial
cross sections and final state distributions measured for the incident
particle types. Details of the nuclear density and the Pauli blocking 
are then taken into account. 

At higher energies, instead, two parton string models,
the quark-gluon string (QGS)  model~\cite{ref:bert,ref:QGSP} and the Fritiof
(FTP) model~\cite{ref:QGSP} are used, in addition to a High Energy
Parametrized model (HEP)
derived from the high energy part of the GHEISHA code used inside
GEANT3~\cite{ref:gheisha}.
The parametrized models of GEANT4 (HEP and LEP) are intended to be fast,
but conserve energy and momentum on average and not event-by-event.

A realistic GEANT4 simulation is built by combining models and physics processes
into what is called a ``physics list'', that is included in the standard
GEANT4 Toolkit release. Each physics list  corresponds to a collection of
models suitable for a given user problem. 

As examples, the QGSP physics list is based on the QGS model, the pre-compound
nucleus model and some of the Low Energy Parametrized (LEP) model~\footnote{
Also this model, at low energy, has its root in the GHEISHA code inside
GEANT3.}, while the LHEP physics list~\cite{ref:lhep} is based on 
the parametrized LEP model and HEP models. Currently the most widely
used physics list in LHC experiments is the so-called QGSP-Bert physics list,
see reference \cite{Apo} for details. 

The MARS code system~\cite{ref:mars} uses as basic model an inclusive
approach multi-particle production originated by R. Feynman. Above 5~GeV
phenomenological particle production models are used, while below 5~GeV
a cascade-exciton model~\cite{ref:casca} combined with the Fermi
break-up model, the coalescence model, an evaporation model and a
multi-fragmentation extension are used instead.  

The comparison, just outlined in our paper,  between data and models
is reasonable,
but some discrepancies are evident for some models especially at lower
energies and small angles. Discrepancies up to a factor of three are seen.

The full set of  HARP data, taken with targets
spanning the full periodic table of elements, with small total errors and full
coverage of the solid angle in a single detector may help the validation
of models used in hadronic simulations in the difficult energy range between
3 and 15~\GeVc of incident momentum, as done e.g. in reference~\cite{ref:gall}
for charged pion production.

\begin{figure}[tbp]
\begin{center}
  \includegraphics[width=.42\textwidth,angle=90]{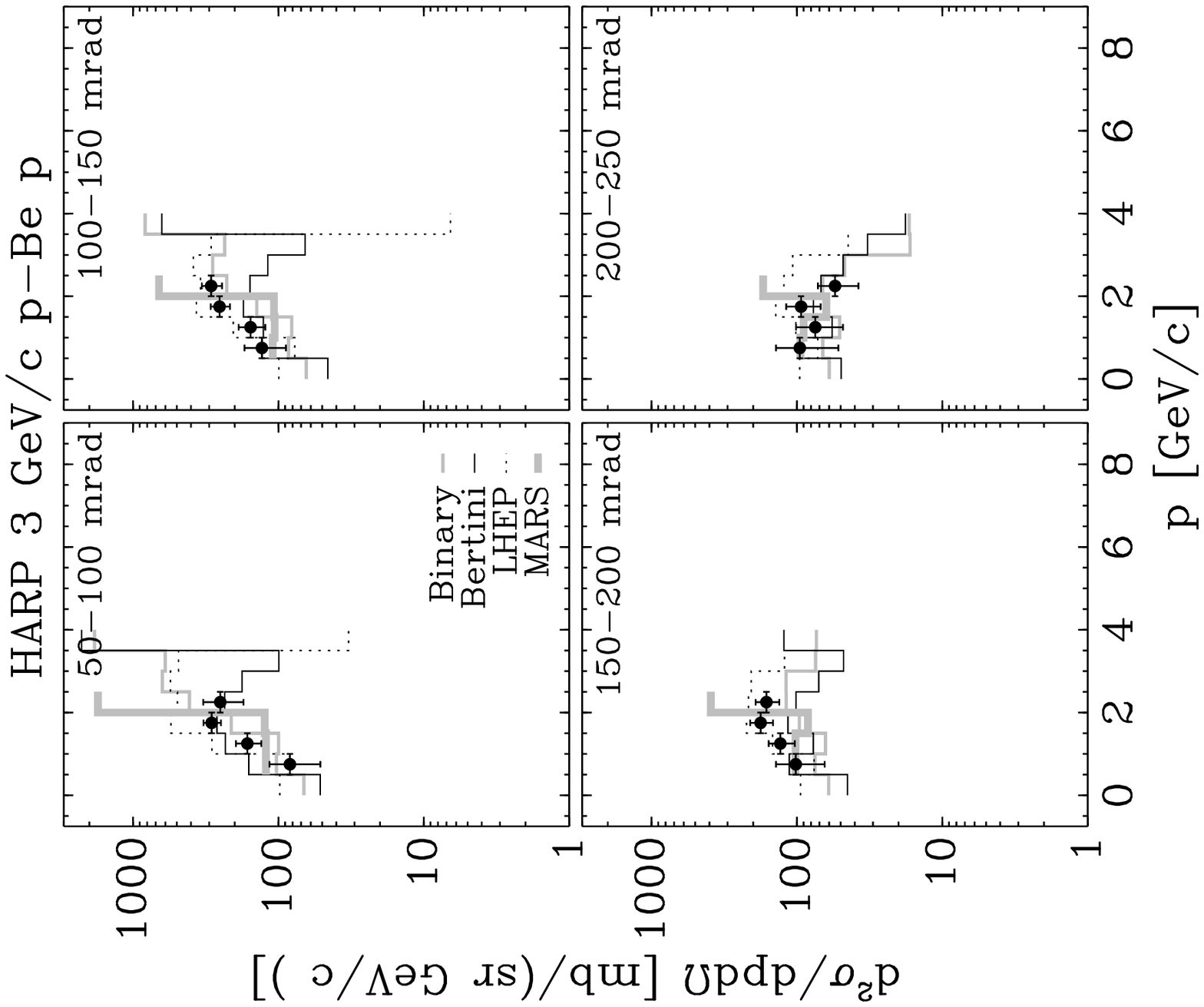}
  \includegraphics[width=.42\textwidth,angle=90]{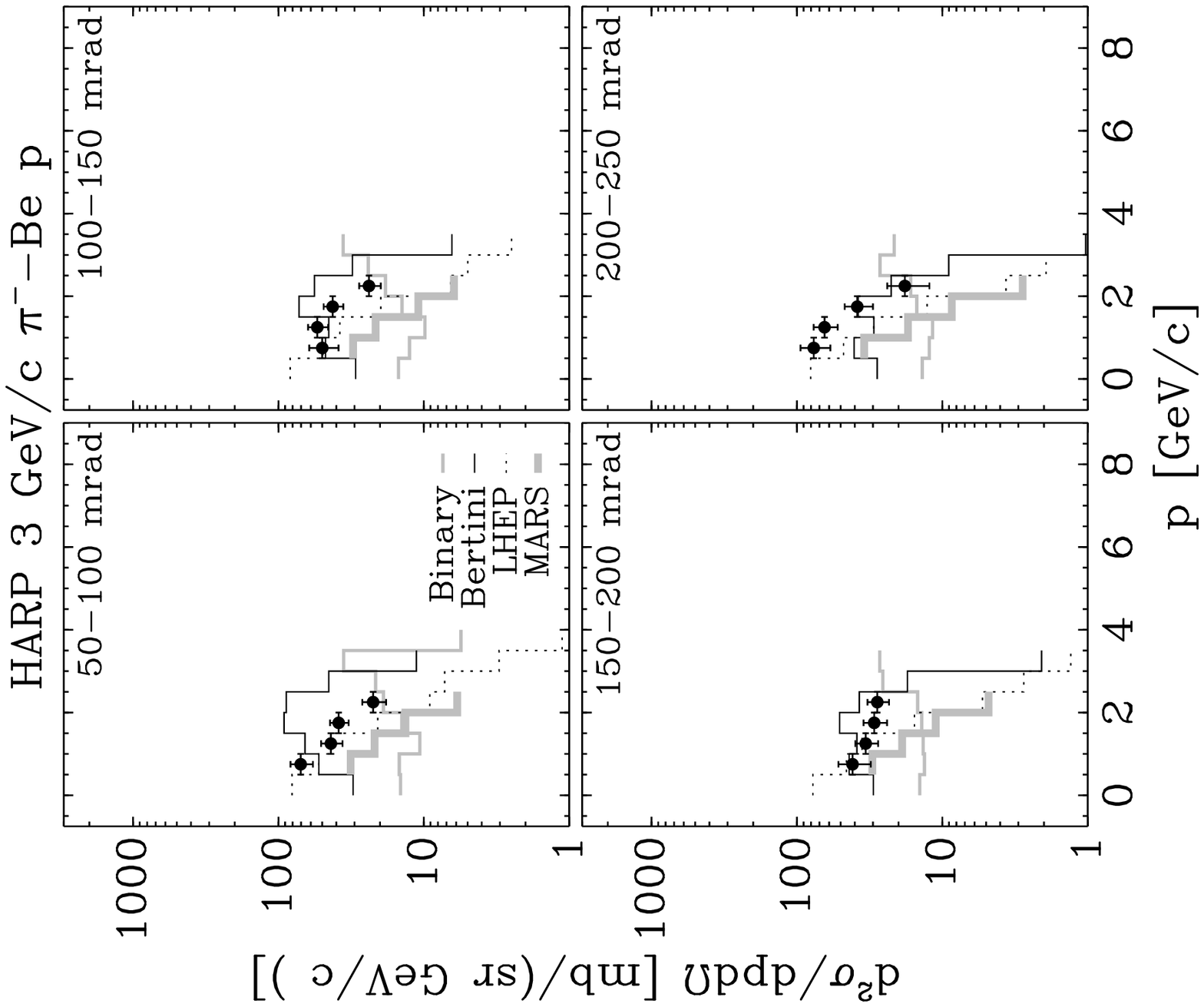}
  \includegraphics[width=.42\textwidth,angle=90]{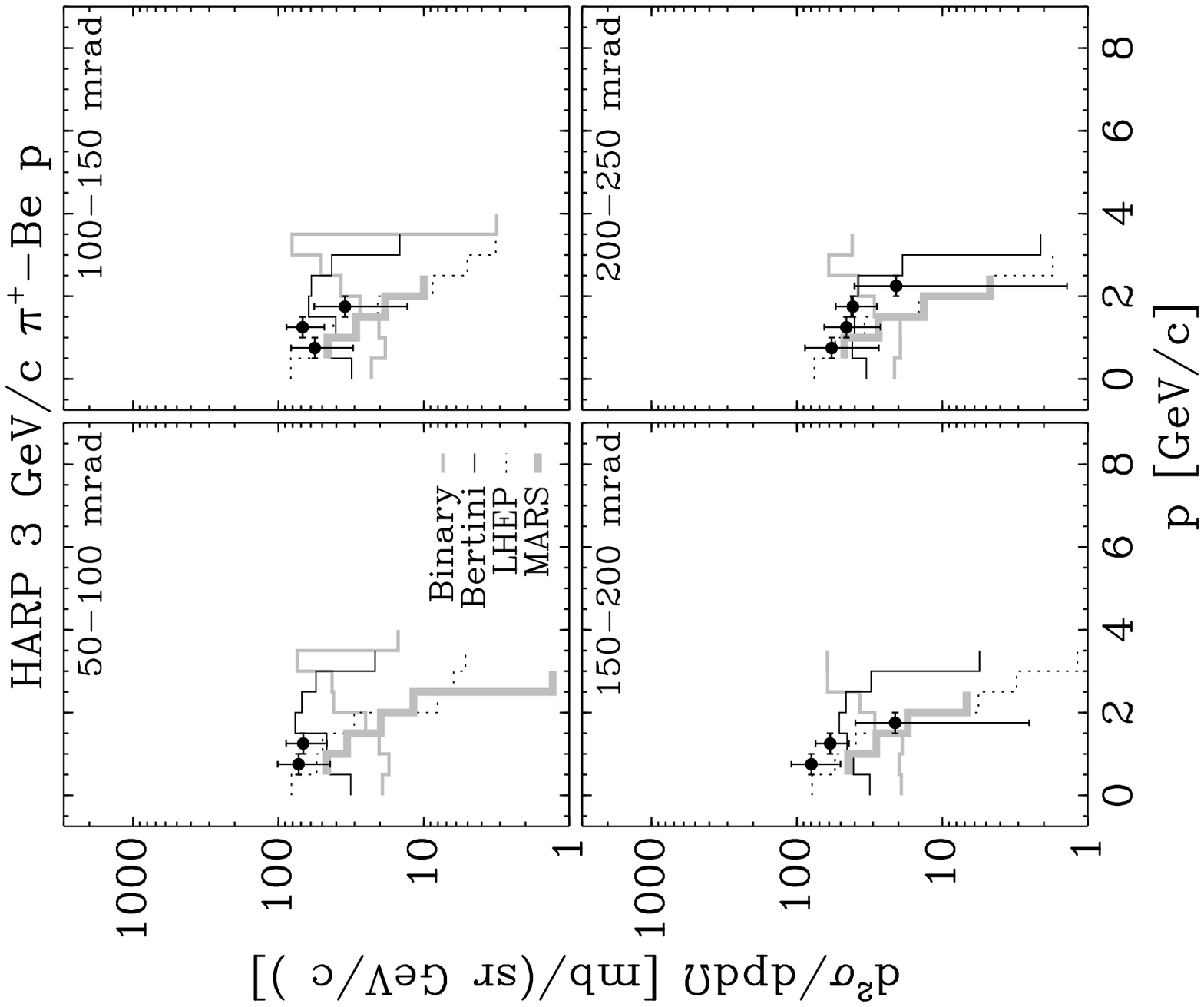}
\end{center}
\caption{
 Comparison of HARP double-differential proton cross sections for p--Be, \pim--Be,
\pip--Be interactions  at 3~\GeVc with
 GEANT4 and MARS MC predictions, using several generator models 
(see text for details).
}
\label{fig:G43a}
\end{figure}

\begin{figure}[tbp]
\begin{center}
\includegraphics[width=.42\textwidth,angle=90]{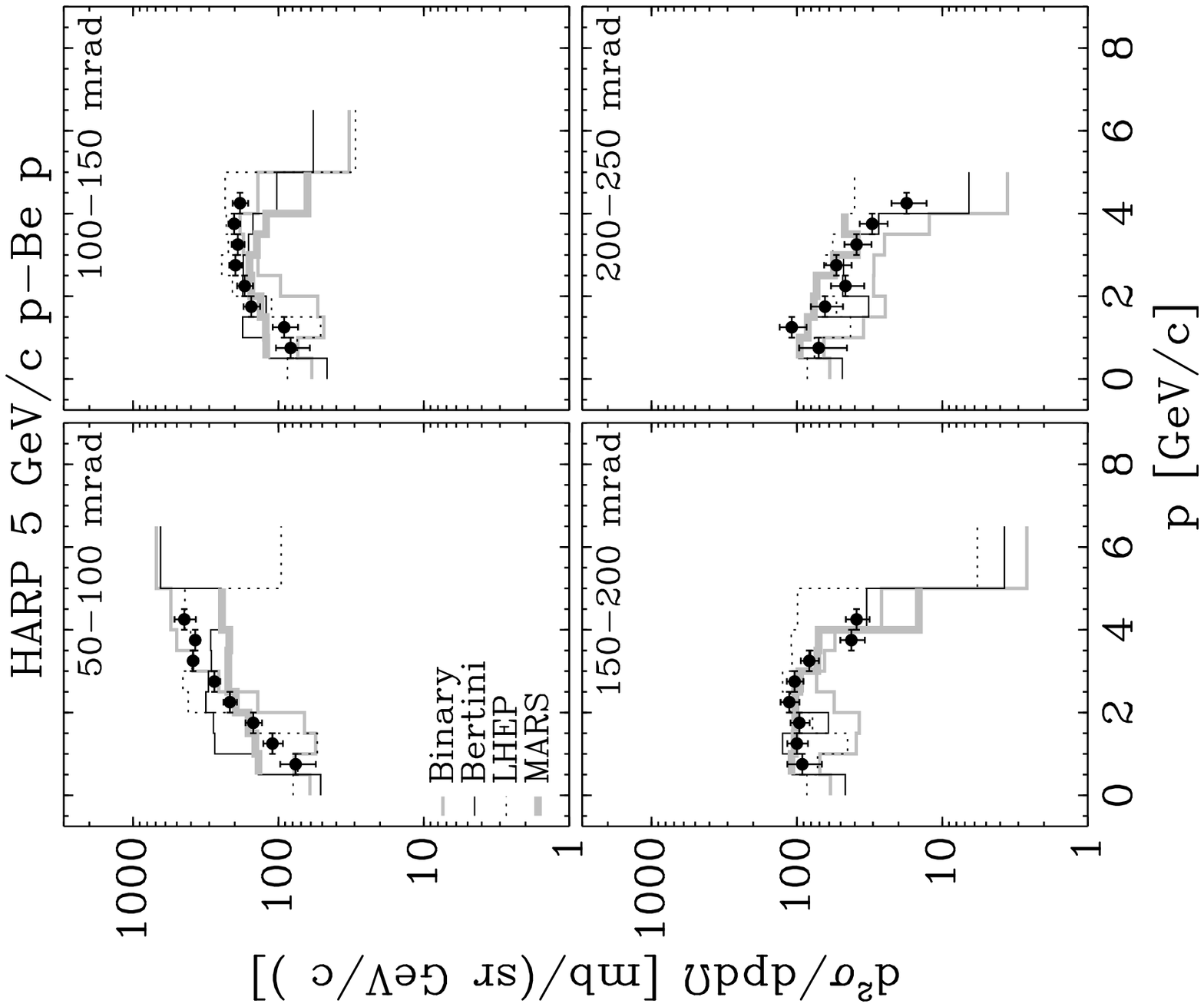}
\includegraphics[width=.42\textwidth,angle=90]{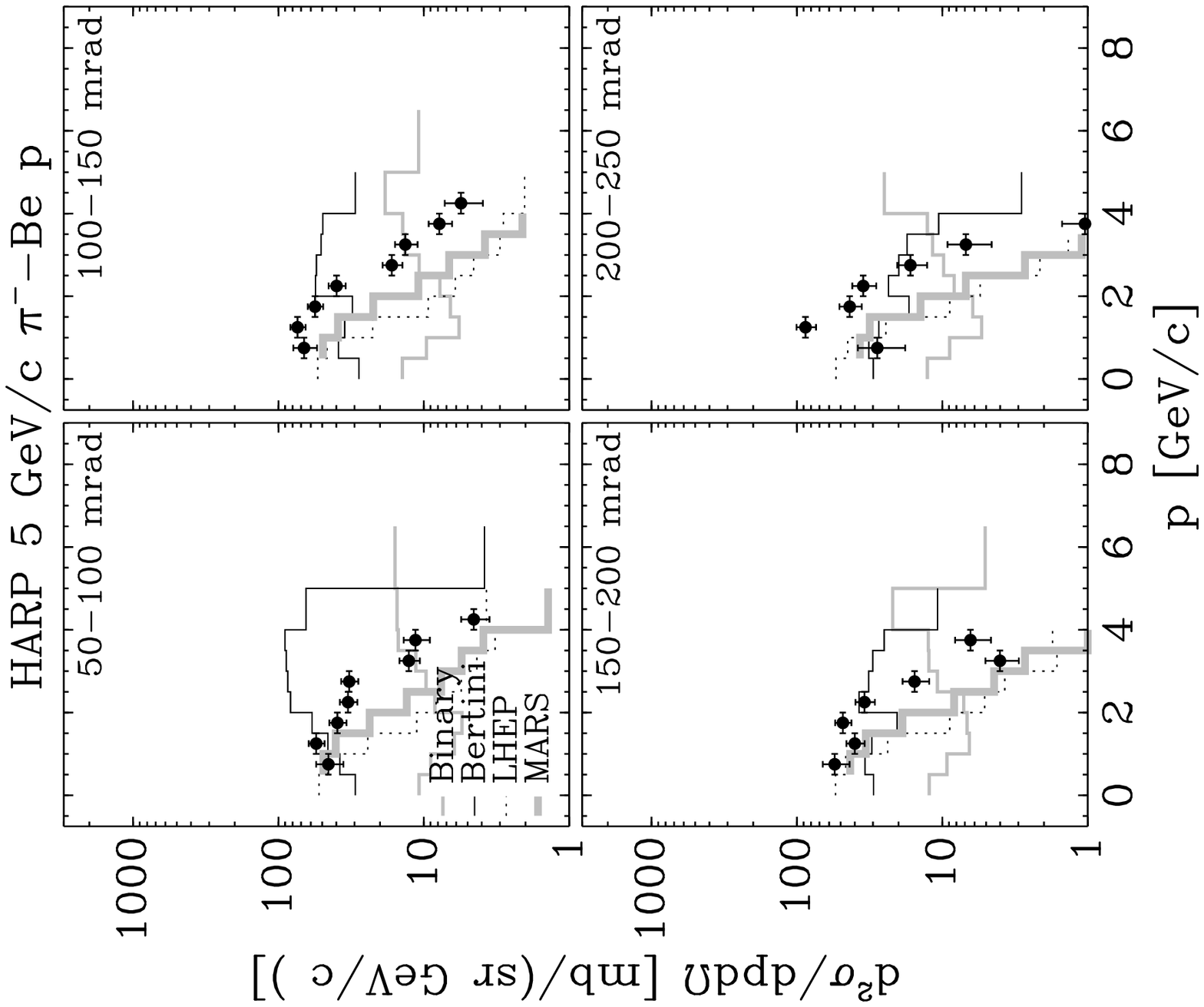}
\includegraphics[width=.42\textwidth,angle=90]{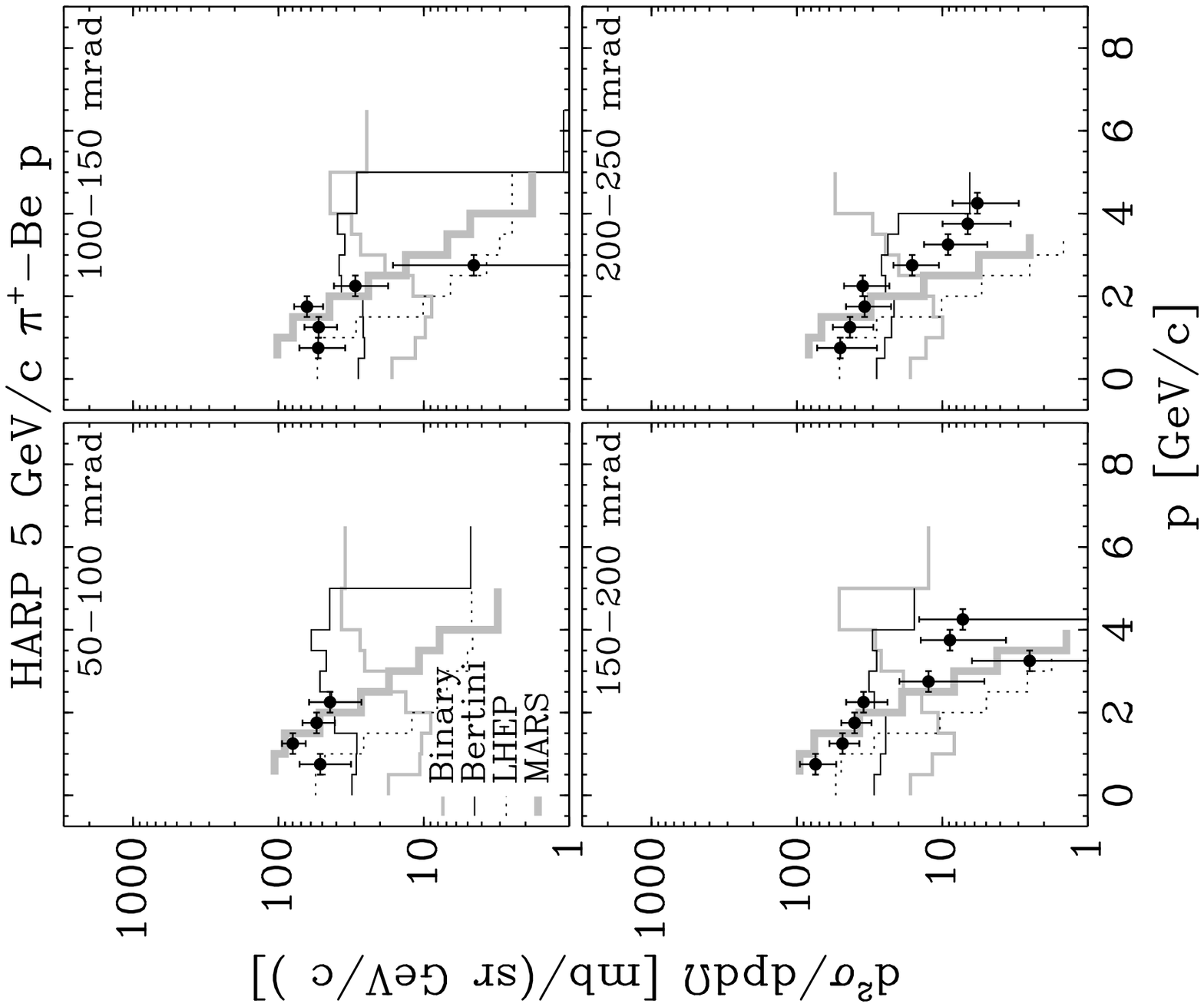}
\end{center}
\caption{
 Comparison of HARP double-differential proton cross sections for p--Be, \pim--Be,
\pip--Be interactions  at 5~\GeVc with
 GEANT4 and MARS MC predictions, using several generator models 
(see text for details).
}
\label{fig:G44a}
\end{figure}
\begin{figure}[tbp]
\begin{center}
\includegraphics[width=.42\textwidth,angle=90]{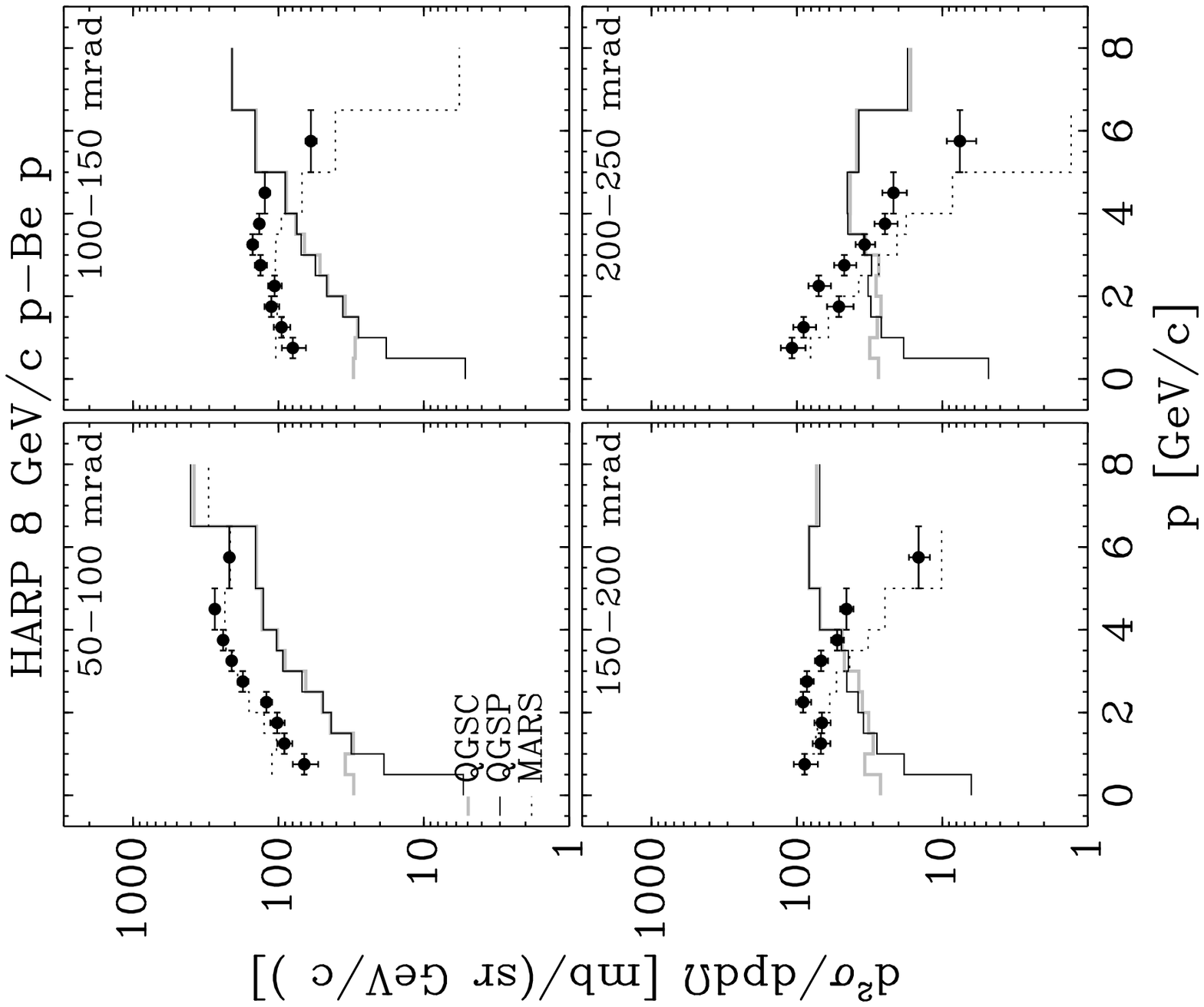}
\includegraphics[width=.42\textwidth,angle=90]{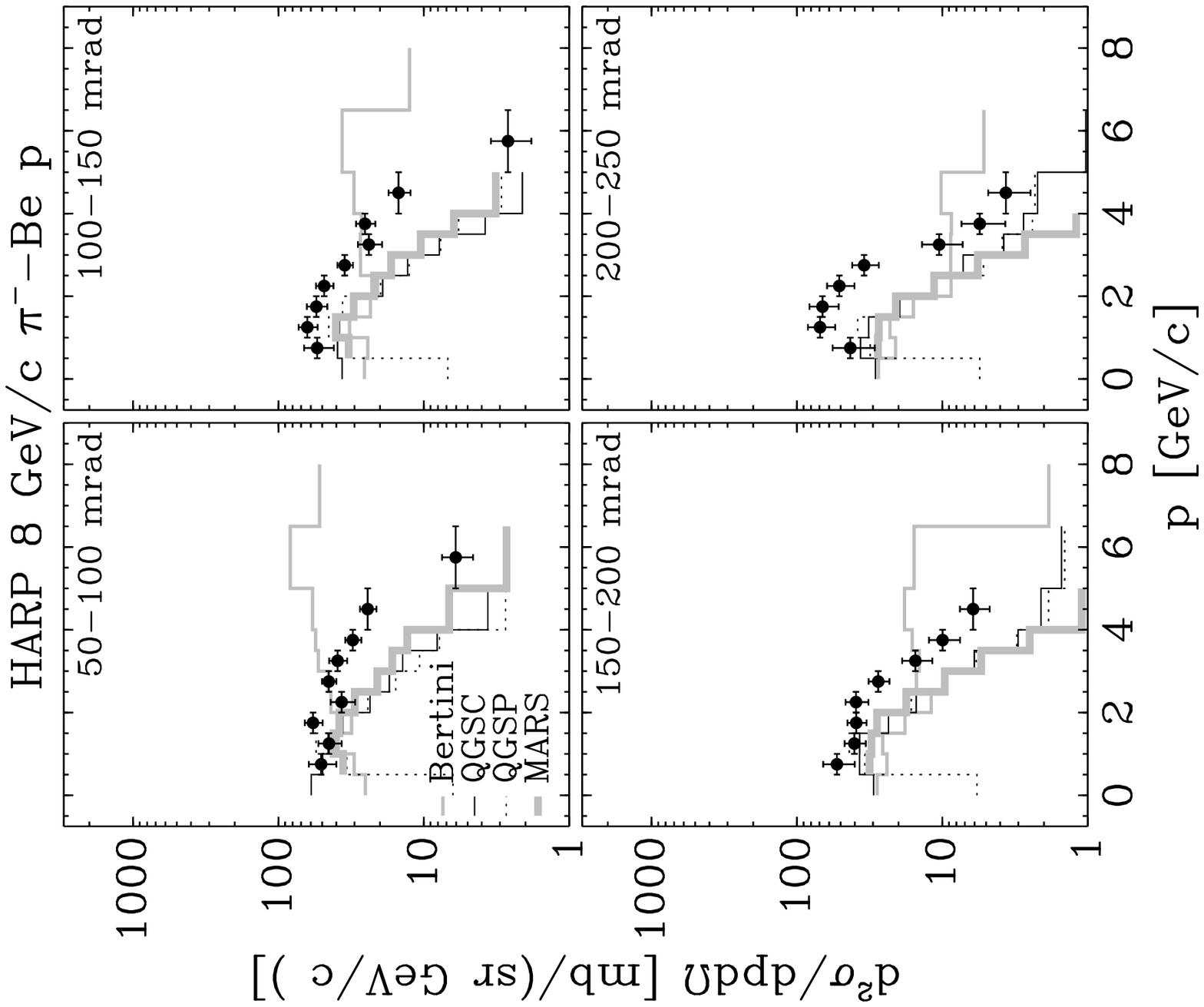}
\includegraphics[width=.42\textwidth,angle=90]{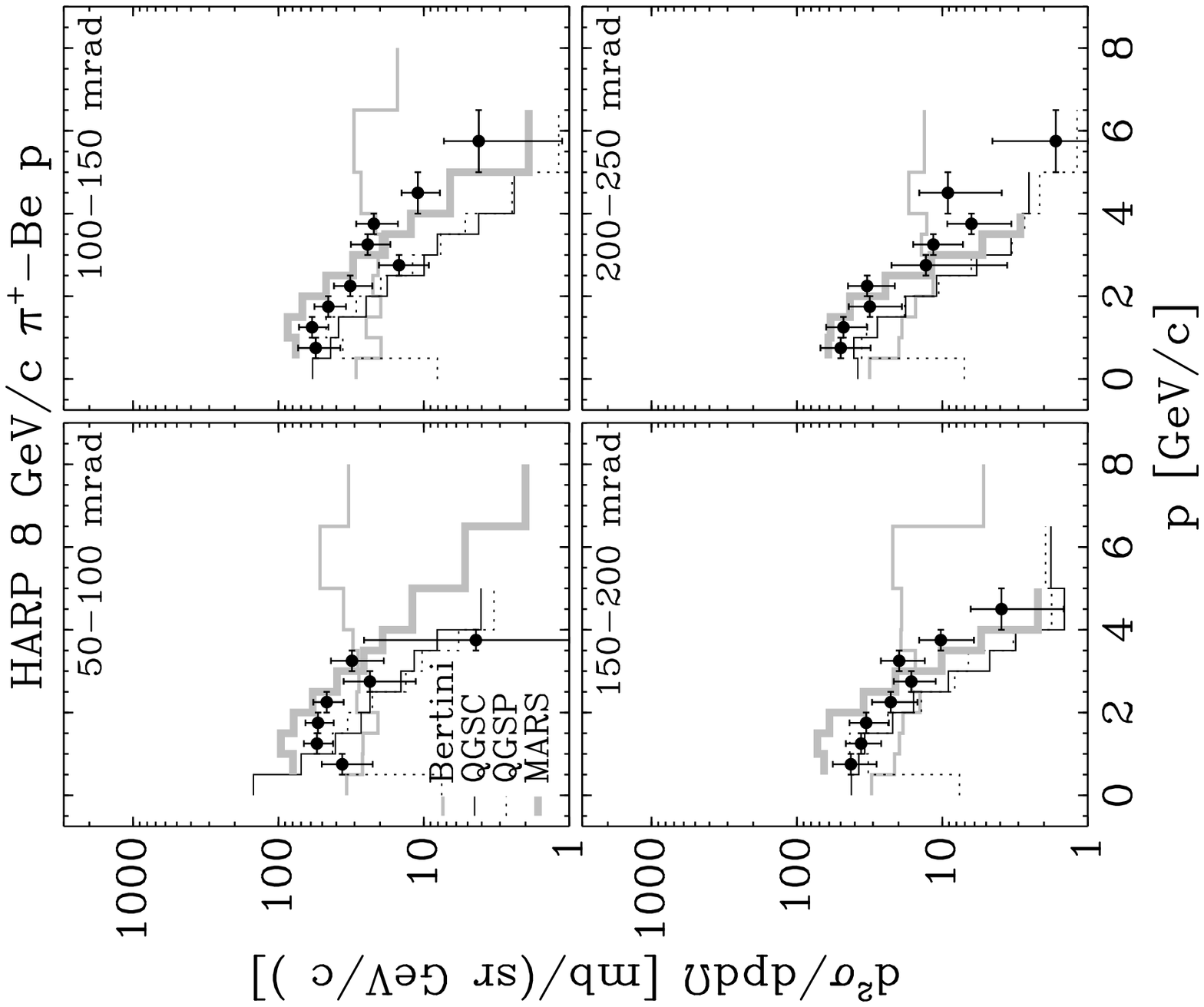}
\end{center}
\caption{
 Comparison of HARP double-differential proton cross sections for p--Be, \pim--Be,
\pip--Be interactions  at 8~\GeVc with
 GEANT4 and MARS MC predictions, using several generator models 
(see text for details).
}
\label{fig:G45a}
\end{figure}

\begin{figure}[tbp]
\begin{center}
\includegraphics[width=.42\textwidth,angle=90]{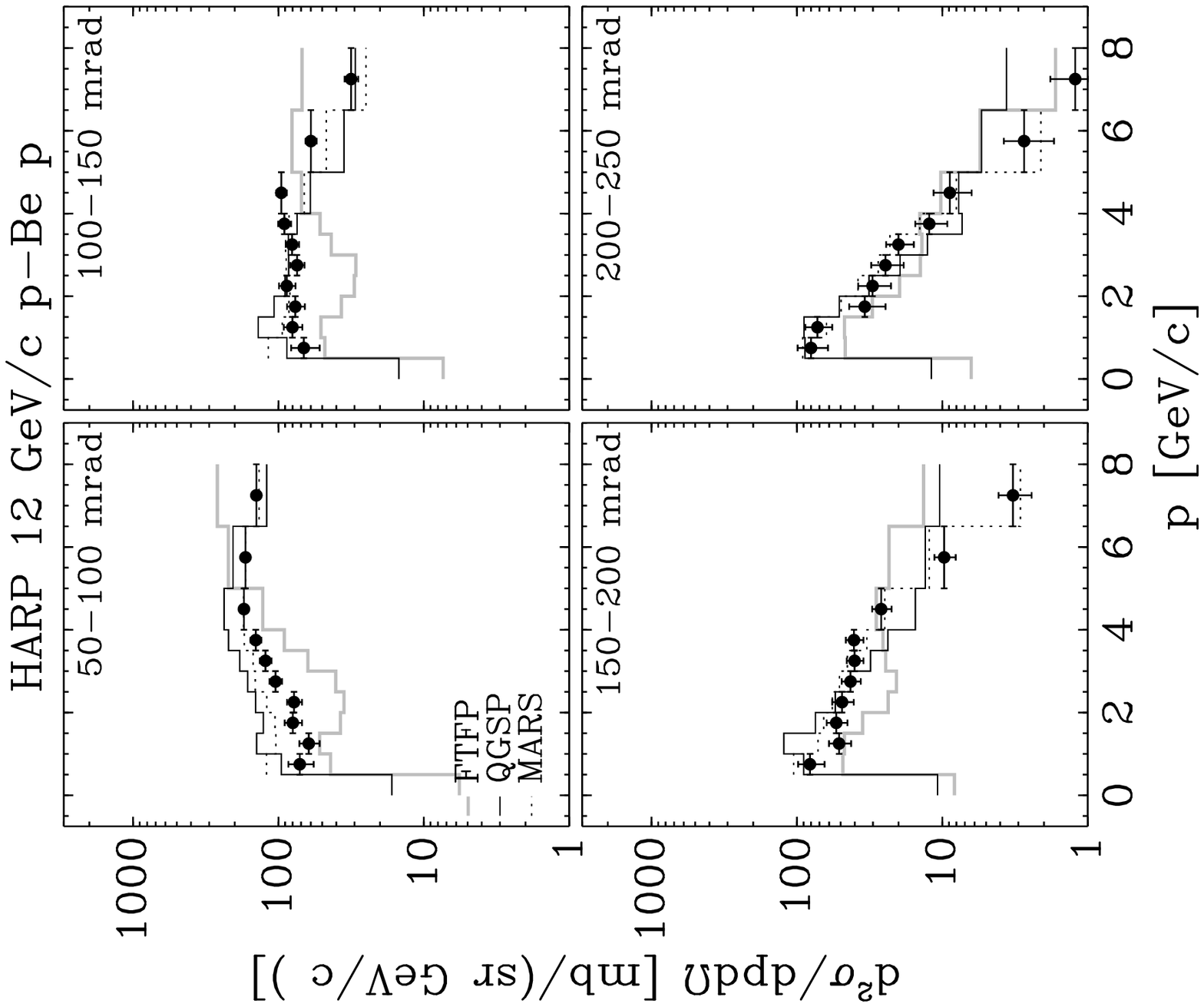}
\includegraphics[width=.42\textwidth,angle=90]{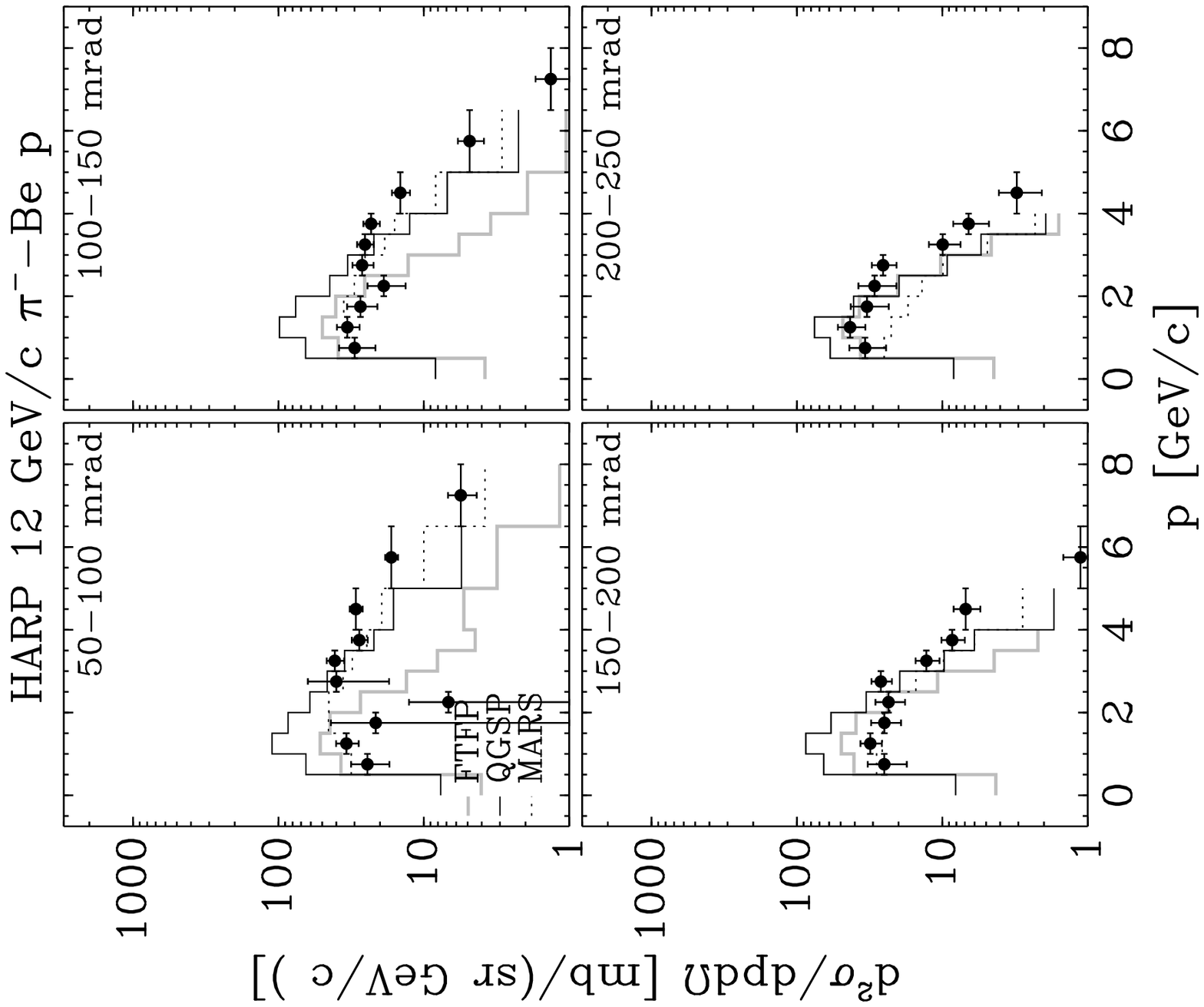}
\includegraphics[width=.42\textwidth,angle=90]{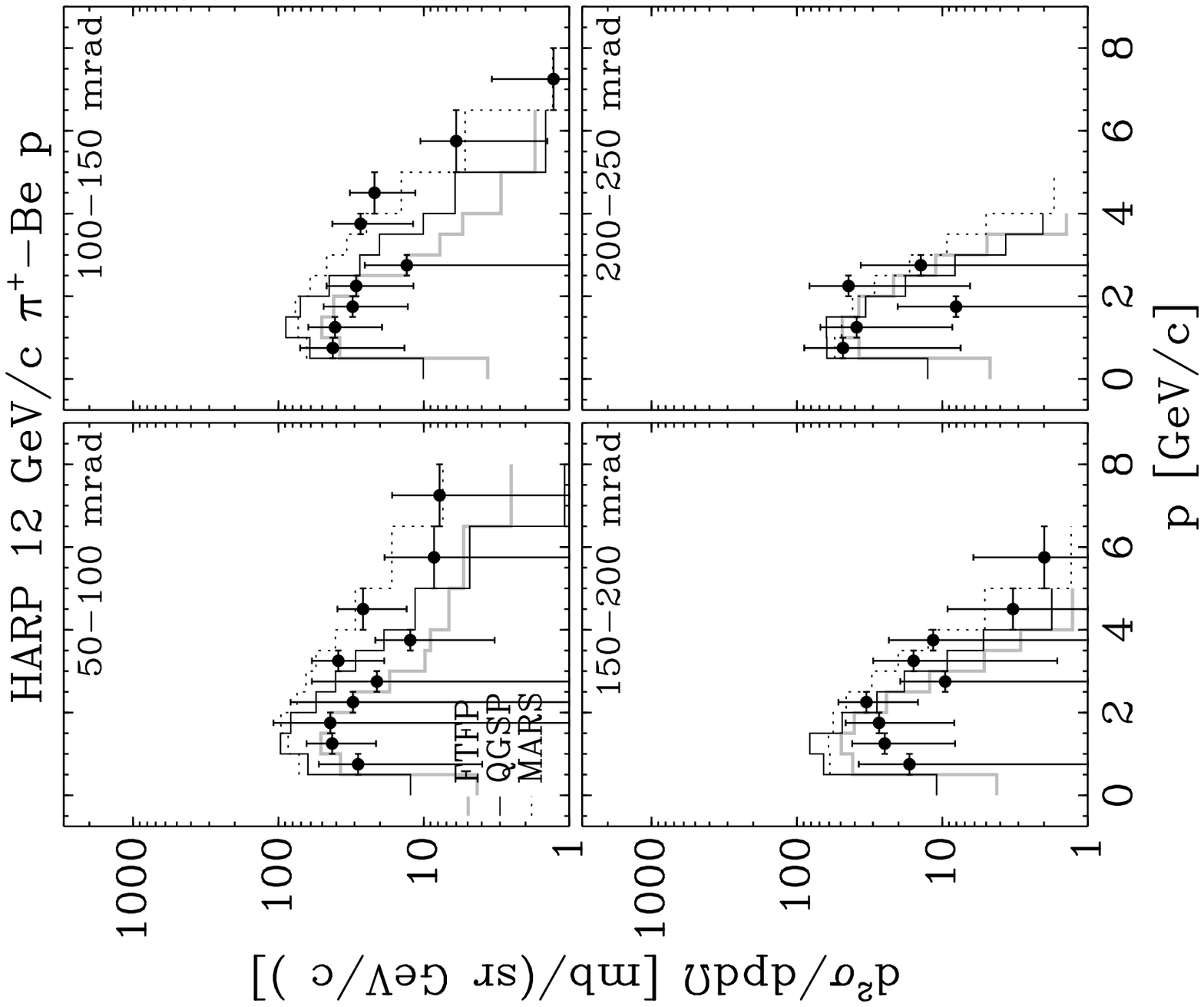}
\end{center}
\caption{
 Comparison of HARP double-differential proton cross sections for p--Be, \pim--Be,
\pip--Be interactions  at 12~\GeVc with
 GEANT4 and MARS MC predictions, using several generator models 
(see text for details). 
}
\label{fig:G46}
\end{figure}

\begin{figure}[tbp]
\begin{center}
\includegraphics[width=.42\textwidth,angle=90]{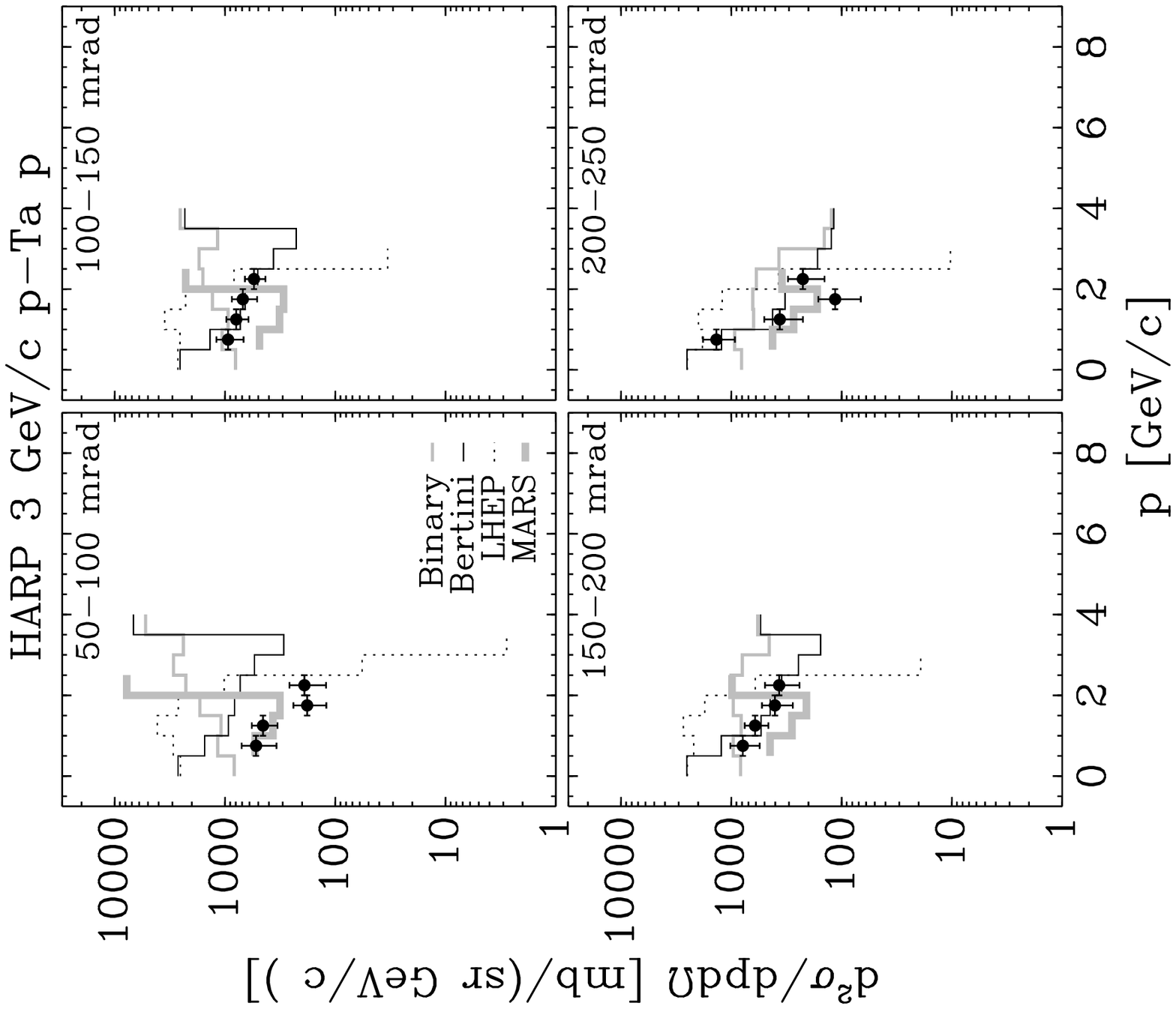}
\includegraphics[width=.42\textwidth,angle=90]{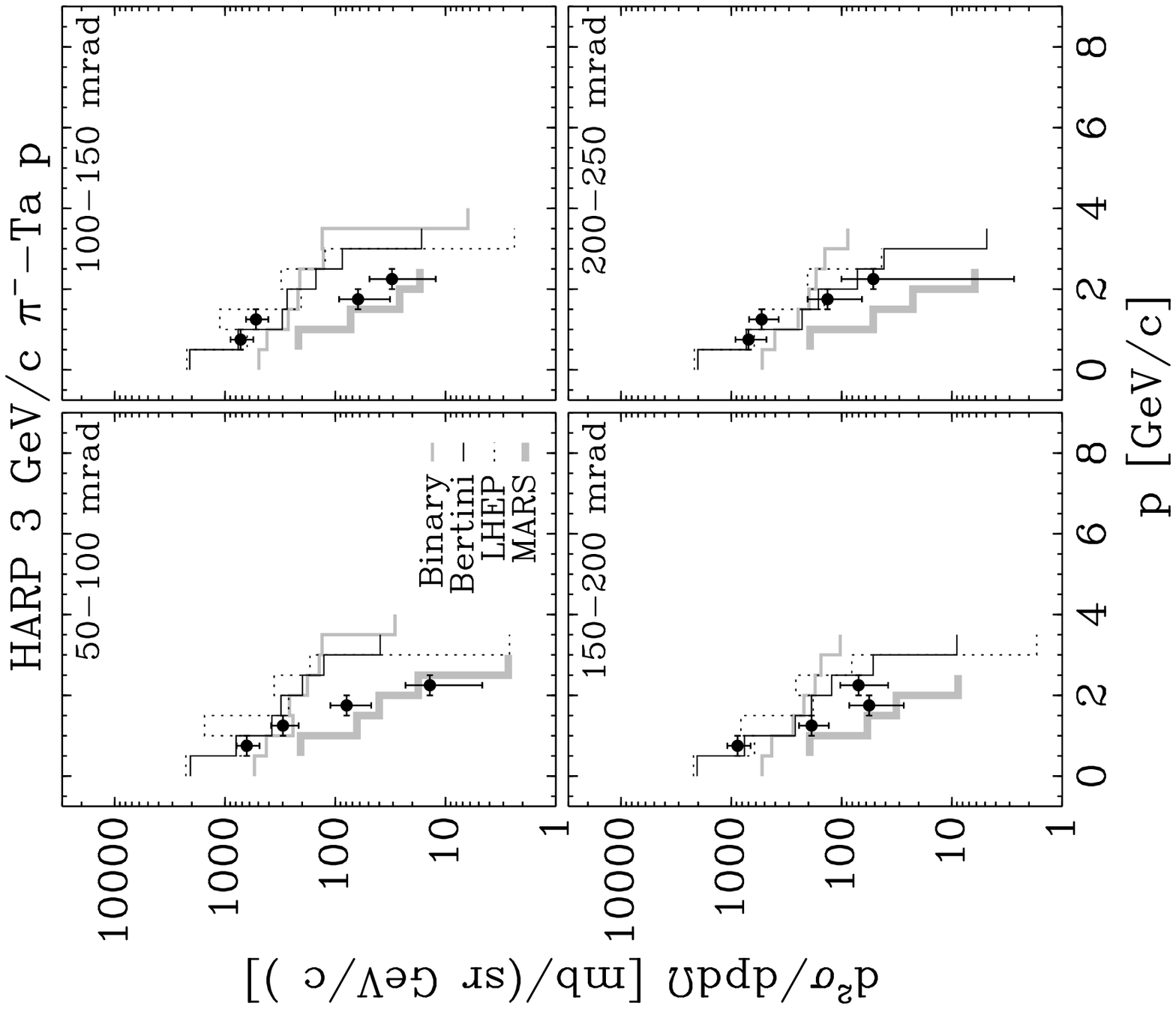}
\includegraphics[width=.42\textwidth,angle=90]{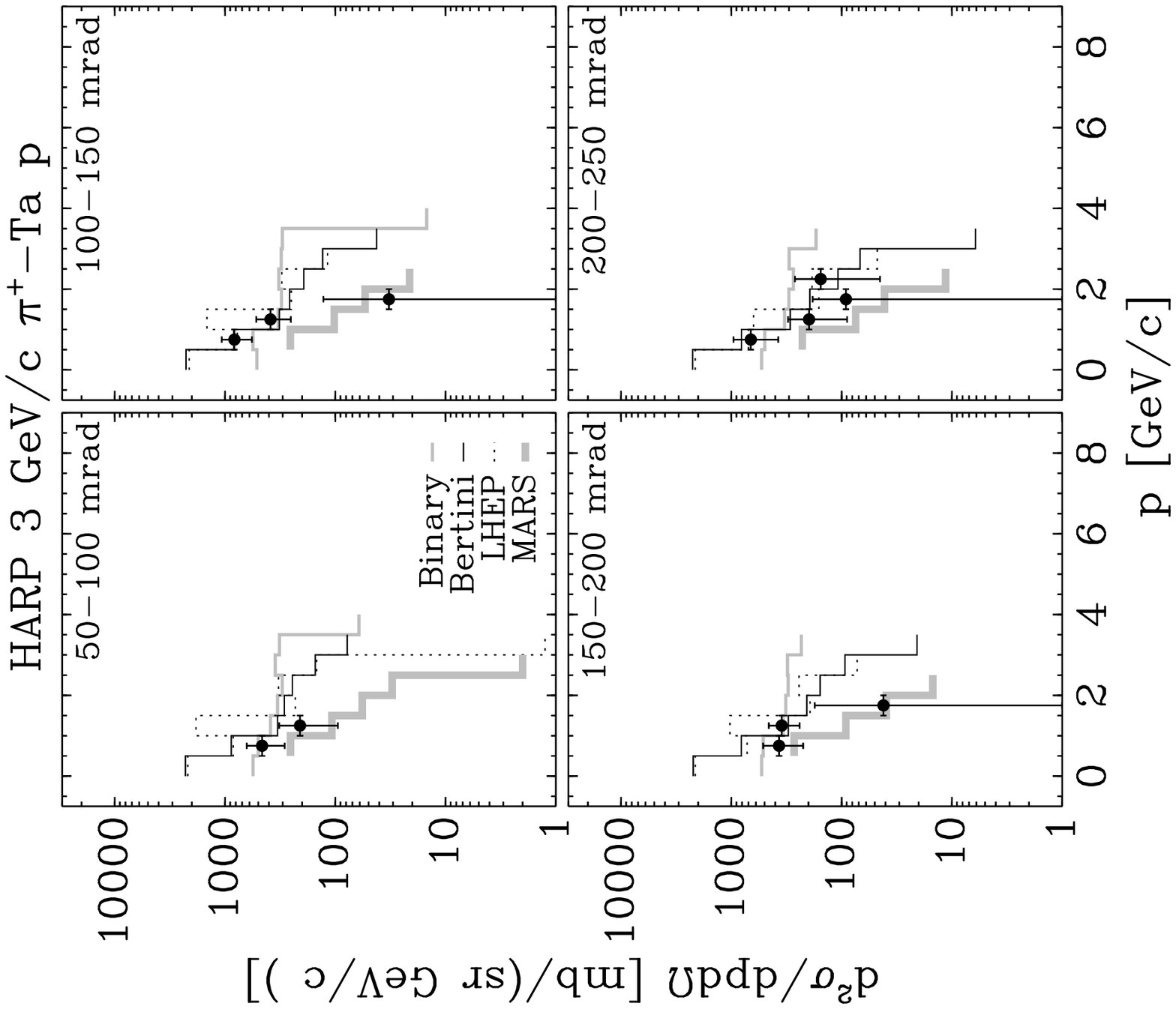}
\end{center}
\caption{
Comparison of HARP double-differential proton cross sections for p--Ta, \pim--Ta,
\pip--Ta interactions  at 3~\GeVc with
 GEANT4 and MARS MC predictions, using several generator models
(see text for details).
}
\label{fig:G53}
\end{figure}

\begin{figure}[tbp]
\begin{center}
\includegraphics[width=.42\textwidth,angle=90]{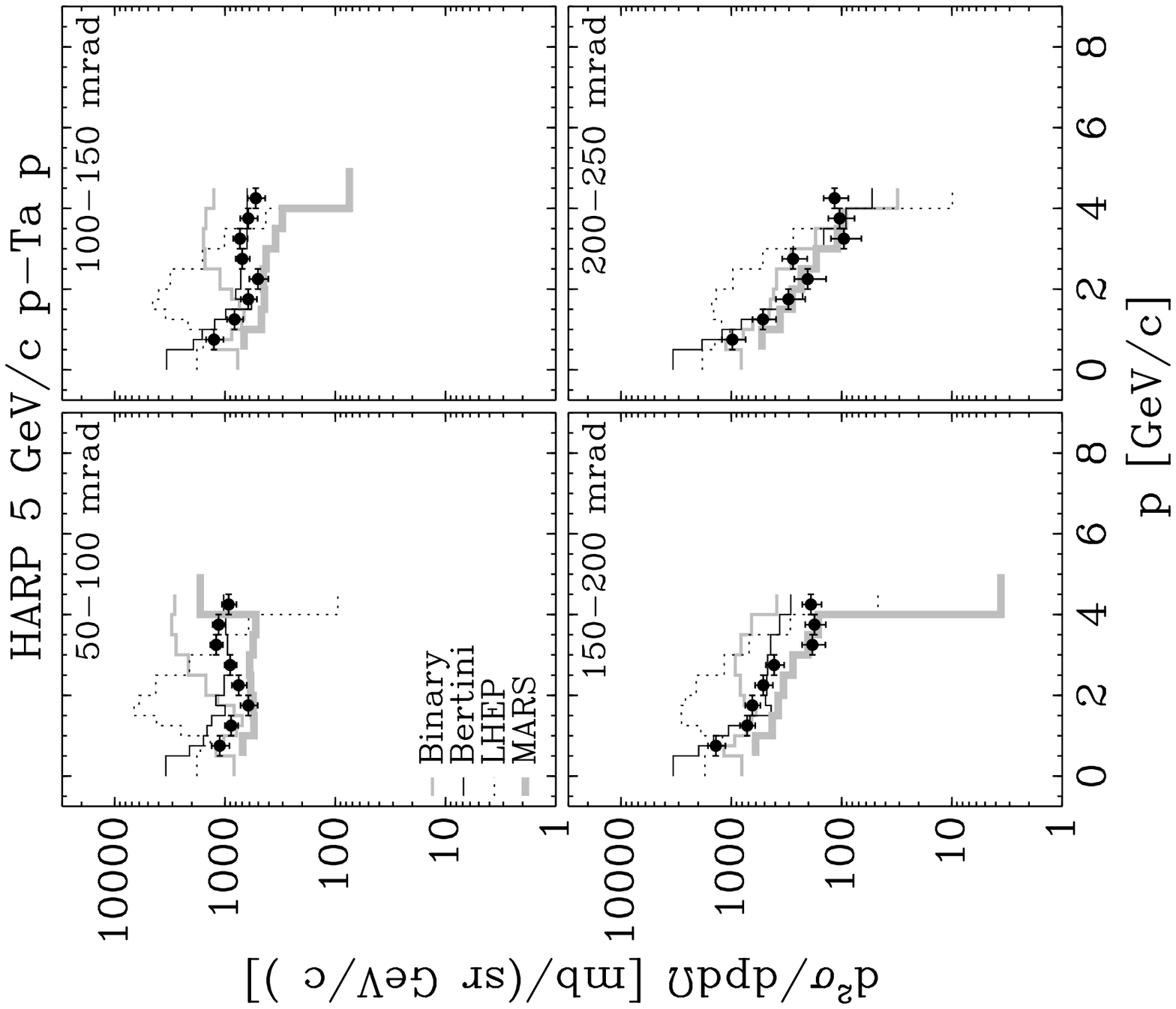}
\includegraphics[width=.42\textwidth,angle=90]{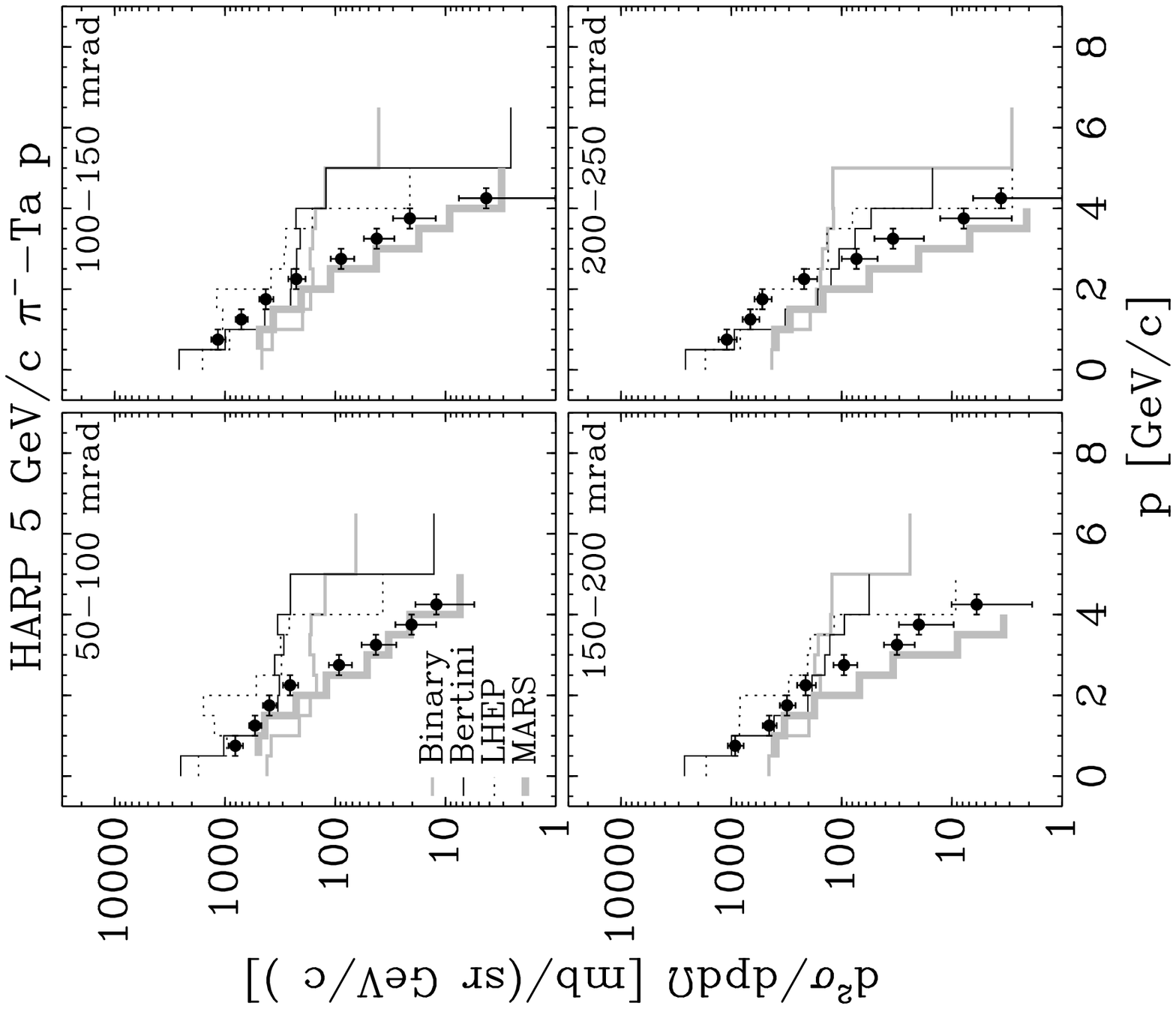}
\includegraphics[width=.42\textwidth,angle=90]{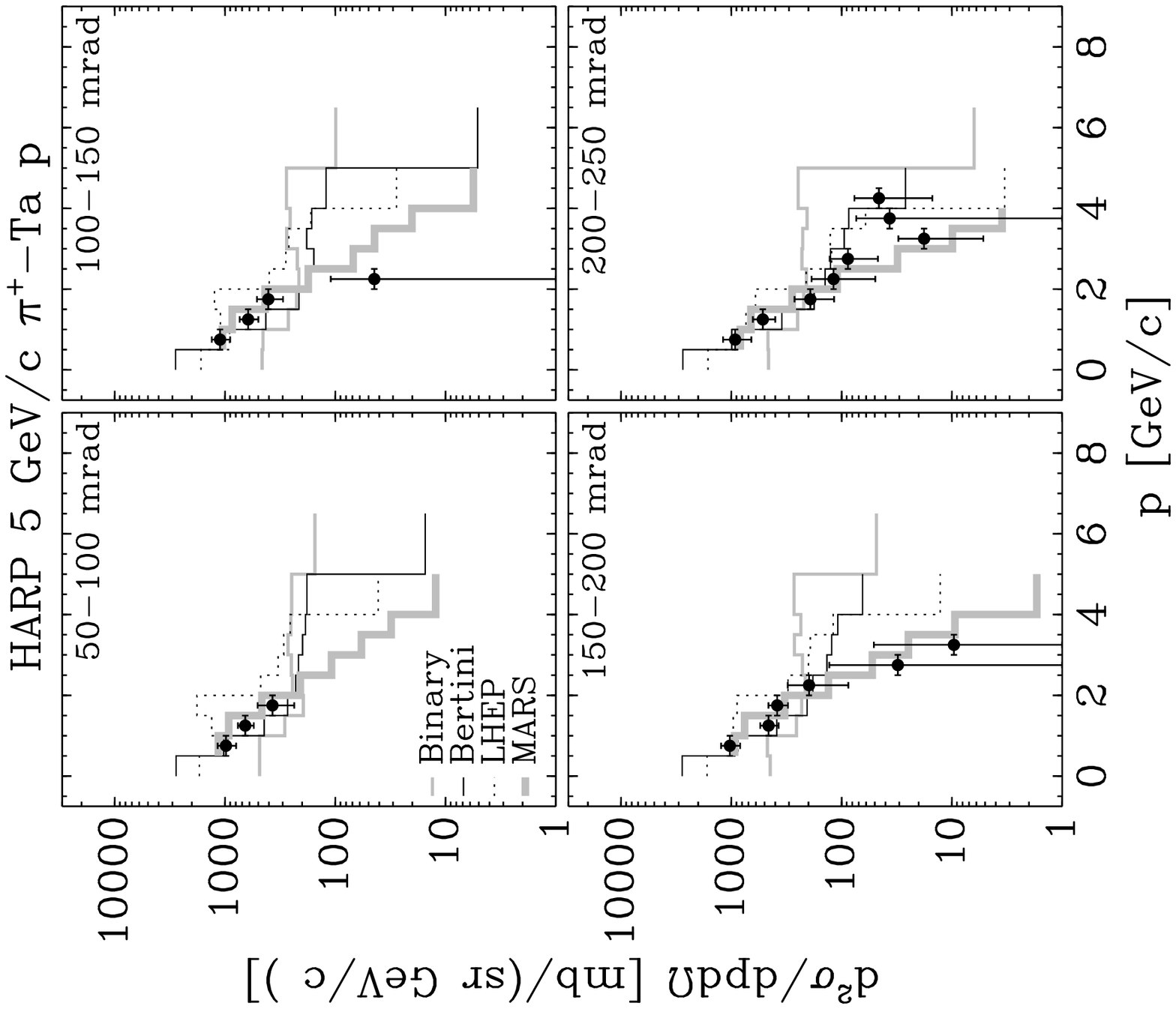}
\end{center}
\caption{
Comparison of HARP double-differential proton cross sections for p--Ta, \pim--Ta,
\pip--Ta interactions  at  5~\GeVc with
 GEANT4 and MARS MC predictions, using several generator models
(see text for details).
}
\label{fig:G54a}
\end{figure}

\begin{figure}[tbp]
\begin{center}
\includegraphics[width=.42\textwidth,angle=90]{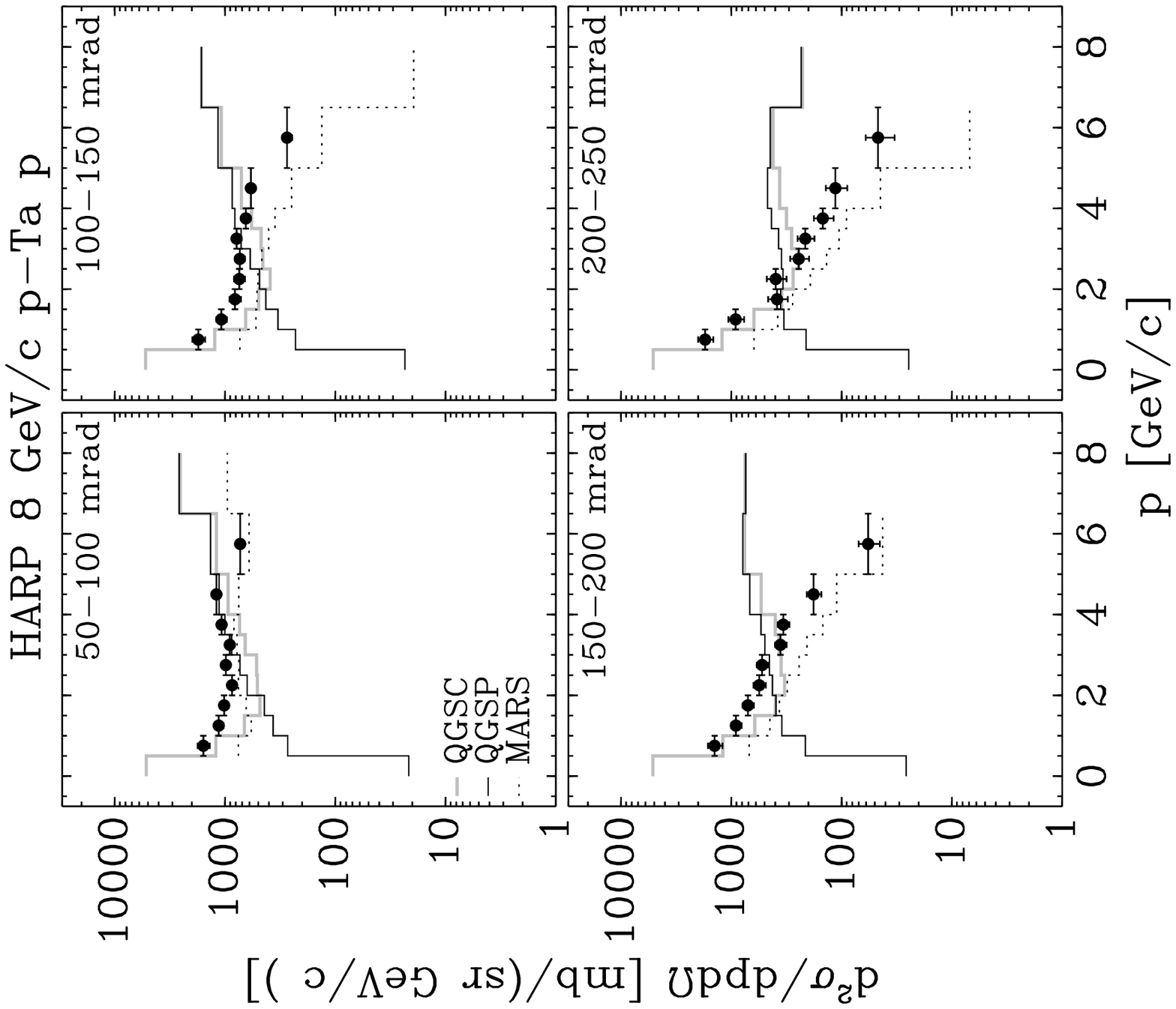}
\includegraphics[width=.42\textwidth,angle=90]{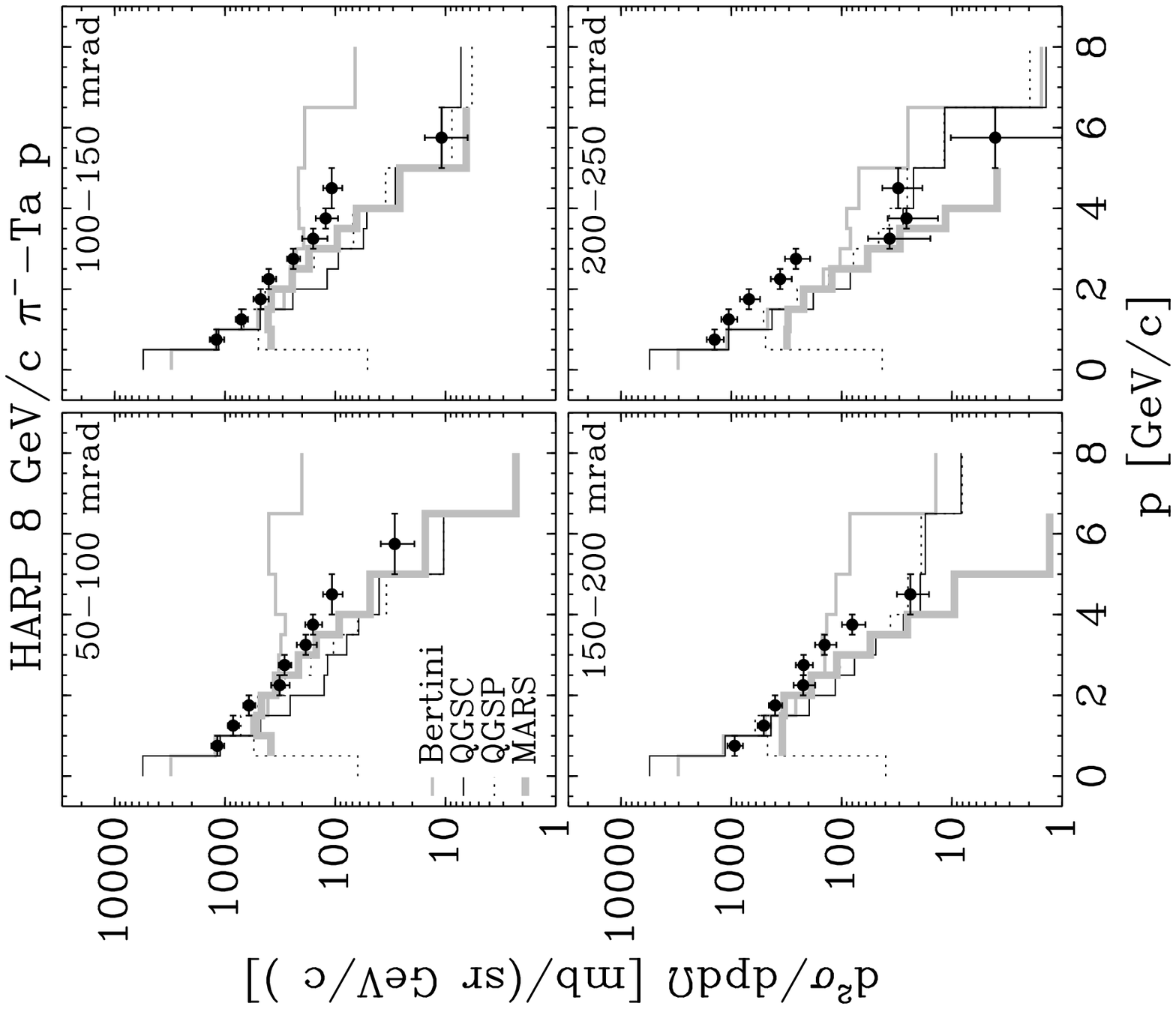}
\includegraphics[width=.42\textwidth,angle=90]{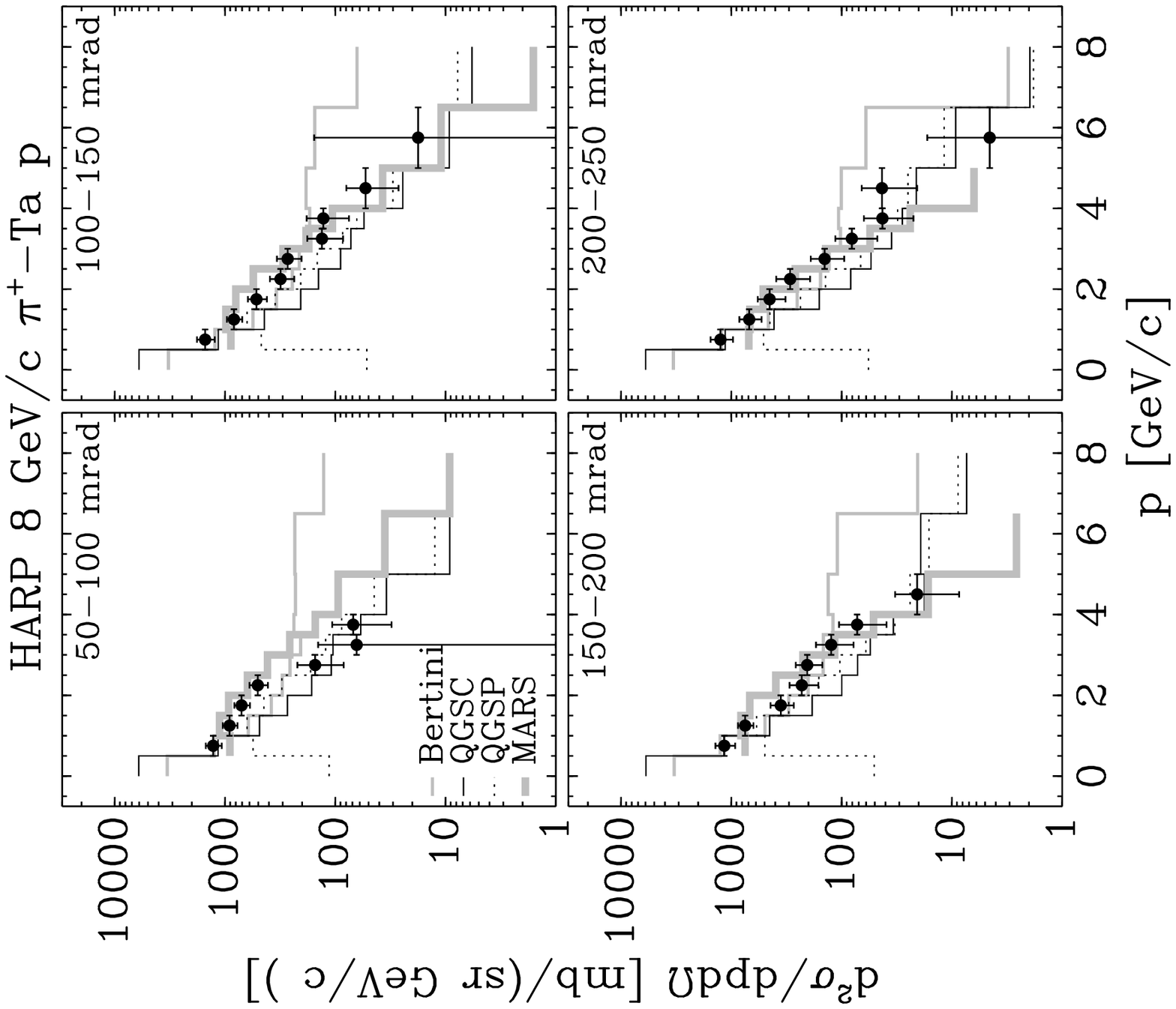}
\end{center}
\caption{
Comparison of HARP double-differential proton cross sections for p--Ta, \pim--Ta,
\pip--Ta interactions  at  8~\GeVc with
 GEANT4 and MARS MC predictions, using several generator models
(see text for details).
}
\label{fig:G55a}
\end{figure}

\begin{figure}[tbp]
\begin{center}
\includegraphics[width=.42\textwidth,angle=90]{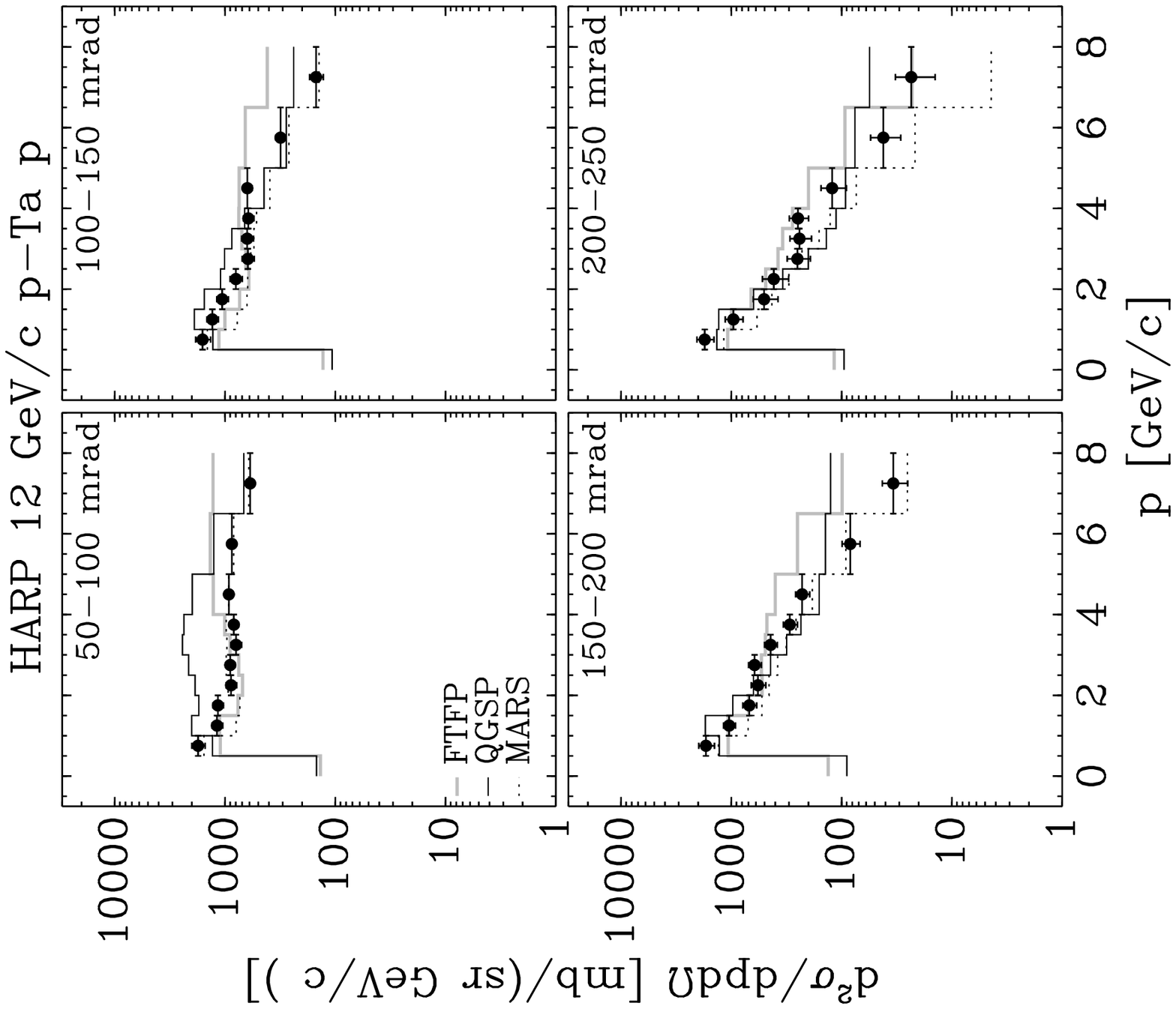}
\includegraphics[width=.42\textwidth,angle=90]{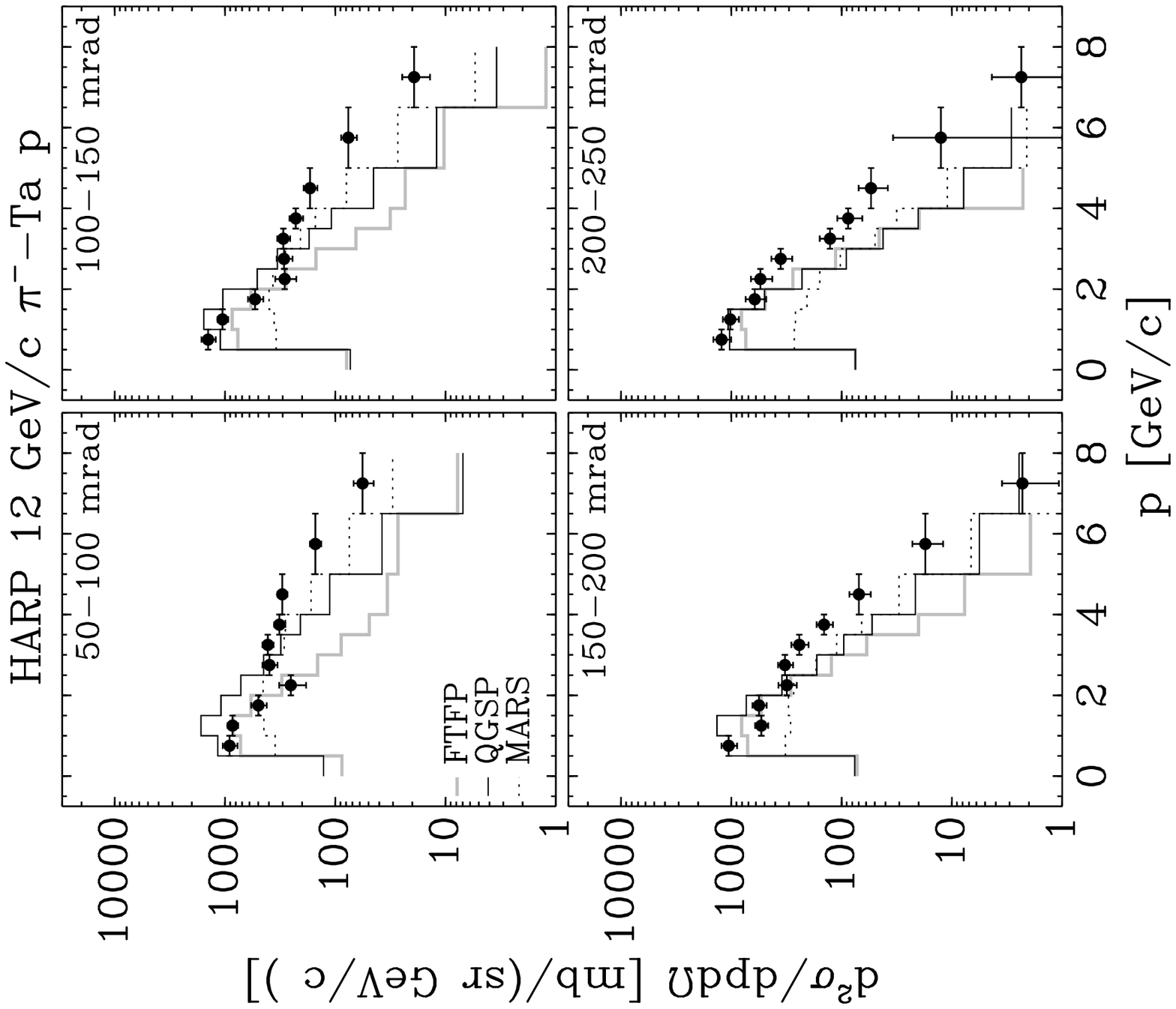}
\includegraphics[width=.42\textwidth,angle=90]{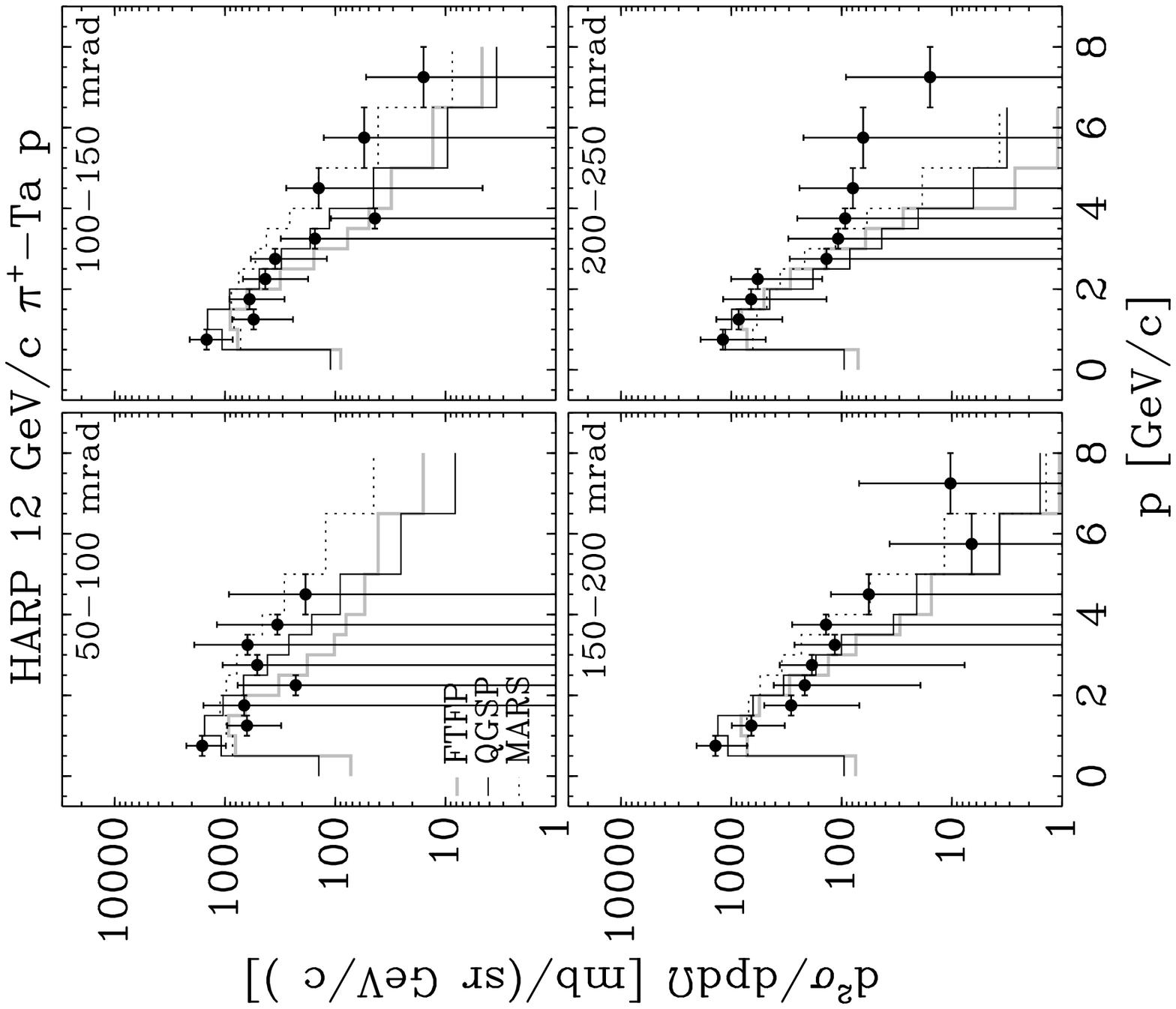}
\end{center}
\caption{
Comparison of HARP double-differential proton cross sections for p--Ta, \pim--Ta,
\pip--Ta interactions  at 12~\GeVc with
 GEANT4 and MARS MC predictions, using several generator models
(see text for details).
}
\label{fig:G56}
\end{figure}

\FloatBarrier

\section{Summary and conclusions}\label{sec:conclusions}

In this paper we report our results on
double-differential cross sections for the forward production
of protons in the kinematic range 
0.5 GeV/c$\leq p_\pi <  8$ GeV/c 
and 0.05 rad $\leq \theta_\pi < $ 0.25 rad
from the collisions
of protons and charged pions of 3, 5, 8 and 12~\GeVc 
on beryllium, carbon, aluminium, copper, 
tin, tantalum and lead targets of 5\% $\lambda_{\mathrm{I}}$ thickness.

The proton yield averaged over different
momentum and angular ranges 
increases smoothly
with the atomic number $A$ of the target and with the
energy of the incoming beam.

Comparisons with GEANT4 and MARS generators are presented. 

We stress that the HARP data presented here are the first 
precision measurements of forward proton production
in this kinematic region and may have a major impact on the
tuning of Monte Carlo generators.

\section*{Acknowledgments}

We gratefully acknowledge the help and support of the PS beam staff
and of the numerous technical collaborators who contributed to the
detector design, construction, commissioning and operation.  
In particular, we would like to thank
G.~Barichello,
R.~Brocard,
K.~Burin,
V.~Carassiti,
F.~Chignoli,
D.~Conventi,
G.~Decreuse,
M.~Delattre,
C.~Detraz,  
A.~Domeniconi,
M.~Dwuznik,   
F.~Evangelisti,
B.~Friend,
A.~Iaciofano,
I.~Krasin, 
D.~Lacroix,
J.-C.~Legrand,
M.~Lobello, 
M.~Lollo,
J.~Loquet,
F.~Marinilli,
R.~Mazza,
J.~Mulon,
L.~Musa,
R.~Nicholson,
A.~Pepato,
P.~Petev, 
X.~Pons,
I.~Rusinov,
M.~Scandurra,
E.~Usenko,
R.~van der Vlugt,
for their support in the construction of the detector
and P. Dini for his contribution to Monte Carlo production. 
The collaboration acknowledges the major contributions and advice of
M.~Baldo-Ceolin, 
L.~Linssen, 
M.T.~Muciaccia and A. Pullia
during the construction of the experiment.
The collaboration is indebted to 
V.~Ableev,
F.~Bergsma,
P.~Binko,
E.~Boter,
M.~Calvi, 
C.~Cavion,
M.Chizov, 
A.~Chukanov,
A.~DeSanto, 
A.~DeMin, 
M.~Doucet,
D.~D\"{u}llmann,
V.~Ermilova, 
W.~Flegel,
Y.~Hayato,
A.~Ichikawa,
O.~Klimov,
T.~Kobayashi,
D.~Kustov, 
M.~Laveder, 
M.~Mass,
H.~Meinhard,
A.~Menegolli, 
T.~Nakaya,
K.~Nishikawa,
M.~Paganoni,
F.~Paleari,
M.~Pasquali,
M.~Placentino,
V.~Serdiouk,
S.~Simone,
P.J.~Soler,
S.~Troquereau,
S.~Ueda,
A.~Valassi and
R.~Veenhof
for their contributions to the experiment.

We acknowledge the contributions of 
V.~Ammosov,
G.~Chelkov,
D.~Dedovich,
F.~Dydak,
M.~Gostkin,
A.~Guskov,
D.~Khartchenko,
V.~Koreshev,
Z.~Kroumchtein,
I.~Nefedov,
A.~Semak,
J.~Wotschack,
V.~Zaets and
A.~Zhemchugov
to the work described in this paper.

 The experiment was made possible by grants from
the Institut Interuniversitaire des Sciences Nucl\'eair\-es and the
Interuniversitair Instituut voor Kernwetenschappen (Belgium), 
Ministerio de Educacion y Ciencia, Grant FPA2003-06921-c02-02 and
Generalitat Valenciana, grant GV00-054-1,
CERN (Geneva, Switzerland), 
the German Bundesministerium f\"ur Bildung und Forschung (Germany), 
the Istituto Na\-zio\-na\-le di Fisica Nucleare (Italy), 
INR RAS (Moscow), the Russian Foundation for Basic Research 
(grant 08-02-00018), the Bulgarian Mational Science Fund 
(contract VU-F-205/2006) 
and the Particle Physics and Astronomy Research Council (UK).
We gratefully acknowledge their support.

\clearpage 

\begin{appendix}
\section{Cross-section data}\label{app:data}
The following tables report the measured differential cross-section
for forward proton production in interactions
of 3, 5, 8 and 12~\GeVc momentum charged pions or protons
on different types of nuclear targets.
The data are presented in the kinematic range of
0.5~\GeVc$\leq p_\pi <  8$~\GeVc
and 0.05~rad $\leq \theta_\pi < $ 0.25~rad.
The overall normalization uncertainty ($ \leq 2 \%$) is not included
in the reported errors.
\clearpage


\begin{table}[!ht]
  \caption{\label{tab:xsec_results_Be_pim}
    HARP results for the double-differential $p$  production
    cross-section in the laboratory system,
    $d^2\sigma^{p}/(dpd\Omega)$, for $\pi^{-}$--Be interactions at 3,5,8,12~\GeVc.
    Each row refers to a
    different $(p_{\hbox{\small min}} \le p<p_{\hbox{\small max}},
    \theta_{\hbox{\small min}} \le \theta<\theta_{\hbox{\small max}})$ bin,
    where $p$ and $\theta$ are the outgoing proton momentum and polar angle, respectively.
    The central value as well as the square-root of the diagonal elements
    of the covariance matrix are given.}

\small{
\begin{tabular}{rrrr|r@{$\pm$}lr@{$\pm$}lr@{$\pm$}lr@{$\pm$}l}
\hline
$\theta_{\hbox{\small min}}$ &
$\theta_{\hbox{\small max}}$ &
$p_{\hbox{\small min}}$ &
$p_{\hbox{\small max}}$ &
\multicolumn{8}{c}{$d^2\sigma^{p}/(dpd\Omega)$}
\\
(rad) & (rad) & (\GeVc) & (\GeVc) &
\multicolumn{8}{c}{(barn/(sr \GeVc ))}
\\
  &  &  &
&\multicolumn{2}{c}{$ \bf{3 \ \GeVc}$}
&\multicolumn{2}{c}{$ \bf{5 \ \GeVc}$}
&\multicolumn{2}{c}{$ \bf{8 \ \GeVc}$}
&\multicolumn{2}{c}{$ \bf{12 \ \GeVc}$}
\\
\hline

0.050 &0.100 & 0.50 & 1.00& 0.070 &  0.012& 0.045 &  0.010& 0.051 &  0.011& 0.024 &  0.007\\ 
      &      & 1.00 & 1.50& 0.044 &  0.007& 0.055 &  0.007& 0.045 &  0.008& 0.034 &  0.006\\ 
      &      & 1.50 & 2.00& 0.039 &  0.006& 0.039 &  0.005& 0.058 &  0.008& 0.021 &  0.022\\ 
      &      & 2.00 & 2.50& 0.022 &  0.004& 0.033 &  0.005& 0.037 &  0.007& 0.007 &  0.006\\ 
      &      & 2.50 & 3.00&       &       & 0.033 &  0.004& 0.045 &  0.005&  0.04 &   0.02\\ 
      &      & 3.00 & 3.50&       &       & 0.013 &  0.002& 0.039 &  0.006& 0.041 &  0.006\\ 
      &      & 3.50 & 4.00&       &       & 0.011 &  0.002& 0.031 &  0.004& 0.028 &  0.004\\ 
      &      & 4.00 & 5.00&       &       & 0.005 &  0.001& 0.024 &  0.003& 0.029 &  0.003\\ 
      &      & 5.00 & 6.50&       &       &       &       & 0.006 &  0.001& 0.017 &  0.002\\ 
      &      & 6.50 & 8.00&       &       &       &       &       &       & 0.006 &  0.001\\ 
0.100 &0.150 & 0.50 & 1.00& 0.050 &  0.011& 0.067 &  0.012& 0.054 &  0.013& 0.030 &  0.008\\ 
      &      & 1.00 & 1.50& 0.054 &  0.008& 0.074 &  0.009& 0.063 &  0.009& 0.034 &  0.006\\ 
      &      & 1.50 & 2.00& 0.042 &  0.007& 0.056 &  0.007& 0.055 &  0.009& 0.027 &  0.006\\ 
      &      & 2.00 & 2.50& 0.024 &  0.004& 0.040 &  0.005& 0.049 &  0.007& 0.019 &  0.006\\ 
      &      & 2.50 & 3.00&       &       & 0.017 &  0.003& 0.035 &  0.004& 0.027 &  0.004\\ 
      &      & 3.00 & 3.50&       &       & 0.013 &  0.002& 0.024 &  0.005& 0.025 &  0.003\\ 
      &      & 3.50 & 4.00&       &       & 0.008 &  0.001& 0.025 &  0.004& 0.023 &  0.003\\ 
      &      & 4.00 & 5.00&       &       & 0.006 &  0.002& 0.015 &  0.003& 0.015 &  0.002\\ 
      &      & 5.00 & 6.50&       &       &       &       & 0.003 &  0.001& 0.005 &  0.001\\ 
      &      & 6.50 & 8.00&       &       &       &       &       &       &       &       \\ 
0.150 &0.200 & 0.50 & 1.00& 0.041 &  0.010& 0.055 &  0.012& 0.053 &  0.013& 0.025 &  0.008\\ 
      &      & 1.00 & 1.50& 0.034 &  0.006& 0.040 &  0.006& 0.040 &  0.007& 0.031 &  0.005\\ 
      &      & 1.50 & 2.00& 0.029 &  0.005& 0.048 &  0.006& 0.039 &  0.006& 0.025 &  0.006\\ 
      &      & 2.00 & 2.50& 0.028 &  0.005& 0.034 &  0.005& 0.039 &  0.007& 0.023 &  0.005\\ 
      &      & 2.50 & 3.00&       &       & 0.016 &  0.003& 0.028 &  0.005& 0.026 &  0.004\\ 
      &      & 3.00 & 3.50&       &       & 0.004 &  0.001& 0.015 &  0.004& 0.013 &  0.002\\ 
      &      & 3.50 & 4.00&       &       & 0.006 &  0.002& 0.010 &  0.002& 0.009 &  0.002\\ 
      &      & 4.00 & 5.00&       &       & 0.001 &  0.001& 0.006 &  0.001& 0.007 &  0.001\\ 
      &      & 5.00 & 6.50&       &       &       &       &       &       &       &       \\ 
      &      & 6.50 & 8.00&       &       &       &       &       &       &       &       \\ 
0.200 &0.250 & 0.50 & 1.00&  0.08 &   0.02& 0.028 &  0.010& 0.043 &  0.014& 0.034 &  0.010\\ 
      &      & 1.00 & 1.50& 0.065 &  0.012& 0.087 &  0.013& 0.069 &  0.015& 0.043 &  0.009\\ 
      &      & 1.50 & 2.00& 0.038 &  0.008& 0.043 &  0.008&  0.07 &   0.02& 0.033 &  0.010\\ 
      &      & 2.00 & 2.50& 0.018 &  0.006& 0.035 &  0.007& 0.051 &  0.011& 0.029 &  0.009\\ 
      &      & 2.50 & 3.00&       &       & 0.017 &  0.004& 0.034 &  0.007& 0.026 &  0.005\\ 
      &      & 3.00 & 3.50&       &       & 0.007 &  0.002& 0.011 &  0.003& 0.010 &  0.002\\ 
      &      & 3.50 & 4.00&       &       &       &       & 0.006 &  0.002& 0.007 &  0.002\\ 
      &      & 4.00 & 5.00&       &       &       &       & 0.004 &  0.001& 0.003 &  0.001\\ 
      &      & 5.00 & 6.50&       &       &       &       & 0.000 &  0.001&       &       \\ 
      &      & 6.50 & 8.00&       &       &       &       &       &       &       &       \\ 
\hline
\end{tabular}
}
\end{table}
\clearpage

\begin{table}[!ht]
  \caption{\label{tab:xsec_results_Be_pip}
    HARP results for the double-differential $p$  production
    cross-section in the laboratory system,
    $d^2\sigma^{p}/(dpd\Omega)$, for $\pi^{+}$--Be interactions at 3,5,8,8.9~\GeVc.
    Each row refers to a
    different $(p_{\hbox{\small min}} \le p<p_{\hbox{\small max}},
    \theta_{\hbox{\small min}} \le \theta<\theta_{\hbox{\small max}})$ bin,
    where $p$ and $\theta$ are the outgoing proton momentum and polar angle, respectively.
    The central value as well as the square-root of the diagonal elements
    of the covariance matrix are given.}

\small{
\begin{tabular}{rrrr|r@{$\pm$}lr@{$\pm$}lr@{$\pm$}lr@{$\pm$}l}
\hline
$\theta_{\hbox{\small min}}$ &
$\theta_{\hbox{\small max}}$ &
$p_{\hbox{\small min}}$ &
$p_{\hbox{\small max}}$ &
\multicolumn{8}{c}{$d^2\sigma^{p}/(dpd\Omega)$}
\\
(rad) & (rad) & (\GeVc) & (\GeVc) &
\multicolumn{8}{c}{(barn/(sr \GeVc ))}
\\
  &  &  &
&\multicolumn{2}{c}{$ \bf{3 \ \GeVc}$}
&\multicolumn{2}{c}{$ \bf{5 \ \GeVc}$}
&\multicolumn{2}{c}{$ \bf{8 \ \GeVc}$}
&\multicolumn{2}{c}{$ \bf{8.9 \ \GeVc}$}
\\
\hline

0.050 &0.100 & 0.50 & 1.00&  0.07 &   0.03&  0.05 &   0.02& 0.036 &  0.014& 0.042 &  0.012\\ 
      &      & 1.00 & 1.50&  0.07 &   0.02& 0.080 &  0.015& 0.054 &  0.012& 0.061 &  0.009\\ 
      &      & 1.50 & 2.00&       &  0.000& 0.055 &  0.014& 0.053 &  0.012& 0.053 &  0.009\\ 
      &      & 2.00 & 2.50&       &   0.02&  0.04 &   0.02& 0.047 &  0.011& 0.047 &  0.006\\ 
      &      & 2.50 & 3.00&       &       &       &       & 0.024 &  0.012& 0.047 &  0.007\\ 
      &      & 3.00 & 3.50&       &       &       &       & 0.031 &  0.012& 0.042 &  0.006\\ 
      &      & 3.50 & 4.00&       &       &       &       & 0.004 &  0.021& 0.048 &  0.005\\ 
      &      & 4.00 & 5.00&       &       &       &       &       &       & 0.042 &  0.004\\ 
      &      & 5.00 & 6.50&       &       &       &       &       &       & 0.020 &  0.003\\ 
0.100 &0.150 & 0.50 & 1.00&  0.06 &   0.03&  0.05 &   0.02&  0.06 &   0.02& 0.049 &  0.015\\ 
      &      & 1.00 & 1.50&  0.07 &   0.02& 0.053 &  0.013& 0.059 &  0.013& 0.047 &  0.009\\ 
      &      & 1.50 & 2.00&  0.03 &   0.02& 0.064 &  0.014& 0.045 &  0.011& 0.053 &  0.007\\ 
      &      & 2.00 & 2.50&       &       & 0.030 &  0.012& 0.032 &  0.009& 0.045 &  0.007\\ 
      &      & 2.50 & 3.00&       &       & 0.005 &  0.012& 0.015 &  0.006& 0.044 &  0.006\\ 
      &      & 3.00 & 3.50&       &       &       &       & 0.024 &  0.007& 0.037 &  0.005\\ 
      &      & 3.50 & 4.00&       &       &       &       & 0.022 &  0.007& 0.035 &  0.004\\ 
      &      & 4.00 & 5.00&       &       &       &       & 0.011 &  0.003& 0.024 &  0.003\\ 
      &      & 5.00 & 6.50&       &       &       &       & 0.004 &  0.003& 0.007 &  0.001\\ 
0.150 &0.200 & 0.50 & 1.00&  0.08 &   0.03&  0.07 &   0.02& 0.042 &  0.014& 0.047 &  0.011\\ 
      &      & 1.00 & 1.50&  0.06 &   0.02& 0.049 &  0.011& 0.036 &  0.010& 0.054 &  0.006\\ 
      &      & 1.50 & 2.00&  0.02 &   0.02& 0.040 &  0.009& 0.033 &  0.010& 0.028 &  0.006\\ 
      &      & 2.00 & 2.50&       &       & 0.035 &  0.011& 0.023 &  0.008& 0.031 &  0.007\\ 
      &      & 2.50 & 3.00&       &       & 0.012 &  0.007& 0.016 &  0.005& 0.024 &  0.005\\ 
      &      & 3.00 & 3.50&       &       & 0.003 &  0.004& 0.020 &  0.007& 0.012 &  0.004\\ 
      &      & 3.50 & 4.00&       &       & 0.009 &  0.005& 0.010 &  0.004& 0.011 &  0.003\\ 
      &      & 4.00 & 5.00&       &       & 0.007 &  0.007& 0.004 &  0.002& 0.009 &  0.002\\ 
      &      & 5.00 & 6.50&       &       &       &       & 0.000 &  0.001&       &       \\ 
0.200 &0.250 & 0.50 & 1.00&  0.06 &   0.03&  0.05 &   0.02&  0.05 &   0.02& 0.042 &  0.013\\ 
      &      & 1.00 & 1.50&  0.05 &   0.02& 0.043 &  0.013&  0.05 &   0.02& 0.048 &  0.009\\ 
      &      & 1.50 & 2.00& 0.041 &  0.013& 0.034 &  0.012& 0.031 &  0.013& 0.038 &  0.007\\ 
      &      & 2.00 & 2.50&  0.02 &   0.02& 0.035 &  0.012& 0.033 &  0.012& 0.028 &  0.006\\ 
      &      & 2.50 & 3.00&       &       & 0.016 &  0.006& 0.013 &  0.009& 0.017 &  0.004\\ 
      &      & 3.00 & 3.50&       &       & 0.009 &  0.004& 0.012 &  0.004& 0.009 &  0.002\\ 
      &      & 3.50 & 4.00&       &       & 0.007 &  0.003& 0.006 &  0.003& 0.006 &  0.002\\ 
      &      & 4.00 & 5.00&       &       & 0.006 &  0.003& 0.009 &  0.005& 0.003 &  0.002\\ 
      &      & 5.00 & 6.50&       &       &       &       & 0.002 &  0.003& 0.001 &  0.001\\ 
\hline
\end{tabular}
}
\end{table}
\clearpage

\begin{table}[!ht]
  \caption{\label{tab:xsec_results_Be_pr}
    HARP results for the double-differential $p$  production
    cross-section in the laboratory system,
    $d^2\sigma^{p}/(dpd\Omega)$, for p--Be interactions at 3,5,8,8.9,12~\GeVc.
    Each row refers to a
    different $(p_{\hbox{\small min}} \le p<p_{\hbox{\small max}},
    \theta_{\hbox{\small min}} \le \theta<\theta_{\hbox{\small max}})$ bin,
    where $p$ and $\theta$ are the outgoing proton momentum and polar angle, respectively.
    The central value as well as the square-root of the diagonal elements
    of the covariance matrix are given.}
\small{
\begin{tabular}{rrrr|r@{$\pm$}lr@{$\pm$}lr@{$\pm$}lr@{$\pm$}lr@{$\pm$}l}
\hline
$\theta_{\hbox{\small min}}$ &
$\theta_{\hbox{\small max}}$ &
$p_{\hbox{\small min}}$ &
$p_{\hbox{\small max}}$ &
\multicolumn{10}{c}{$d^2\sigma^{p}/(dpd\Omega)$}
\\
(rad) & (rad) & (\GeVc) & (\GeVc) &
\multicolumn{10}{c}{(barn/(sr \GeVc))}
\\
  &  &  &
&\multicolumn{2}{c}{$ \bf{3 \ \GeVc}$}
&\multicolumn{2}{c}{$ \bf{5 \ \GeVc}$}
&\multicolumn{2}{c}{$ \bf{8 \ \GeVc}$}
&\multicolumn{2}{c}{$ \bf{8.9 \ \GeVc}$}
&\multicolumn{2}{c}{$ \bf{12 \ \GeVc}$}
\\
\hline

0.050 &0.100 & 0.50 & 1.00&  0.08 &   0.03&  0.08 &   0.02& 0.067 &  0.013& 0.079 &  0.013& 0.071 &  0.014\\ 
      &      & 1.00 & 1.50&  0.16 &   0.03&  0.11 &   0.02& 0.091 &  0.011& 0.080 &  0.010& 0.062 &  0.010\\ 
      &      & 1.50 & 2.00&  0.29 &   0.04&  0.15 &   0.02& 0.102 &  0.011& 0.100 &  0.010& 0.080 &  0.011\\ 
      &      & 2.00 & 2.50&  0.25 &   0.08&  0.22 &   0.02& 0.121 &  0.010& 0.116 &  0.010& 0.078 &  0.009\\ 
      &      & 2.50 & 3.00&       &       &  0.28 &   0.02& 0.176 &  0.015& 0.134 &  0.010& 0.105 &  0.010\\ 
      &      & 3.00 & 3.50&       &       &  0.39 &   0.03& 0.210 &  0.015& 0.170 &  0.013& 0.123 &  0.011\\ 
      &      & 3.50 & 4.00&       &       &  0.37 &   0.03&  0.24 &   0.02& 0.197 &  0.013& 0.143 &  0.012\\ 
      &      & 4.00 & 5.00&       &       &  0.44 &   0.07& 0.274  &  0.013 & 0.223 &  0.010& 0.173 &  0.010\\ 
      &      & 5.00 & 6.50&       &       &       &       & 0.218  &  0.008 & 0.206 &  0.008& 0.169 &  0.008\\ 
      &      & 6.50 & 8.00&       &       &       &       &       &       &       &       & 0.142 &  0.006\\ 
0.100 &0.150 & 0.50 & 1.00&  0.13 &   0.04&  0.08 &   0.02& 0.080 &  0.015& 0.082 &  0.014& 0.067 &  0.015\\ 
      &      & 1.00 & 1.50&  0.16 &   0.03&  0.09 &   0.02& 0.095 &  0.012& 0.093 &  0.011& 0.080 &  0.012\\ 
      &      & 1.50 & 2.00&  0.25 &   0.04&  0.15 &   0.02& 0.112 &  0.013& 0.092 &  0.010& 0.077 &  0.011\\ 
      &      & 2.00 & 2.50&  0.29 &   0.05&  0.17 &   0.02& 0.106 &  0.011& 0.092 &  0.009& 0.088 &  0.011\\ 
      &      & 2.50 & 3.00&       &       &  0.20 &   0.02& 0.133 &  0.013& 0.113 &  0.011& 0.075 &  0.008\\ 
      &      & 3.00 & 3.50&       &       &  0.19 &   0.02& 0.151 &  0.012& 0.121 &  0.009& 0.081 &  0.008\\ 
      &      & 3.50 & 4.00&       &       &  0.20 &   0.02& 0.135 &  0.010& 0.112 &  0.008& 0.091 &  0.009\\ 
      &      & 4.00 & 5.00&       &       &  0.18 &   0.02& 0.124 &  0.010& 0.109 &  0.008& 0.096 &  0.008\\ 
      &      & 5.00 & 6.50&       &       &       &       & 0.060 &  0.005& 0.058 &  0.005& 0.060 &  0.005\\ 
      &      & 6.50 & 8.00&       &       &       &       &       &       &       &       & 0.032 &  0.003\\ 
0.150 &0.200 & 0.50 & 1.00&  0.10 &   0.04&  0.09 &   0.02&  0.09 &   0.02& 0.078 &  0.014&  0.08 &   0.02\\ 
      &      & 1.00 & 1.50&  0.13 &   0.03&  0.10 &   0.02& 0.068 &  0.009& 0.067 &  0.007& 0.051 &  0.009\\ 
      &      & 1.50 & 2.00&  0.18 &   0.03& 0.10  &  0.01 & 0.067 &  0.009& 0.060 &  0.008& 0.054 &  0.009\\ 
      &      & 2.00 & 2.50&  0.16 &   0.03&  0.11 &   0.02& 0.091 &  0.011& 0.072 &  0.008& 0.049 &  0.008\\ 
      &      & 2.50 & 3.00&       &       & 0.104 &  0.014& 0.085 &  0.009& 0.058 &  0.006& 0.043 &  0.006\\ 
      &      & 3.00 & 3.50&       &       & 0.082 &  0.012& 0.068 &  0.007& 0.062 &  0.006& 0.040 &  0.005\\ 
      &      & 3.50 & 4.00&       &       & 0.042 &  0.008& 0.053 &  0.005& 0.047 &  0.005& 0.040 &  0.006\\ 
      &      & 4.00 & 5.00&       &       & 0.039 &  0.007& 0.046 &  0.005& 0.033 &  0.003& 0.026 &  0.004\\ 
      &      & 5.00 & 6.50&       &       &       &       & 0.015 &  0.002& 0.014 &  0.002& 0.010 &  0.002\\ 
      &      & 6.50 & 8.00&       &       &       &       &       &       &       &       & 0.003 &  0.001\\ 
0.200 &0.250 & 0.50 & 1.00&  0.10 &   0.04&  0.07 &   0.03&  0.11 &   0.02& 0.083 &  0.015&  0.08 &   0.02\\ 
      &      & 1.00 & 1.50&  0.07 &   0.03&  0.11 &   0.02&  0.09 &   0.02& 0.071 &  0.010&  0.07 &   0.02\\ 
      &      & 1.50 & 2.00&  0.09 &   0.02&  0.06 &   0.02& 0.051 &  0.011& 0.056 &  0.009& 0.034 &  0.010\\ 
      &      & 2.00 & 2.50&  0.05 &   0.02& 0.046 &  0.012& 0.071 &  0.012& 0.050 &  0.007& 0.030 &  0.008\\ 
      &      & 2.50 & 3.00&       &       & 0.054 &  0.011& 0.047 &  0.008& 0.040 &  0.006& 0.025 &  0.006\\ 
      &      & 3.00 & 3.50&       &       & 0.039 &  0.008& 0.034 &  0.005& 0.022 &  0.003& 0.020 &  0.004\\ 
      &      & 3.50 & 4.00&       &       & 0.030 &  0.007& 0.025 &  0.004& 0.017 &  0.003& 0.012 &  0.003\\ 
      &      & 4.00 & 5.00&       &       & 0.018 &  0.005& 0.022 &  0.004& 0.018 &  0.003& 0.009 &  0.003\\ 
      &      & 5.00 & 6.50&       &       &       &       & 0.008 &  0.002& 0.006 &  0.001& 0.003 &  0.001\\ 
      &      & 6.50 & 8.00&       &       &       &       &       &       &       &       & 0.001 &  0.001\\ 

\hline
\end{tabular}
}
\end{table}
\clearpage
\begin{table}[!ht]
  \caption{\label{tab:xsec_results_C_pim}
    HARP results for the double-differential $p$  production
    cross-section in the laboratory system,
    $d^2\sigma^{p}/(dpd\Omega)$, for $\pi^{-}$--C interactions at 3,5,8,12~\GeVc.
    Each row refers to a
    different $(p_{\hbox{\small min}} \le p<p_{\hbox{\small max}},
    \theta_{\hbox{\small min}} \le \theta<\theta_{\hbox{\small max}})$ bin,
    where $p$ and $\theta$ are the outgoing proton momentum and polar angle, respectively.
    The central value as well as the square-root of the diagonal elements
    of the covariance matrix are given.}

\small{
\begin{tabular}{rrrr|r@{$\pm$}lr@{$\pm$}lr@{$\pm$}lr@{$\pm$}l}
\hline
$\theta_{\hbox{\small min}}$ &
$\theta_{\hbox{\small max}}$ &
$p_{\hbox{\small min}}$ &
$p_{\hbox{\small max}}$ &
\multicolumn{8}{c}{$d^2\sigma^{p}/(dpd\Omega)$}
\\
(rad) & (rad) & (\GeVc) & (\GeVc) &
\multicolumn{8}{c}{(barn/(sr \GeVc ))}
\\
  &  &  &
&\multicolumn{2}{c}{$ \bf{3 \ \GeVc}$}
&\multicolumn{2}{c}{$ \bf{5 \ \GeVc}$}
&\multicolumn{2}{c}{$ \bf{8 \ \GeVc}$}
&\multicolumn{2}{c}{$ \bf{12 \ \GeVc}$}
\\
\hline

0.050 &0.100 & 0.50 & 1.00&  0.11 &   0.02&  0.08 &   0.02&  0.08 &   0.02&  0.05 &   0.02\\ 
      &      & 1.00 & 1.50& 0.056 &  0.014& 0.068 &  0.011& 0.057 &  0.012& 0.060 &  0.013\\ 
      &      & 1.50 & 2.00& 0.054 &  0.011& 0.069 &  0.010& 0.077 &  0.012&  0.05 &   0.03\\ 
      &      & 2.00 & 2.50& 0.031 &  0.008& 0.045 &  0.007& 0.051 &  0.010& 0.005 &  0.006\\ 
      &      & 2.50 & 3.00&       &       & 0.026 &  0.005& 0.052 &  0.007& 0.035 &  0.011\\ 
      &      & 3.00 & 3.50&       &       & 0.018 &  0.005& 0.041 &  0.007& 0.056 &  0.009\\ 
      &      & 3.50 & 4.00&       &       & 0.012 &  0.003& 0.043 &  0.006& 0.052 &  0.007\\ 
      &      & 4.00 & 5.00&       &       & 0.008 &  0.003& 0.020 &  0.004& 0.055 &  0.006\\ 
      &      & 5.00 & 6.50&       &       &       &       & 0.007 &  0.002& 0.026 &  0.004\\ 
      &      & 6.50 & 8.00&       &       &       &       &       &       & 0.011 &  0.002\\ 
0.100 &0.150 & 0.50 & 1.00&  0.08 &   0.02&  0.07 &   0.02&  0.07 &   0.02&  0.05 &   0.02\\ 
      &      & 1.00 & 1.50&  0.07 &   0.02& 0.077 &  0.013& 0.069 &  0.012& 0.065 &  0.014\\ 
      &      & 1.50 & 2.00& 0.071 &  0.015& 0.060 &  0.010& 0.061 &  0.011& 0.036 &  0.011\\ 
      &      & 2.00 & 2.50& 0.022 &  0.007& 0.047 &  0.008& 0.048 &  0.008& 0.034 &  0.012\\ 
      &      & 2.50 & 3.00&       &       & 0.029 &  0.006& 0.051 &  0.006& 0.036 &  0.008\\ 
      &      & 3.00 & 3.50&       &       & 0.011 &  0.003& 0.025 &  0.006& 0.048 &  0.008\\ 
      &      & 3.50 & 4.00&       &       & 0.011 &  0.003& 0.022 &  0.004& 0.047 &  0.007\\ 
      &      & 4.00 & 5.00&       &       & 0.005 &  0.002& 0.011 &  0.003& 0.031 &  0.005\\ 
      &      & 5.00 & 6.50&       &       &       &       & 0.003 &  0.001& 0.009 &  0.002\\ 
      &      & 6.50 & 8.00&       &       &       &       &       &       & 0.003 &  0.001\\ 
0.150 &0.200 & 0.50 & 1.00&  0.09 &   0.03&  0.08 &   0.02&  0.06 &   0.02&  0.07 &   0.02\\ 
      &      & 1.00 & 1.50& 0.036 &  0.010& 0.048 &  0.009& 0.044 &  0.009& 0.041 &  0.010\\ 
      &      & 1.50 & 2.00& 0.037 &  0.010& 0.060 &  0.010& 0.053 &  0.008& 0.047 &  0.012\\ 
      &      & 2.00 & 2.50& 0.031 &  0.009& 0.034 &  0.008& 0.048 &  0.010& 0.022 &  0.008\\ 
      &      & 2.50 & 3.00&       &       & 0.021 &  0.005& 0.025 &  0.005& 0.033 &  0.008\\ 
      &      & 3.00 & 3.50&       &       & 0.010 &  0.003& 0.021 &  0.005& 0.041 &  0.009\\ 
      &      & 3.50 & 4.00&       &       & 0.003 &  0.002& 0.013 &  0.003& 0.019 &  0.006\\ 
      &      & 4.00 & 5.00&       &       &       &       & 0.007 &  0.002& 0.007 &  0.002\\ 
      &      & 5.00 & 6.50&       &       &       &       &       &       & 0.002 &  0.001\\ 
      &      & 6.50 & 8.00&       &       &       &       &       &       &       &       \\ 
0.200 &0.250 & 0.50 & 1.00&  0.09 &   0.03&  0.09 &   0.02&  0.06 &   0.02&  0.05 &   0.02\\ 
      &      & 1.00 & 1.50&  0.11 &   0.03&  0.07 &   0.02&  0.09 &   0.02&  0.06 &   0.02\\ 
      &      & 1.50 & 2.00&  0.05 &   0.02& 0.064 &  0.014&  0.06 &   0.02&  0.05 &   0.02\\ 
      &      & 2.00 & 2.50& 0.034 &  0.013& 0.036 &  0.009&  0.08 &   0.02&  0.04 &   0.02\\ 
      &      & 2.50 & 3.00&       &       & 0.022 &  0.007& 0.035 &  0.008& 0.030 &  0.010\\ 
      &      & 3.00 & 3.50&       &       & 0.009 &  0.004& 0.013 &  0.004& 0.015 &  0.006\\ 
      &      & 3.50 & 4.00&       &       & 0.002 &  0.001& 0.010 &  0.003& 0.007 &  0.003\\ 
      &      & 4.00 & 5.00&       &       &       &       & 0.005 &  0.002& 0.004 &  0.002\\ 
      &      & 5.00 & 6.50&       &       &       &       & 0.001 &  0.001&       &       \\ 
      &      & 6.50 & 8.00&       &       &       &       &       &       &       &        \\ 
\hline
\end{tabular}
}
\end{table}
\clearpage

\begin{table}[!ht]
  \caption{\label{tab:xsec_results_C_pip}
    HARP results for the double-differential $p$  production
    cross-section in the laboratory system,
    $d^2\sigma^{p}/(dpd\Omega)$, for $\pi^{+}$--C interactions at 3,5,8,12~\GeVc.
    Each row refers to a
    different $(p_{\hbox{\small min}} \le p<p_{\hbox{\small max}},
    \theta_{\hbox{\small min}} \le \theta<\theta_{\hbox{\small max}})$ bin,
    where $p$ and $\theta$ are the outgoing proton momentum and polar angle, respectively.
    The central value as well as the square-root of the diagonal elements
    of the covariance matrix are given.}

\small{
\begin{tabular}{rrrr|r@{$\pm$}lr@{$\pm$}lr@{$\pm$}lr@{$\pm$}l}
\hline
$\theta_{\hbox{\small min}}$ &
$\theta_{\hbox{\small max}}$ &
$p_{\hbox{\small min}}$ &
$p_{\hbox{\small max}}$ &
\multicolumn{8}{c}{$d^2\sigma^{p}/(dpd\Omega)$}
\\
(rad) & (rad) & (\GeVc) & (\GeVc) &
\multicolumn{8}{c}{(barn/(sr \GeVc ))}
\\
  &  &  &
&\multicolumn{2}{c}{$ \bf{3 \ \GeVc}$}
&\multicolumn{2}{c}{$ \bf{5 \ \GeVc}$}
&\multicolumn{2}{c}{$ \bf{8 \ \GeVc}$}
&\multicolumn{2}{c}{$ \bf{12 \ \GeVc}$}
\\
\hline
0.050 &0.100 & 0.50 & 1.00&  0.10 &   0.04&  0.07 &   0.02&  0.06 &   0.02&  0.06 &   0.05\\ 
      &      & 1.00 & 1.50&  0.09 &   0.03&  0.08 &  0.02&  0.07 &   0.02&  0.07 &   0.04\\ 
      &      & 1.50 & 2.00&       &       &  0.07 &   0.02&  0.08 &   0.02&  0.01 &   0.03\\ 
      &      & 2.00 & 2.50&       &       & 0.015 &  0.016& 0.063 &  0.014&  0.03 &   0.08\\ 
      &      & 2.50 & 3.00&       &       &       &       & 0.036 &  0.013& 0.055 &  0.084\\ 
      &      & 3.00 & 3.50&       &       &       &       & 0.002 &  0.008& 0.026 &  0.039\\ 
      &      & 3.50 & 4.00&       &       &       &       &  0.05 &   0.02& 0.001 &  0.012\\ 
      &      & 4.00 & 5.00&       &       &       &       &       &       &       &       \\ 
      &      & 5.00 & 6.50&       &       &       &       &       &       &       &     \\ 
      &      & 6.50 & 8.00&       &       &       &       &       &       & 0.013 &  0.008\\ 
0.100 &0.150 & 0.50 & 1.00&  0.09 &   0.04&  0.07 &   0.02&  0.06 &   0.02&  0.10 &   0.07\\ 
      &      & 1.00 & 1.50&  0.10 &   0.03& 0.06  &   0.01& 0.06  &  0.02 &       &      \\ 
      &      & 1.50 & 2.00&  0.05 &   0.03&  0.06 &   0.02&  0.06 &   0.02&  0.09 &   0.05\\ 
      &      & 2.00 & 2.50&       &       &  0.03 &   0.02& 0.039 &  0.012&  0.03 &   0.03\\ 
      &      & 2.50 & 3.00&       &       &       &       & 0.040 &  0.010&  0.01 &   0.02\\ 
      &      & 3.00 & 3.50&       &       &       &       & 0.041 &  0.011& 0.024 &  0.024\\ 
      &      & 3.50 & 4.00&       &       &       &       & 0.018 &  0.007&  &  \\ 
      &      & 4.00 & 5.00&       &       &       &       & 0.011 &  0.005& &  \\ 
      &      & 5.00 & 6.50&       &       &       &       & 0.006 &  0.019&  &  \\ 
      &      & 6.50 & 8.00&       &       &       &       &       &       & &  \\ 
0.150 &0.200 & 0.50 & 1.00&  0.10 &   0.04&  0.10 &   0.02&  0.05 &   0.02&  0.09 &   0.07\\ 
      &      & 1.00 & 1.50&  0.08 &   0.02& 0.070 &  0.013& 0.051 &  0.014&  0.02 &   0.03\\ 
      &      & 1.50 & 2.00& 0.018 &  0.023& 0.058 &  0.011& 0.036 &  0.012&  0.05 &   0.04\\ 
      &      & 2.00 & 2.50&       &       & 0.029 &  0.011& 0.029 &  0.010& 0.018 &  0.024\\ 
      &      & 2.50 & 3.00&       &       & 0.012 &  0.008& 0.020 &  0.007& 0.014 &  0.022\\ 
      &      & 3.00 & 3.50&       &       & 0.005 &  0.006& 0.011 &  0.005& 0.020 &  0.046\\ 
      &      & 3.50 & 4.00&       &       & 0.011 &  0.007& 0.015 &  0.006& 0.005 &  0.022\\ 
      &      & 4.00 & 5.00&       &       & 0.006 &  0.006& 0.006 &  0.003& 0.005 &  0.022\\ 
      &      & 5.00 & 6.50&       &       &       &       & 0.001 &  0.001& 0.001 &  0.006\\ 
      &      & 6.50 & 8.00&       &       &       &       &       &       &       &       \\ 
0.200 &0.250 & 0.50 & 1.00&  0.09 &   0.04&  0.07 &   0.03&  0.05 &   0.02& 0.06  &  0.06 \\ 
      &      & 1.00 & 1.50&  0.04 &   0.02&  0.06 &   0.02&  0.09 &   0.03&  0.03 &   0.06\\ 
      &      & 1.50 & 2.00&  0.04 &   0.02& 0.036 &  0.012&  0.07 &   0.02&  0.02 &   0.05\\ 
      &      & 2.00 & 2.50&  0.02 &   0.02& 0.039 &  0.013& 0.028 &  0.013&  0.06 &   0.09\\ 
      &      & 2.50 & 3.00&       &       & 0.024 &  0.007& 0.019 &  0.008&  0.01 &   0.02\\ 
      &      & 3.00 & 3.50&       &       & 0.012 &  0.005& 0.024 &  0.008& 0.029 &  0.058\\ 
      &      & 3.50 & 4.00&       &       & 0.007 &  0.003& 0.015 &  0.005& 0.012 &  0.025\\ 
      &      & 4.00 & 5.00&       &       & 0.004 &  0.002& 0.014 &  0.006& 0.002 &  0.011\\ 
      &      & 5.00 & 6.50&       &       &       &       & 0.004 &  0.003& 0.000 &  0.004\\ 
      &      & 6.50 & 8.00&       &       &       &       &       &       & 0.000 &  0.005\\ 
\hline
\end{tabular}
}
\end{table}
\clearpage
\begin{table}[!ht]
  \caption{\label{tab:xsec_results_C_pr}
    HARP results for the double-differential $p$  production
    cross-section in the laboratory system,
    $d^2\sigma^{p}/(dpd\Omega)$, for $p$--C interactions at 3,5,8,12~\GeVc.
    Each row refers to a
    different $(p_{\hbox{\small min}} \le p<p_{\hbox{\small max}},
    \theta_{\hbox{\small min}} \le \theta<\theta_{\hbox{\small max}})$ bin,
    where $p$ and $\theta$ are the outgoing proton momentum and polar angle, respectively.
    The central value as well as the square-root of the diagonal elements
    of the covariance matrix are given.}

\small{
\begin{tabular}{rrrr|r@{$\pm$}lr@{$\pm$}lr@{$\pm$}lr@{$\pm$}l}
\hline
$\theta_{\hbox{\small min}}$ &
$\theta_{\hbox{\small max}}$ &
$p_{\hbox{\small min}}$ &
$p_{\hbox{\small max}}$ &
\multicolumn{8}{c}{$d^2\sigma^{p}/(dpd\Omega)$}
\\
(rad) & (rad) & (\GeVc) & (\GeVc) &
\multicolumn{8}{c}{(barn/(sr \GeVc ))}
\\
  &  &  &
&\multicolumn{2}{c}{$ \bf{3 \ \GeVc}$}
&\multicolumn{2}{c}{$ \bf{5 \ \GeVc}$}
&\multicolumn{2}{c}{$ \bf{8 \ \GeVc}$}
&\multicolumn{2}{c}{$ \bf{12 \ \GeVc}$}
\\
\hline

0.050 &0.100 & 0.50 & 1.00&  0.12 &   0.05&  0.13 &   0.03&  0.10 &   0.02&  0.09 &   0.02\\ 
      &      & 1.00 & 1.50&  0.21 &   0.04&  0.13 &   0.02& 0.122 &  0.015& 0.088 &  0.015\\ 
      &      & 1.50 & 2.00&  0.31 &   0.05&  0.16 &   0.02&  0.16 &   0.02& 0.090 &  0.015\\ 
      &      & 2.00 & 2.50&  0.27 &   0.09&  0.22 &   0.02& 0.16  &  0.02 &  0.12 &   0.02\\ 
      &      & 2.50 & 3.00&       &       &  0.33 &   0.03&  0.20 &   0.02& 0.15  &  0.01 \\ 
      &      & 3.00 & 3.50&       &       &  0.41 &   0.03&  0.27 &   0.02&  0.14 &   0.02\\ 
      &      & 3.50 & 4.00&       &       &  0.40 &   0.03&  0.30 &   0.02&  0.18 &   0.02\\ 
      &      & 4.00 & 5.00&       &       &  0.50 &   0.07&  0.33 &   0.02& 0.196 &  0.013\\ 
      &      & 5.00 & 6.50&       &       &       &       & 0.24  &  0.01& 0.191 &  0.010\\ 
      &      & 6.50 & 8.00&       &       &       &       &       &       & 0.159 &  0.008\\ 
0.100 &0.150 & 0.50 & 1.00&  0.12 &   0.05&  0.13 &   0.03&  0.13 &   0.02&  0.13 &   0.03\\ 
      &      & 1.00 & 1.50&  0.22 &   0.04&  0.12 &   0.02& 0.11  &  0.02 &  0.14 &   0.02\\ 
      &      & 1.50 & 2.00&  0.32 &   0.05&  0.18 &   0.02&  0.12 &   0.02& 0.088 &  0.014\\ 
      &      & 2.00 & 2.50&  0.30 &   0.05&  0.19 &   0.02&  0.14 &   0.02& 0.080 &  0.013\\ 
      &      & 2.50 & 3.00&       &       &  0.22 &   0.02&  0.18 &   0.02& 0.106 &  0.014\\ 
      &      & 3.00 & 3.50&       &       &  0.22 &   0.02& 0.177 &  0.015& 0.099 &  0.012\\ 
      &      & 3.50 & 4.00&       &       &  0.18 &   0.02& 0.155 &  0.012& 0.117 &  0.012\\ 
      &      & 4.00 & 5.00&       &       &  0.19 &   0.03& 0.137 &  0.011& 0.112 &  0.010\\ 
      &      & 5.00 & 6.50&       &       &       &       & 0.068 &  0.006& 0.064 &  0.006\\ 
      &      & 6.50 & 8.00&       &       &       &       &       &       & 0.038 &  0.004\\ 
0.150 &0.200 & 0.50 & 1.00&  0.08 &   0.04&  0.13 &   0.03&  0.07 &   0.02&  0.10 &   0.02\\ 
      &      & 1.00 & 1.50&  0.17 &   0.04&  0.13 &   0.02& 0.085 &  0.013& 0.063 &  0.012\\ 
      &      & 1.50 & 2.00&  0.15 &   0.03&  0.14 &   0.02& 0.088 &  0.012& 0.076 &  0.014\\ 
      &      & 2.00 & 2.50&  0.11 &   0.03&  0.14 &   0.02& 0.111 &  0.013& 0.061 &  0.011\\ 
      &      & 2.50 & 3.00&       &       & 0.122 &  0.014& 0.098 &  0.011& 0.063 &  0.010\\ 
      &      & 3.00 & 3.50&       &       & 0.093 &  0.012& 0.083 &  0.009& 0.054 &  0.008\\ 
      &      & 3.50 & 4.00&       &       & 0.079 &  0.011& 0.067 &  0.007& 0.042 &  0.006\\ 
      &      & 4.00 & 5.00&       &       & 0.058 &  0.007& 0.052 &  0.006& 0.034 &  0.005\\ 
      &      & 5.00 & 6.50&       &       &       &       & 0.015 &  0.003& 0.015 &  0.003\\ 
      &      & 6.50 & 8.00&       &       &       &       &       &       & 0.004 &  0.001\\ 
0.200 &0.250 & 0.50 & 1.00&  0.15 &   0.07&  0.09 &   0.03&  0.12 &   0.03&  0.10 &   0.03\\ 
      &      & 1.00 & 1.50&  0.14 &   0.04&  0.10 &   0.02&  0.10 &   0.02&  0.07 &   0.02\\ 
      &      & 1.50 & 2.00&  0.09 &   0.03&  0.07 &   0.02&  0.07 &   0.02& 0.036 &  0.011\\ 
      &      & 2.00 & 2.50&  0.08 &   0.03& 0.059 &  0.013&  0.09 &   0.02& 0.054 &  0.015\\ 
      &      & 2.50 & 3.00&       &       & 0.056 &  0.011& 0.079 &  0.014& 0.019 &  0.006\\ 
      &      & 3.00 & 3.50&       &       & 0.031 &  0.007& 0.060 &  0.009& 0.020 &  0.005\\ 
      &      & 3.50 & 4.00&       &       & 0.026 &  0.005& 0.043 &  0.008& 0.022 &  0.005\\ 
      &      & 4.00 & 5.00&       &       & 0.026 &  0.005& 0.027 &  0.005& 0.014 &  0.004\\ 
      &      & 5.00 & 6.50&       &       &       &       & 0.007 &  0.002& 0.005 &  0.002\\ 
      &      & 6.50 & 8.00&       &       &       &       &       &       & 0.002 &  0.001\\ 
\hline
\end{tabular}
}
\end{table}
\clearpage
\begin{table}[!ht]
  \caption{\label{tab:xsec_results_Al_pim}
    HARP results for the double-differential $p$  production
    cross-section in the laboratory system,
    $d^2\sigma^{p}/(dpd\Omega)$, for $\pi^{-}$--Al interactions at 3,5,8,12~\GeVc.
    Each row refers to a
    different $(p_{\hbox{\small min}} \le p<p_{\hbox{\small max}},
    \theta_{\hbox{\small min}} \le \theta<\theta_{\hbox{\small max}})$ bin,
    where $p$ and $\theta$ are the outgoing proton momentum and polar angle, respectively.
    The central value as well as the square-root of the diagonal elements
    of the covariance matrix are given.}

\small{
\begin{tabular}{rrrr|r@{$\pm$}lr@{$\pm$}lr@{$\pm$}lr@{$\pm$}l}
\hline
$\theta_{\hbox{\small min}}$ &
$\theta_{\hbox{\small max}}$ &
$p_{\hbox{\small min}}$ &
$p_{\hbox{\small max}}$ &
\multicolumn{8}{c}{$d^2\sigma^{p}/(dpd\Omega)$}
\\
(rad) & (rad) & (\GeVc) & (\GeVc) &
\multicolumn{8}{c}{(barn/(sr \GeVc ))}
\\
  &  &  &
&\multicolumn{2}{c}{$ \bf{3 \ \GeVc}$}
&\multicolumn{2}{c}{$ \bf{5 \ \GeVc}$}
&\multicolumn{2}{c}{$ \bf{8 \ \GeVc}$}
&\multicolumn{2}{c}{$ \bf{12 \ \GeVc}$}
\\
\hline

0.050 &0.100 & 0.50 & 1.00&  0.16 &   0.03&  0.15 &   0.04&  0.15 &   0.03&  0.12 &   0.03\\ 
      &      & 1.00 & 1.50&  0.12 &   0.02&  0.13 &   0.02&  0.14 &   0.02&  0.11 &   0.02\\ 
      &      & 1.50 & 2.00& 0.075 &  0.013&  0.12 &   0.02&  0.14 &   0.02&  0.08 &   0.02\\ 
      &      & 2.00 & 2.50& 0.043 &  0.010&  0.09 &   0.02&  0.08 &   0.02& 0.010 &  0.013\\ 
      &      & 2.50 & 3.00&       &       & 0.047 &  0.011& 0.104 &  0.013&  0.08 &   0.02\\ 
      &      & 3.00 & 3.50&       &       & 0.016 &  0.006& 0.067 &  0.011& 0.096 &  0.013\\ 
      &      & 3.50 & 4.00&       &       & 0.033 &  0.009& 0.067 &  0.009& 0.086 &  0.011\\ 
      &      & 4.00 & 5.00&       &       & 0.012 &  0.006& 0.045 &  0.007& 0.104 &  0.010\\ 
      &      & 5.00 & 6.50&       &       &       &       & 0.010 &  0.003& 0.042 &  0.006\\ 
      &      & 6.50 & 8.00&       &       &       &       &       &       & 0.022 &  0.004\\ 
0.100 &0.150 & 0.50 & 1.00&  0.20 &   0.04&  0.18 &   0.04&  0.17 &   0.03&  0.14 &   0.03\\ 
      &      & 1.00 & 1.50&  0.14 &   0.02&  0.19 &   0.03&  0.15 &   0.02&  0.15 &   0.02\\ 
      &      & 1.50 & 2.00&  0.09 &   0.02&  0.16 &   0.03&  0.12 &   0.02&  0.09 &   0.02\\ 
      &      & 2.00 & 2.50& 0.047 &  0.010&  0.10 &   0.02&  0.11 &   0.02&  0.06 &   0.02\\ 
      &      & 2.50 & 3.00&       &       & 0.048 &  0.011& 0.087 &  0.011& 0.070 &  0.013\\ 
      &      & 3.00 & 3.50&       &       & 0.021 &  0.007& 0.056 &  0.011& 0.097 &  0.015\\ 
      &      & 3.50 & 4.00&       &       & 0.017 &  0.006& 0.039 &  0.007& 0.082 &  0.012\\ 
      &      & 4.00 & 5.00&       &       & 0.008 &  0.004& 0.025 &  0.005& 0.053 &  0.008\\ 
      &      & 5.00 & 6.50&       &       &       &       & 0.005 &  0.002& 0.015 &  0.003\\ 
      &      & 6.50 & 8.00&       &       &       &       &       &       & 0.004 &  0.001\\ 
0.150 &0.200 & 0.50 & 1.00&  0.16 &   0.04&  0.09 &   0.03&  0.16 &   0.03&  0.15 &   0.03\\ 
      &      & 1.00 & 1.50&  0.08 &   0.02&  0.12 &   0.02&  0.10 &   0.02&  0.08 &   0.02\\ 
      &      & 1.50 & 2.00& 0.061 &  0.013&  0.10 &   0.02&  0.12 &   0.02&  0.09 &   0.02\\ 
      &      & 2.00 & 2.50& 0.052 &  0.011&  0.08 &   0.02&  0.08 &   0.02&  0.07 &   0.02\\ 
      &      & 2.50 & 3.00&       &       & 0.027 &  0.008& 0.068 &  0.012& 0.054 &  0.011\\ 
      &      & 3.00 & 3.50&       &       & 0.029 &  0.009& 0.044 &  0.010& 0.049 &  0.009\\ 
      &      & 3.50 & 4.00&       &       & 0.002 &  0.002& 0.026 &  0.006& 0.045 &  0.008\\ 
      &      & 4.00 & 5.00&       &       & 0.000 &  0.001& 0.011 &  0.003& 0.024 &  0.005\\ 
      &      & 5.00 & 6.50&       &       &       &       & 0.001 &  0.001& 0.004 &  0.001\\ 
      &      & 6.50 & 8.00&       &       &       &       &       &       &       &       \\ 
0.200 &0.250 & 0.50 & 1.00&  0.16 &   0.04&  0.16 &   0.05&  0.18 &   0.04&  0.17 &   0.04\\ 
      &      & 1.00 & 1.50&  0.14 &   0.03&  0.18 &   0.04&  0.21 &   0.04&  0.15 &   0.03\\ 
      &      & 1.50 & 2.00&  0.09 &   0.02&  0.13 &   0.03&  0.16 &   0.04&  0.11 &   0.03\\ 
      &      & 2.00 & 2.50& 0.030 &  0.012&  0.09 &   0.02&  0.13 &   0.03&  0.08 &   0.03\\ 
      &      & 2.50 & 3.00&       &       &  0.06 &   0.02&  0.06 &   0.02&  0.06 &   0.02\\ 
      &      & 3.00 & 3.50&       &       & 0.019 &  0.009& 0.024 &  0.008& 0.036 &  0.009\\ 
      &      & 3.50 & 4.00&       &       & 0.005 &  0.003& 0.008 &  0.004& 0.021 &  0.006\\ 
      &      & 4.00 & 5.00&       &       & 0.001 &  0.002& 0.003 &  0.002& 0.012 &  0.005\\ 
      &      & 5.00 & 6.50&       &       &       &       & 0.000 &  0.002& 0.001 &  0.001\\ 
      &      & 6.50 & 8.00&       &       &       &       &       &       & 0.000 &  0.001\\ 
\hline
\end{tabular}
}
\end{table}
\clearpage
\begin{table}[!ht]
  \caption{\label{tab:xsec_results_Al_pip}
    HARP results for the double-differential $p$  production
    cross-section in the laboratory system,
    $d^2\sigma^{p}/(dpd\Omega)$, for $\pi^{+}$--Al interactions at 3,5,8,12,12.9~\GeVc.
    Each row refers to a
    different $(p_{\hbox{\small min}} \le p<p_{\hbox{\small max}},
    \theta_{\hbox{\small min}} \le \theta<\theta_{\hbox{\small max}})$ bin,
    where $p$ and $\theta$ are the ougoing proton momentum and polar angle, respectively.
    The central value as well as the square-root of the diagonal elements
    of the covariance matrix are given.}
\small{
\begin{tabular}{rrrr|r@{$\pm$}lr@{$\pm$}lr@{$\pm$}lr@{$\pm$}lr@{$\pm$}l}
\hline
$\theta_{\hbox{\small min}}$ &
$\theta_{\hbox{\small max}}$ &
$p_{\hbox{\small min}}$ &
$p_{\hbox{\small max}}$ &
\multicolumn{10}{c}{$d^2\sigma^{p}/(dpd\Omega)$}
\\
(rad) & (rad) & (\GeVc) & (\GeVc) &
\multicolumn{10}{c}{(barn/(sr \GeVc))}
\\
  &  &  &
&\multicolumn{2}{c}{$ \bf{3 \ \GeVc}$}
&\multicolumn{2}{c}{$ \bf{5 \ \GeVc}$}
&\multicolumn{2}{c}{$ \bf{8 \ \GeVc}$}
&\multicolumn{2}{c}{$ \bf{12 \ \GeVc}$}
&\multicolumn{2}{c}{$ \bf{12.9 \ \GeVc}$}
\\
\hline

0.050 &0.100 & 0.50 & 1.00&  0.16 &   0.06&  0.17 &   0.04&  0.14 &   0.04&  0.17 &   0.19&  0.18 &   0.05\\ 
      &      & 1.00 & 1.50&  0.16 &   0.05&  0.17 &   0.03&  0.13 &   0.03&  0.10 &   0.09&  0.12 &   0.03\\ 
      &      & 1.50 & 2.00&       &       &  0.13 &   0.03&  0.11 &   0.03&  0.05 &   0.08&  0.14 &   0.03\\ 
      &      & 2.00 & 2.50&       &       &  0.09 &   0.03&  0.09 &   0.02&  0.07 &   0.12&  0.13 &   0.02\\ 
      &      & 2.50 & 3.00&       &       &       &       &  0.07 &   0.02&  0.07 &   0.08&  0.12 &   0.02\\ 
      &      & 3.00 & 3.50&       &       &       &       &  0.03 &   0.02&  0.08 &   0.10&  0.11 &   0.02\\ 
      &      & 3.50 & 4.00&       &       &       &       &       &       &  0.02 &   0.08&  0.10 &   0.02\\ 
      &      & 4.00 & 5.00&       &       &       &       &       &       &  0.03 &   0.15&  0.14 &   0.02\\ 
      &      & 5.00 & 6.50&       &       &       &       &       &       & 0.005 &  0.092& 0.075 &  0.010\\ 
      &      & 6.50 & 8.00&       &       &       &       &       &       & 0.002 &  0.127& 0.039 &  0.007\\ 
0.100 &0.150 & 0.50 & 1.00&  0.21 &   0.07&  0.21 &   0.05&  0.24 &   0.06&  0.01 &   0.03&  0.14 &   0.05\\ 
      &      & 1.00 & 1.50&  0.15 &   0.04&  0.13 &   0.03&  0.13 &   0.03&  0.11 &   0.08&  0.14 &   0.03\\ 
      &      & 1.50 & 2.00&  0.07 &   0.05&  0.13 &   0.03&  0.10 &   0.03&  0.07 &   0.08&  0.11 &   0.03\\ 
      &      & 2.00 & 2.50&       &       &  0.05 &   0.07&  0.08 &   0.02&  0.06 &   0.06&  0.09 &   0.02\\ 
      &      & 2.50 & 3.00&       &       &       &       &  0.09 &   0.02&  0.07 &   0.07&  0.08 &   0.02\\ 
      &      & 3.00 & 3.50&       &       &       &       &  0.05 &   0.02&  0.02 &   0.03&  0.09 &   0.02\\ 
      &      & 3.50 & 4.00&       &       &       &       & 0.019 &  0.009& 0.039 &  0.045& 0.075 &  0.013\\ 
      &      & 4.00 & 5.00&       &       &       &       & 0.017 &  0.007& 0.064 &  0.075& 0.068 &  0.012\\ 
      &      & 5.00 & 6.50&       &       &       &       & 0.011 &  0.006& 0.011 &  0.026& 0.030 &  0.006\\ 
      &      & 6.50 & 8.00&       &       &       &       &       &       & 0.005 &  0.016& 0.010 &  0.003\\ 
0.150 &0.200 & 0.50 & 1.00&  0.19 &   0.07&  0.20 &   0.04&  0.20 &   0.05&  0.10 &   0.14&  0.08 &   0.04\\ 
      &      & 1.00 & 1.50&  0.18 &   0.04&  0.12 &   0.02&  0.12 &   0.03&  0.16 &   0.11&  0.12 &   0.03\\ 
      &      & 1.50 & 2.00&  0.05 &   0.04&  0.11 &   0.02&  0.07 &   0.02&  0.15 &   0.11&  0.08 &   0.02\\ 
      &      & 2.00 & 2.50&       &       &  0.08 &   0.02&  0.06 &   0.02&  0.06 &   0.07&  0.07 &   0.02\\ 
      &      & 2.50 & 3.00&       &       &  0.02 &   0.02&  0.08 &   0.02& 0.013 &  0.032& 0.051 &  0.015\\ 
      &      & 3.00 & 3.50&       &       & 0.001 &  0.006& 0.035 &  0.012& 0.051 &  0.076& 0.049 &  0.014\\ 
      &      & 3.50 & 4.00&       &       & 0.013 &  0.011& 0.013 &  0.008& 0.028 &  0.044& 0.037 &  0.010\\ 
      &      & 4.00 & 5.00&       &       & 0.012 &  0.013& 0.009 &  0.008& 0.011 &  0.024& 0.018 &  0.006\\ 
      &      & 5.00 & 6.50&       &       &       &       & 0.002 &  0.002& 0.006 &  0.015& 0.010 &  0.003\\ 
      &      & 6.50 & 8.00&       &       &       &       &       &       &       &       & 0.002 &  0.001\\ 
0.200 &0.250 & 0.50 & 1.00&  0.13 &   0.07&  0.17 &   0.06&  0.17 &   0.05&  0.10 &   0.17&  0.17 &   0.06\\ 
      &      & 1.00 & 1.50&  0.14 &   0.05&  0.11 &   0.03&  0.11 &   0.03&  0.08 &   0.13&  0.08 &   0.02\\ 
      &      & 1.50 & 2.00&  0.06 &   0.03&  0.06 &   0.02&  0.10 &   0.03&  0.02 &   0.09&  0.06 &   0.03\\ 
      &      & 2.00 & 2.50&  0.04 &   0.03&  0.06 &   0.02&  0.09 &   0.03&  0.13 &   0.22&  0.06 &   0.02\\ 
      &      & 2.50 & 3.00&       &       & 0.042 &  0.014&  0.06 &   0.02&  0.01 &   0.02&  0.04 &   0.02\\ 
      &      & 3.00 & 3.50&       &       & 0.019 &  0.009&  0.04 &   0.02& 0.079 &  0.116& 0.031 &  0.009\\ 
      &      & 3.50 & 4.00&       &       & 0.012 &  0.006& 0.026 &  0.009& 0.004 &  0.016& 0.019 &  0.008\\ 
      &      & 4.00 & 5.00&       &       & 0.009 &  0.004& 0.021 &  0.009& 0.000 &  0.004& 0.008 &  0.005\\ 
      &      & 5.00 & 6.50&       &       &       &       & 0.008 &  0.008& 0.001 &  0.013& 0.002 &  0.002\\ 
      &      & 6.50 & 8.00&       &       &       &       &       &       & 0.000 &  0.004& 0.001 &  0.001\\ 
\hline
\end{tabular}
}
\end{table}
\clearpage

\begin{table}[!ht]
  \caption{\label{tab:xsec_results_Al_pr}
    HARP results for the double-differential $p$  production
    cross-section in the laboratory system,
    $d^2\sigma^{p}/(dpd\Omega)$, for p--Al interactions at 3,5,8,12,12.9~\GeVc.
    Each row refers to a
    different $(p_{\hbox{\small min}} \le p<p_{\hbox{\small max}},
    \theta_{\hbox{\small min}} \le \theta<\theta_{\hbox{\small max}})$ bin,
    where $p$ and $\theta$ are the outgoing proton momentum and polar angle, respectively.
    The central value as well as the square-root of the diagonal elements
    of the covariance matrix are given.}
\small{
\begin{tabular}{rrrr|r@{$\pm$}lr@{$\pm$}lr@{$\pm$}lr@{$\pm$}lr@{$\pm$}l}
\hline
$\theta_{\hbox{\small min}}$ &
$\theta_{\hbox{\small max}}$ &
$p_{\hbox{\small min}}$ &
$p_{\hbox{\small max}}$ &
\multicolumn{10}{c}{$d^2\sigma^{p`}/(dpd\Omega)$}
\\
(rad) & (rad) & (\GeVc) & (\GeVc) &
\multicolumn{10}{c}{(barn/(sr \GeVc))}
\\
  &  &  &
&\multicolumn{2}{c}{$ \bf{3 \ \GeVc}$}
&\multicolumn{2}{c}{$ \bf{5 \ \GeVc}$}
&\multicolumn{2}{c}{$ \bf{8 \ \GeVc}$}
&\multicolumn{2}{c}{$ \bf{12 \ \GeVc}$}
&\multicolumn{2}{c}{$ \bf{12.9 \ \GeVc}$}
\\
\hline

0.050 &0.100 & 0.50 & 1.00&  0.20 &   0.08&  0.22 &   0.05&  0.22 &   0.04&  0.22 &   0.05&  0.23 &   0.03\\ 
      &      & 1.00 & 1.50&  0.32 &   0.07&  0.27 &   0.03&  0.24 &   0.03&  0.23 &   0.04&  0.21 &   0.02\\ 
      &      & 1.50 & 2.00&  0.41 &   0.08&  0.33 &   0.04&  0.23 &   0.02&  0.26 &   0.04&  0.21 &   0.02\\ 
      &      & 2.00 & 2.50&  0.39 &   0.12&  0.34 &   0.04&  0.26 &   0.03&  0.17 &   0.03&  0.22 &   0.02\\ 
      &      & 2.50 & 3.00&       &       &  0.53 &   0.04&  0.35 &   0.03&  0.26 &   0.03&  0.25 &   0.02\\ 
      &      & 3.00 & 3.50&       &       &  0.53 &   0.05&  0.37 &   0.02&  0.28 &   0.03&  0.26 &   0.02\\ 
      &      & 3.50 & 4.00&       &       &  0.59 &   0.05&  0.35 &   0.02&  0.30 &   0.03&  0.30 &   0.02\\ 
      &      & 4.00 & 5.00&       &       &  0.62 &   0.08&  0.41 &   0.02&  0.34 &   0.03&  0.34 &   0.02\\ 
      &      & 5.00 & 6.50&       &       &       &       &  0.35 &   0.02&  0.31 &   0.02& 0.312 &  0.011\\ 
      &      & 6.50 & 8.00&       &       &       &       &       &       &  0.29 &   0.02& 0.272 &  0.010\\ 
0.100 &0.150 & 0.50 & 1.00&  0.28 &   0.09&  0.26 &   0.05&  0.23 &   0.04&  0.24 &   0.05&  0.24 &   0.03\\ 
      &      & 1.00 & 1.50&  0.30 &   0.07&  0.29 &   0.04&  0.18 &   0.03&  0.26 &   0.04&  0.22 &   0.02\\ 
      &      & 1.50 & 2.00&  0.53 &   0.09&  0.27 &   0.03&  0.24 &   0.03&  0.22 &   0.03&  0.21 &   0.02\\ 
      &      & 2.00 & 2.50&  0.44 &   0.08&  0.35 &   0.04&  0.23 &   0.03&  0.18 &   0.03&  0.21 &   0.02\\ 
      &      & 2.50 & 3.00&       &       &  0.33 &   0.03&  0.29 &   0.03&  0.20 &   0.03&  0.20 &   0.02\\ 
      &      & 3.00 & 3.50&       &       &  0.34 &   0.04&  0.28 &   0.02&  0.20 &   0.03&  0.20 &   0.02\\ 
      &      & 3.50 & 4.00&       &       &  0.32 &   0.03&  0.22 &   0.02&  0.17 &   0.02& 0.189 &  0.015\\ 
      &      & 4.00 & 5.00&       &       &  0.32 &   0.05&  0.20 &   0.02&  0.20 &   0.02& 0.190 &  0.013\\ 
      &      & 5.00 & 6.50&       &       &       &       & 0.116 &  0.009& 0.117 &  0.012& 0.121 &  0.009\\ 
      &      & 6.50 & 8.00&       &       &       &       &       &       & 0.076 &  0.009& 0.063 &  0.006\\ 
0.150 &0.200 & 0.50 & 1.00&  0.19 &   0.08&  0.27 &   0.06&  0.18 &   0.04&  0.22 &   0.05&  0.21 &   0.03\\ 
      &      & 1.00 & 1.50&  0.27 &   0.06&  0.21 &   0.03&  0.17 &   0.02&  0.18 &   0.03&  0.18 &   0.02\\ 
      &      & 1.50 & 2.00&  0.25 &   0.06&  0.22 &   0.03&  0.15 &   0.02&  0.15 &   0.03&  0.15 &   0.02\\ 
      &      & 2.00 & 2.50&  0.24 &   0.06&  0.23 &   0.03&  0.16 &   0.02&  0.15 &   0.03& 0.139 &  0.015\\ 
      &      & 2.50 & 3.00&       &       &  0.17 &   0.03&  0.15 &   0.02&  0.10 &   0.02& 0.127 &  0.012\\ 
      &      & 3.00 & 3.50&       &       &  0.12 &   0.02&  0.14 &   0.02&  0.14 &   0.02& 0.109 &  0.010\\ 
      &      & 3.50 & 4.00&       &       & 0.11 &  0.02& 0.094 &  0.011& 0.082 &  0.014& 0.086 &  0.008\\ 
      &      & 4.00 & 5.00&       &       &  0.11 &   0.02& 0.074 &  0.009& 0.052 &  0.009& 0.065 &  0.006\\ 
      &      & 5.00 & 6.50&       &       &       &       & 0.036 &  0.005& 0.022 &  0.004& 0.027 &  0.003\\ 
      &      & 6.50 & 8.00&       &       &       &       &       &       & 0.010 &  0.003& 0.011 &  0.002\\ 
0.200 &0.250 & 0.50 & 1.00&  0.31 &   0.13&  0.26 &   0.06&  0.28 &   0.05&  0.17 &   0.05&  0.21 &   0.03\\ 
      &      & 1.00 & 1.50&  0.12 &   0.05&  0.18 &   0.04&  0.20 &   0.04&  0.12 &   0.04&  0.16 &   0.02\\ 
      &      & 1.50 & 2.00&  0.15 &   0.05&  0.14 &   0.03&  0.16 &   0.03&  0.11 &   0.03&  0.10 &   0.02\\ 
      &      & 2.00 & 2.50&  0.10 &   0.04&  0.12 &   0.03&  0.14 &   0.03&  0.08 &   0.03&  0.09 &   0.02\\ 
      &      & 2.50 & 3.00&       &       &  0.11 &   0.02&  0.12 &   0.02&  0.06 &   0.02& 0.056 &  0.010\\ 
      &      & 3.00 & 3.50&       &       &  0.08 &   0.02& 0.064 &  0.011& 0.042 &  0.012& 0.039 &  0.006\\ 
      &      & 3.50 & 4.00&       &       & 0.050 &  0.011& 0.049 &  0.009& 0.032 &  0.010& 0.034 &  0.005\\ 
      &      & 4.00 & 5.00&       &       & 0.031 &  0.008& 0.045 &  0.008& 0.023 &  0.009& 0.030 &  0.005\\ 
      &      & 5.00 & 6.50&       &       &       &       & 0.028 &  0.007& 0.006 &  0.004& 0.013 &  0.002\\ 
      &      & 6.50 & 8.00&       &       &       &       &       &       & 0.007 &  0.004& 0.007 &  0.001\\ 
\hline
\end{tabular}
}
\end{table}
\clearpage

\begin{table}[!ht]
  \caption{\label{tab:xsec_results_Cu_pim}
    HARP results for the double-differential $p$  production
    cross-section in the laboratory system,
    $d^2\sigma^{p}/(dpd\Omega)$, for $\pi^{-}$--Cu interactions at 3,5,8,12~\GeVc.
    Each row refers to a
    different $(p_{\hbox{\small min}} \le p<p_{\hbox{\small max}},
    \theta_{\hbox{\small min}} \le \theta<\theta_{\hbox{\small max}})$ bin,
    where $p$ and $\theta$ are the outgoing proton  momentum and polar angle, respectively.
    The central value as well as the square-root of the diagonal elements
    of the covariance matrix are given.}

\small{
\begin{tabular}{rrrr|r@{$\pm$}lr@{$\pm$}lr@{$\pm$}lr@{$\pm$}l}
\hline
$\theta_{\hbox{\small min}}$ &
$\theta_{\hbox{\small max}}$ &
$p_{\hbox{\small min}}$ &
$p_{\hbox{\small max}}$ &
\multicolumn{8}{c}{$d^2\sigma^{p}/(dpd\Omega)$}
\\
(rad) & (rad) & (\GeVc) & (\GeVc) &
\multicolumn{8}{c}{(barn/(sr \GeVc ))}
\\
  &  &  &
&\multicolumn{2}{c}{$ \bf{3 \ \GeVc}$}
&\multicolumn{2}{c}{$ \bf{5 \ \GeVc}$}
&\multicolumn{2}{c}{$ \bf{8 \ \GeVc}$}
&\multicolumn{2}{c}{$ \bf{12 \ \GeVc}$}
\\
\hline
0.050 &0.100 & 0.50 & 1.00&  0.31 &   0.05&  0.35 &   0.05&  0.42 &   0.06&  0.33 &   0.07\\ 
      &      & 1.00 & 1.50&  0.18 &   0.03&  0.29 &   0.03&  0.32 &   0.04&  0.31 &   0.05\\ 
      &      & 1.50 & 2.00&  0.13 &   0.02&  0.21 &   0.03&  0.31 &   0.04&  0.26 &   0.05\\ 
      &      & 2.00 & 2.50& 0.074 &  0.013&  0.14 &   0.02&  0.20 &   0.03&  0.09 &   0.03\\ 
      &      & 2.50 & 3.00&       &       & 0.075 &  0.013&  0.20 &   0.02&  0.17 &   0.03\\ 
      &      & 3.00 & 3.50&       &       & 0.033 &  0.008&  0.13 &   0.02&  0.22 &   0.03\\ 
      &      & 3.50 & 4.00&       &       & 0.021 &  0.006&  0.13 &   0.02&  0.17 &   0.03\\ 
      &      & 4.00 & 5.00&       &       & 0.016 &  0.005& 0.062 &  0.011&  0.15 &   0.02\\ 
      &      & 5.00 & 6.50&       &       &       &       & 0.018 &  0.005& 0.071 &  0.010\\ 
      &      & 6.50 & 8.00&       &       &       &       &       &       & 0.034 &  0.007\\ 
0.100 &0.150 & 0.50 & 1.00&  0.34 &   0.05&  0.45 &   0.07&  0.49 &   0.08&  0.45 &   0.09\\ 
      &      & 1.00 & 1.50&  0.24 &   0.03&  0.35 &   0.04&  0.33 &   0.04&  0.41 &   0.06\\ 
      &      & 1.50 & 2.00&  0.12 &   0.02&  0.25 &   0.03&  0.28 &   0.04&  0.21 &   0.04\\ 
      &      & 2.00 & 2.50& 0.080 &  0.013&  0.17 &   0.02&  0.20 &   0.03&  0.12 &   0.03\\ 
      &      & 2.50 & 3.00&       &       & 0.068 &  0.012&  0.19 &   0.02&  0.12 &   0.03\\ 
      &      & 3.00 & 3.50&       &       & 0.050 &  0.009&  0.10 &   0.02&  0.16 &   0.03\\ 
      &      & 3.50 & 4.00&       &       & 0.016 &  0.007& 0.081 &  0.013&  0.12 &   0.02\\ 
      &      & 4.00 & 5.00&       &       & 0.011 &  0.004& 0.046 &  0.009&  0.09 &   0.02\\ 
      &      & 5.00 & 6.50&       &       &       &       & 0.010 &  0.003& 0.026 &  0.006\\ 
      &      & 6.50 & 8.00&       &       &       &       &       &       & 0.013 &  0.004\\ 
0.150 &0.200 & 0.50 & 1.00&  0.30 &   0.05&  0.38 &   0.06&  0.40 &   0.07&  0.36 &   0.08\\ 
      &      & 1.00 & 1.50&  0.17 &   0.02&  0.20 &   0.03&  0.29 &   0.03&  0.25 &   0.04\\ 
      &      & 1.50 & 2.00&  0.08 &   0.02&  0.16 &   0.02&  0.23 &   0.03&  0.19 &   0.04\\ 
      &      & 2.00 & 2.50& 0.071 &  0.013&  0.13 &   0.02&  0.16 &   0.03&  0.17 &   0.04\\ 
      &      & 2.50 & 3.00&       &       & 0.063 &  0.011&  0.11 &   0.02&  0.12 &   0.02\\ 
      &      & 3.00 & 3.50&       &       & 0.033 &  0.008&  0.10 &   0.02&  0.12 &   0.02\\ 
      &      & 3.50 & 4.00&       &       & 0.009 &  0.004& 0.052 &  0.011&  0.08 &   0.02\\ 
      &      & 4.00 & 5.00&       &       & 0.003 &  0.002& 0.020 &  0.005& 0.041 &  0.010\\ 
      &      & 5.00 & 6.50&       &       &       &       & 0.000 &  0.001& 0.009 &  0.003\\ 
      &      & 6.50 & 8.00&       &       &       &       &       &       & 0.003 &  0.002\\ 
0.200 &0.250 & 0.50 & 1.00&  0.34 &   0.07&  0.32 &   0.07&  0.48 &   0.09&  0.38 &   0.09\\ 
      &      & 1.00 & 1.50&  0.26 &   0.04&  0.33 &   0.05&  0.48 &   0.08&  0.45 &   0.09\\ 
      &      & 1.50 & 2.00&  0.16 &   0.03&  0.32 &   0.05&  0.27 &   0.06&  0.29 &   0.07\\ 
      &      & 2.00 & 2.50&  0.07 &   0.02&  0.19 &   0.03&  0.31 &   0.06&  0.31 &   0.08\\ 
      &      & 2.50 & 3.00&       &       &  0.06 &   0.02&  0.16 &   0.03&  0.19 &   0.04\\ 
      &      & 3.00 & 3.50&       &       & 0.024 &  0.008& 0.047 &  0.013&  0.09 &   0.02\\ 
      &      & 3.50 & 4.00&       &       & 0.006 &  0.003& 0.030 &  0.009& 0.039 &  0.012\\ 
      &      & 4.00 & 5.00&       &       & 0.003 &  0.002& 0.016 &  0.005& 0.017 &  0.007\\ 
      &      & 5.00 & 6.50&       &       &       &       & 0.001 &  0.003& 0.004 &  0.003\\ 
      &      & 6.50 & 8.00&       &       &       &       &       &       & 0.002 &  0.002\\

\hline
\end{tabular}
}
\end{table}
\clearpage

\begin{table}[!ht]
  \caption{\label{tab:xsec_results_Cu_pip}
    HARP results for the double-differential $p$  production
    cross-section in the laboratory system,
    $d^2\sigma^{p}/(dpd\Omega)$, for $\pi^{+}$--Cu interactions at 3,5,8,12~\GeVc.
    Each row refers to a
    different $(p_{\hbox{\small min}} \le p<p_{\hbox{\small max}},
    \theta_{\hbox{\small min}} \le \theta<\theta_{\hbox{\small max}})$ bin,
    where $p$ and $\theta$ are the outgoing proton momentum and polar angle, respectively.
    The central value as well as the square-root of the diagonal elements
    of the covariance matrix are given.}

\small{
\begin{tabular}{rrrr|r@{$\pm$}lr@{$\pm$}lr@{$\pm$}lr@{$\pm$}l}
\hline
$\theta_{\hbox{\small min}}$ &
$\theta_{\hbox{\small max}}$ &
$p_{\hbox{\small min}}$ &
$p_{\hbox{\small max}}$ &
\multicolumn{8}{c}{$d^2\sigma^{p}/(dpd\Omega)$}
\\
(rad) & (rad) & (\GeVc) & (\GeVc) &
\multicolumn{8}{c}{(barn/(sr \GeVc ))}
\\
  &  &  &
&\multicolumn{2}{c}{$ \bf{3 \ \GeVc}$}
&\multicolumn{2}{c}{$ \bf{5 \ \GeVc}$}
&\multicolumn{2}{c}{$ \bf{8 \ \GeVc}$}
&\multicolumn{2}{c}{$ \bf{12 \ \GeVc}$}
\\
\hline
0.050 &0.100 & 0.50 & 1.00&  0.25 &   0.12&  0.39 &   0.08&  0.43 &   0.08&  0.17 &   0.16\\ 
      &      & 1.00 & 1.50&  0.28 &   0.10&  0.36 &   0.05&  0.32 &   0.06&  0.48 &   0.21\\ 
      &      & 1.50 & 2.00&       &       &  0.25 &   0.05&  0.30 &   0.05&  0.30 &   0.41\\ 
      &      & 2.00 & 2.50&       &       &  0.05 &   0.05&  0.16 &   0.04&  0.19 &   0.32\\ 
      &      & 2.50 & 3.00&       &       &       &       &  0.13 &   0.04&  0.12 &   0.16\\ 
      &      & 3.00 & 3.50&       &       &       &       &  0.09 &   0.04&  0.07 &   0.07\\ 
      &      & 3.50 & 4.00&       &       &       &       &  0.03 &   0.04&  0.16 &   0.13\\ 
      &      & 4.00 & 5.00&       &       &       &       &       &       &  0.10 &   0.12\\ 
      &      & 5.00 & 6.50&       &       &       &       &       &       &  0.02 &   0.06\\ 
      &      & 6.50 & 8.00&       &       &       &       &       &       &  0.01 &   0.07\\ 
0.100 &0.150 & 0.50 & 1.00&  0.30 &   0.13&  0.47 &   0.09&  0.53 &   0.11&  0.69 &   0.36\\ 
      &      & 1.00 & 1.50&  0.27 &   0.09&  0.30 &   0.05&  0.41 &   0.06&  0.44 &   0.20\\ 
      &      & 1.50 & 2.00&  0.07 &   0.08&  0.26 &   0.05&  0.24 &   0.05&  0.32 &   0.17\\ 
      &      & 2.00 & 2.50&       &       &  0.16 &   0.09&  0.14 &   0.04&  0.20 &   0.13\\ 
      &      & 2.50 & 3.00&       &       &       &       &  0.13 &   0.03&  0.15 &   0.11\\ 
      &      & 3.00 & 3.50&       &       &       &       &  0.12 &   0.03&  0.04 &   0.05\\ 
      &      & 3.50 & 4.00&       &       &       &       &  0.04 &   0.02&  0.11 &   0.09\\ 
      &      & 4.00 & 5.00&       &       &       &       & 0.04  &  0.01 &  0.11 &   0.07\\ 
      &      & 5.00 & 6.50&       &       &       &       &  0.02 &   0.02&  0.04 &   0.04\\ 
      &      & 6.50 & 8.00&       &       &       &       &       &       &  0.02 &   0.03\\ 
0.150 &0.200 & 0.50 & 1.00&  0.27 &   0.12&  0.53 &   0.10&  0.46 &   0.10&  0.21 &   0.20\\ 
      &      & 1.00 & 1.50&  0.27 &   0.08&  0.23 &   0.04&  0.30 &   0.05&  0.34 &   0.18\\ 
      &      & 1.50 & 2.00&  0.05 &   0.08&  0.19 &   0.04&  0.20 &   0.04&  0.29 &   0.17\\ 
      &      & 2.00 & 2.50&       &       &  0.07 &   0.03&  0.15 &   0.04&  0.23 &   0.14\\ 
      &      & 2.50 & 3.00&       &       &  0.03 &   0.02&  0.12 &   0.03&  0.15 &   0.11\\ 
      &      & 3.00 & 3.50&       &       &  0.01 &   0.02&  0.10 &   0.03&  0.08 &   0.09\\ 
      &      & 3.50 & 4.00&       &       &  0.02 &   0.02&  0.05 &   0.02&  0.11 &   0.11\\ 
      &      & 4.00 & 5.00&       &       & 0.01  &  0.02 & 0.02 &  0.01&  0.02 &   0.04\\ 
      &      & 5.00 & 6.50&       &       &       &       & 0.003 &  0.003& 0.001 &  0.003\\ 
      &      & 6.50 & 8.00&       &       &       &       &       &       & 0.006 &  0.025\\ 
0.200 &0.250 & 0.50 & 1.00&  0.32 &   0.17&  0.33 &   0.10&  0.39 &   0.10& 0.05  &  0.12 \\ 
      &      & 1.00 & 1.50&  0.17 &   0.09&  0.30 &   0.06&  0.44 &   0.09&  0.09 &   0.17\\ 
      &      & 1.50 & 2.00&  0.08 &   0.05&  0.12 &   0.04&  0.16 &   0.05&  0.04 &   0.14\\ 
      &      & 2.00 & 2.50&  0.07 &   0.07&  0.10 &   0.04&  0.14 &   0.05&  0.26 &   0.31\\ 
      &      & 2.50 & 3.00&       &       &  0.06 &   0.02&  0.09 &   0.03&  0.09 &   0.22\\ 
      &      & 3.00 & 3.50&       &       &  0.04 &   0.02&  0.04 &   0.03&  0.01 &   0.11\\ 
      &      & 3.50 & 4.00&       &       & 0.03 &  0.01&  0.02 &   0.02&       &       \\ 
      &      & 4.00 & 5.00&       &       & 0.02 &  0.01& 0.024 &  0.035&       &      \\ 
      &      & 5.00 & 6.50&       &       &       &       & 0.009 &  0.012&       &      \\ 
      &      & 6.50 & 8.00&       &       &       &       &       &       &       &      \\

\hline
\end{tabular}
}
\end{table}
\clearpage

\begin{table}[!ht]
  \caption{\label{tab:xsec_results_Cu_pr}
    HARP results for the double-differential $p$  production
    cross-section in the laboratory system,
    $d^2\sigma^{p}/(dpd\Omega)$, for $p$--Cu interactions at 3,5,8,12~\GeVc.
    Each row refers to a
    different $(p_{\hbox{\small min}} \le p<p_{\hbox{\small max}},
    \theta_{\hbox{\small min}} \le \theta<\theta_{\hbox{\small max}})$ bin,
    where $p$ and $\theta$ are the outgoing proton momentum and polar angle, respectively.
    The central value as well as the square-root of the diagonal elements
    of the covariance matrix are given.}

\small{
\begin{tabular}{rrrr|r@{$\pm$}lr@{$\pm$}lr@{$\pm$}lr@{$\pm$}l}
\hline
$\theta_{\hbox{\small min}}$ &
$\theta_{\hbox{\small max}}$ &
$p_{\hbox{\small min}}$ &
$p_{\hbox{\small max}}$ &
\multicolumn{8}{c}{$d^2\sigma^{p}/(dpd\Omega)$}
\\
(rad) & (rad) & (\GeVc) & (\GeVc) &
\multicolumn{8}{c}{(barn/(sr \GeVc ))}
\\
  &  &  &
&\multicolumn{2}{c}{$ \bf{3 \ \GeVc}$}
&\multicolumn{2}{c}{$ \bf{5 \ \GeVc}$}
&\multicolumn{2}{c}{$ \bf{8 \ \GeVc}$}
&\multicolumn{2}{c}{$ \bf{12 \ \GeVc}$}
\\
\hline
0.050 &0.100 & 0.50 & 1.00&  0.48 &   0.19&  0.45 &   0.09&  0.53 &   0.08&  0.51 &   0.09\\ 
      &      & 1.00 & 1.50&  0.36 &   0.11&  0.49 &   0.06&  0.52 &   0.05&  0.54 &   0.07\\ 
      &      & 1.50 & 2.00&  0.40 &   0.11&  0.54 &   0.07&  0.54 &   0.05&  0.46 &   0.06\\ 
      &      & 2.00 & 2.50&  0.46 &   0.15&  0.57 &   0.06&  0.50 &   0.04&  0.44 &   0.05\\ 
      &      & 2.50 & 3.00&       &       &  0.76 &   0.08&  0.55 &   0.05&  0.57 &   0.05\\ 
      &      & 3.00 & 3.50&       &       &  0.87 &   0.07&  0.64 &   0.04&  0.49 &   0.05\\ 
      &      & 3.50 & 4.00&       &       &  0.77 &   0.07&  0.66 &   0.05&  0.52 &   0.05\\ 
      &      & 4.00 & 5.00&       &       &  0.86 &   0.12&  0.80 &   0.04&  0.62 &   0.04\\ 
      &      & 5.00 & 6.50&       &       &       &       &  0.54 &   0.03&  0.51 &   0.03\\ 
      &      & 6.50 & 8.00&       &       &       &       &       &       &  0.41 &   0.02\\ 
0.100 &0.150 & 0.50 & 1.00&  0.40 &   0.17&  0.54 &   0.10&  0.67 &   0.09&  0.69 &   0.12\\ 
      &      & 1.00 & 1.50&  0.59 &   0.15&  0.49 &   0.07&  0.48 &   0.05&  0.48 &   0.07\\ 
      &      & 1.50 & 2.00&  0.62 &   0.15&  0.45 &   0.06&  0.47 &   0.05&  0.43 &   0.06\\ 
      &      & 2.00 & 2.50&  0.52 &   0.12&  0.50 &   0.07&  0.41 &   0.04&  0.44 &   0.06\\ 
      &      & 2.50 & 3.00&       &       &  0.52 &   0.06&  0.43 &   0.05&  0.39 &   0.05\\ 
      &      & 3.00 & 3.50&       &       &  0.50 &   0.06&  0.49 &   0.04&  0.35 &   0.04\\ 
      &      & 3.50 & 4.00&       &       &  0.45 &   0.05&  0.42 &   0.03&  0.40 &   0.04\\ 
      &      & 4.00 & 5.00&       &       &  0.50 &   0.06&  0.34 &   0.03&  0.35 &   0.03\\ 
      &      & 5.00 & 6.50&       &       &       &       &  0.19 &   0.02&  0.20 &   0.02\\ 
      &      & 6.50 & 8.00&       &       &       &       &       &       & 0.104 &  0.013\\ 
0.150 &0.200 & 0.50 & 1.00&  0.37 &   0.16&  0.54 &   0.11&  0.61 &   0.10&  0.57 &   0.11\\ 
      &      & 1.00 & 1.50&  0.51 &   0.13&  0.40 &   0.06&  0.40 &   0.05&  0.35 &   0.05\\ 
      &      & 1.50 & 2.00&  0.45 &   0.13&  0.43 &   0.05&  0.32 &   0.04&  0.29 &   0.05\\ 
      &      & 2.00 & 2.50&  0.36 &   0.11&  0.42 &   0.06&  0.38 &   0.04&  0.29 &   0.05\\ 
      &      & 2.50 & 3.00&       &       &  0.23 &   0.04&  0.28 &   0.03&  0.17 &   0.03\\ 
      &      & 3.00 & 3.50&       &       &  0.16 &   0.03&  0.21 &   0.03&  0.22 &   0.03\\ 
      &      & 3.50 & 4.00&       &       &  0.17 &   0.03&  0.16 &   0.02&  0.16 &   0.02\\ 
      &      & 4.00 & 5.00&       &       &  0.16 &   0.02&  0.14 &   0.02&  0.12 &   0.02\\ 
      &      & 5.00 & 6.50&       &       &       &       & 0.051 &  0.008& 0.034 &  0.007\\ 
      &      & 6.50 & 8.00&       &       &       &       &       &       & 0.016 &  0.004\\ 
0.200 &0.250 & 0.50 & 1.00&  0.65 &   0.29&  0.44 &   0.11&  0.59 &   0.10&  0.62 &   0.13\\ 
      &      & 1.00 & 1.50&  0.23 &   0.12&  0.43 &   0.08&  0.40 &   0.07&  0.36 &   0.08\\ 
      &      & 1.50 & 2.00&  0.21 &   0.09&  0.29 &   0.06&  0.32 &   0.06&  0.20 &   0.06\\ 
      &      & 2.00 & 2.50&  0.27 &   0.11&  0.18 &   0.04&  0.26 &   0.05&  0.18 &   0.05\\ 
      &      & 2.50 & 3.00&       &       &  0.19 &   0.04&  0.18 &   0.03&  0.12 &   0.04\\ 
      &      & 3.00 & 3.50&       &       &  0.08 &   0.02&  0.13 &   0.02&  0.09 &   0.02\\ 
      &      & 3.50 & 4.00&       &       & 0.07  &  0.02 & 0.067 &  0.014&  0.05 &   0.02\\ 
      &      & 4.00 & 5.00&       &       &  0.06 &   0.02& 0.063 &  0.013& 0.043 &  0.014\\ 
      &      & 5.00 & 6.50&       &       &       &       & 0.024 &  0.006& 0.027 &  0.010\\ 
      &      & 6.50 & 8.00&       &       &       &       &       &       & 0.019 &  0.008\\

\hline
\end{tabular}
}
\end{table}
\clearpage

\begin{table}[!ht]
  \caption{\label{tab:xsec_results_Sn_pim}
    HARP results for the double-differential $p$  production
    cross-section in the laboratory system,
    $d^2\sigma^{p}/(dpd\Omega)$, for $\pi^{-}$--Sn interactions at 3,5,8,12~\GeVc.
    Each row refers to a
    different $(p_{\hbox{\small min}} \le p<p_{\hbox{\small max}},
    \theta_{\hbox{\small min}} \le \theta<\theta_{\hbox{\small max}})$ bin,
    where $p$ and $\theta$ are the outgoing proton momentum and polar angle, respectively.
    The central value as well as the square-root of the diagonal elements
    of the covariance matrix are given.}

\small{
\begin{tabular}{rrrr|r@{$\pm$}lr@{$\pm$}lr@{$\pm$}lr@{$\pm$}l}
\hline
$\theta_{\hbox{\small min}}$ &
$\theta_{\hbox{\small max}}$ &
$p_{\hbox{\small min}}$ &
$p_{\hbox{\small max}}$ &
\multicolumn{8}{c}{$d^2\sigma^{p}/(dpd\Omega)$}
\\
(rad) & (rad) & (\GeVc) & (\GeVc) &
\multicolumn{8}{c}{(barn/(sr \GeVc ))}
\\
  &  &  &
&\multicolumn{2}{c}{$ \bf{3 \ \GeVc}$}
&\multicolumn{2}{c}{$ \bf{5 \ \GeVc}$}
&\multicolumn{2}{c}{$ \bf{8 \ \GeVc}$}
&\multicolumn{2}{c}{$ \bf{12 \ \GeVc}$}
\\
\hline
0.050 &0.100 & 0.50 & 1.00&  0.52 &   0.09&  0.67 &   0.10&  0.79 &   0.11&  0.75 &   0.11\\ 
      &      & 1.00 & 1.50&  0.26 &   0.05&  0.45 &   0.05&  0.51 &   0.07&  0.55 &   0.06\\ 
      &      & 1.50 & 2.00&  0.15 &   0.03&  0.30 &   0.04&  0.47 &   0.06&  0.41 &   0.08\\ 
      &      & 2.00 & 2.50&  0.05 &   0.02&  0.19 &   0.03&  0.24 &   0.05&  0.11 &   0.34\\ 
      &      & 2.50 & 3.00&       &       &  0.09 &   0.02&  0.26 &   0.03&  0.27 &   0.07\\ 
      &      & 3.00 & 3.50&       &       & 0.043 &  0.013&  0.12 &   0.03&  0.33 &   0.07\\ 
      &      & 3.50 & 4.00&       &       & 0.029 &  0.009&  0.14 &   0.02&  0.25 &   0.04\\ 
      &      & 4.00 & 5.00&       &       & 0.014 &  0.007&  0.07 &   0.02&  0.22 &   0.03\\ 
      &      & 5.00 & 6.50&       &       &       &       & 0.019 &  0.007& 0.121 &  0.015\\ 
      &      & 6.50 & 8.00&       &       &       &       &       &       & 0.046 &  0.010\\ 
0.100 &0.150 & 0.50 & 1.00&  0.53 &   0.09&  0.62 &   0.10&  0.82 &   0.13&  0.81 &   0.13\\ 
      &      & 1.00 & 1.50&  0.33 &   0.06&  0.55 &   0.07&  0.61 &   0.08&  0.72 &   0.08\\ 
      &      & 1.50 & 2.00&  0.16 &   0.03&  0.32 &   0.04&  0.38 &   0.06&  0.42 &   0.07\\ 
      &      & 2.00 & 2.50&  0.06 &   0.02&  0.21 &   0.03&  0.32 &   0.04&  0.22 &   0.07\\ 
      &      & 2.50 & 3.00&       &       &  0.10 &   0.02&  0.21 &   0.03&  0.30 &   0.04\\ 
      &      & 3.00 & 3.50&       &       & 0.050 &  0.012&  0.15 &   0.03&  0.22 &   0.04\\ 
      &      & 3.50 & 4.00&       &       & 0.019 &  0.007&  0.12 &   0.02&  0.17 &   0.02\\ 
      &      & 4.00 & 5.00&       &       & 0.017 &  0.007& 0.041 &  0.011&  0.14 &   0.02\\ 
      &      & 5.00 & 6.50&       &       &       &       & 0.009 &  0.003& 0.052 &  0.010\\ 
      &      & 6.50 & 8.00&       &       &       &       &       &       & 0.011 &  0.003\\ 
0.150 &0.200 & 0.50 & 1.00&  0.51 &   0.10&  0.82 &   0.13&  0.66 &   0.12&  0.58 &   0.11\\ 
      &      & 1.00 & 1.50&  0.19 &   0.04&  0.37 &   0.05&  0.44 &   0.06&  0.58 &   0.07\\ 
      &      & 1.50 & 2.00&  0.09 &   0.03&  0.28 &   0.04&  0.37 &   0.05&  0.24 &   0.05\\ 
      &      & 2.00 & 2.50&  0.04 &   0.02&  0.15 &   0.03&  0.21 &   0.05&  0.22 &   0.10\\ 
      &      & 2.50 & 3.00&       &       &  0.07 &   0.02&  0.17 &   0.03&  0.27 &   0.07\\ 
      &      & 3.00 & 3.50&       &       & 0.024 &  0.008&  0.13 &   0.03&  0.15 &   0.03\\ 
      &      & 3.50 & 4.00&       &       & 0.007 &  0.004&  0.08 &   0.02&  0.09 &   0.02\\ 
      &      & 4.00 & 5.00&       &       & 0.003 &  0.003& 0.020 &  0.006& 0.071 &  0.014\\ 
      &      & 5.00 & 6.50&       &       &       &       & 0.001 &  0.001& 0.011 &  0.004\\ 
      &      & 6.50 & 8.00&       &       &       &       &       &       & 0.002 &  0.001\\ 
0.200 &0.250 & 0.50 & 1.00&  0.57 &   0.13&  0.69 &   0.13&  0.83 &   0.16&  0.73 &   0.14\\ 
      &      & 1.00 & 1.50&  0.29 &   0.07&  0.60 &   0.09&  0.86 &   0.14&  0.70 &   0.12\\ 
      &      & 1.50 & 2.00&  0.14 &   0.04&  0.35 &   0.06&  0.48 &   0.10&  0.53 &   0.11\\ 
      &      & 2.00 & 2.50&  0.07 &   0.03&  0.18 &   0.04&  0.43 &   0.08&  0.37 &   0.10\\ 
      &      & 2.50 & 3.00&       &       &  0.11 &   0.03&  0.17 &   0.05&  0.27 &   0.06\\ 
      &      & 3.00 & 3.50&       &       & 0.025 &  0.012&  0.04 &   0.02&  0.12 &   0.04\\ 
      &      & 3.50 & 4.00&       &       & 0.005 &  0.003& 0.016 &  0.009&  0.06 &   0.02\\ 
      &      & 4.00 & 5.00&       &       & 0.003 &  0.002& 0.007 &  0.004&  0.04 &   0.02\\ 
      &      & 5.00 & 6.50&       &       &       &       & 0.001 &  0.006& 0.008 &  0.004\\ 
      &      & 6.50 & 8.00&       &       &       &       &       &       & 0.002 &  0.002\\ 

\hline
\end{tabular}
}
\end{table}
\clearpage

\begin{table}[!ht]
  \caption{\label{tab:xsec_results_Sn_pip}
    HARP results for the double-differential $p$  production
    cross-section in the laboratory system,
    $d^2\sigma^{p}/(dpd\Omega)$, for $\pi^{+}$--Sn interactions at 3,5,8,12~\GeVc.
    Each row refers to a
    different $(p_{\hbox{\small min}} \le p<p_{\hbox{\small max}},
    \theta_{\hbox{\small min}} \le \theta<\theta_{\hbox{\small max}})$ bin,
    where $p$ and $\theta$ are the outgoing proton momentum and polar angle, respectively.
    The central value as well as the square-root of the diagonal elements
    of the covariance matrix are given.}

\small{
\begin{tabular}{rrrr|r@{$\pm$}lr@{$\pm$}lr@{$\pm$}lr@{$\pm$}l}
\hline
$\theta_{\hbox{\small min}}$ &
$\theta_{\hbox{\small max}}$ &
$p_{\hbox{\small min}}$ &
$p_{\hbox{\small max}}$ &
\multicolumn{8}{c}{$d^2\sigma^{p}/(dpd\Omega)$}
\\
(rad) & (rad) & (\GeVc) & (\GeVc) &
\multicolumn{8}{c}{(barn/(sr \GeVc ))}
\\
  &  &  &
&\multicolumn{2}{c}{$ \bf{3 \ \GeVc}$}
&\multicolumn{2}{c}{$ \bf{5 \ \GeVc}$}
&\multicolumn{2}{c}{$ \bf{8 \ \GeVc}$}
&\multicolumn{2}{c}{$ \bf{12 \ \GeVc}$}
\\
\hline
0.050 &0.100 & 0.50 & 1.00&  0.49 &   0.16&  0.70 &   0.13&  0.66 &   0.12&  0.75 &   0.27\\ 
      &      & 1.00 & 1.50&  0.23 &   0.10&  0.54 &   0.07&  0.62 &   0.09&  0.54 &   0.17\\ 
      &      & 1.50 & 2.00&       &       &  0.37 &   0.12&  0.50 &   0.08&  0.45 &   0.60\\ 
      &      & 2.00 & 2.50&       &       &  0.01 &   0.03&  0.36 &   0.07&  0.21 &   0.32\\ 
      &      & 2.50 & 3.00&       &       &       &       &  0.15 &   0.06&  0.26 &   0.25\\ 
      &      & 3.00 & 3.50&       &       &       &       &  0.05 &   0.11&  0.21 &   0.11\\ 
      &      & 3.50 & 4.00&       &       &       &       &  0.01 &   0.33&  0.21 &   0.10\\ 
      &      & 4.00 & 5.00&       &       &       &       &       &       &  0.12 &   0.08\\ 
      &      & 5.00 & 6.50&       &       &       &       &       &       &       &       \\ 
      &      & 6.50 & 8.00&       &       &       &       &       &       &       &       \\ 
0.100 &0.150 & 0.50 & 1.00&  0.55 &   0.17&  0.84 &   0.15&  1.03 &   0.19&  0.88 &   0.30\\ 
      &      & 1.00 & 1.50&  0.43 &   0.12&  0.51 &   0.08&  0.58 &   0.09&  0.76 &   0.21\\ 
      &      & 1.50 & 2.00&  0.14 &   0.12&  0.41 &   0.08&  0.48 &   0.08&  0.36 &   0.14\\ 
      &      & 2.00 & 2.50&       &       &  0.10 &   0.06&  0.22 &   0.05&  0.33 &   0.14\\ 
      &      & 2.50 & 3.00&       &       &       &       &  0.20 &   0.04&  0.19 &   0.10\\ 
      &      & 3.00 & 3.50&       &       &       &       &  0.21 &   0.06&  0.08 &   0.06\\ 
      &      & 3.50 & 4.00&       &       &       &       &  0.08 &   0.04&  0.15 &   0.09\\ 
      &      & 4.00 & 5.00&       &       &       &       &  0.08 &   0.04&  0.14 &   0.07\\ 
      &      & 5.00 & 6.50&       &       &       &       &  0.01 &   0.02&  0.05 &   0.06\\ 
      &      & 6.50 & 8.00&       &       &       &       &       &       &  0.02 &   0.04\\ 
0.150 &0.200 & 0.50 & 1.00&  0.64 &   0.19&  0.96 &   0.16&  0.67 &   0.14&  0.74 &   0.30\\ 
      &      & 1.00 & 1.50&  0.29 &   0.08&  0.42 &   0.06&  0.48 &   0.08&  0.63 &   0.19\\ 
      &      & 1.50 & 2.00&       &       &  0.30 &   0.05&  0.38 &   0.07&  0.43 &   0.16\\ 
      &      & 2.00 & 2.50&       &       &  0.10 &   0.05&  0.23 &   0.05&  0.25 &   0.12\\ 
      &      & 2.50 & 3.00&       &       &  0.00 &   0.00&  0.13 &   0.03&  0.04 &   0.04\\ 
      &      & 3.00 & 3.50&       &       &       &       &  0.08 &   0.03&  0.09 &   0.06\\ 
      &      & 3.50 & 4.00&       &       &       &       &  0.06 &   0.02&  0.17 &   0.09\\ 
      &      & 4.00 & 5.00&       &       &       &       & 0.041 &  0.013&  0.05 &   0.04\\ 
      &      & 5.00 & 6.50&       &       &       &       & 0.005 &  0.003& 0.002 &  0.005\\ 
      &      & 6.50 & 8.00&       &       &       &       &       &       & 0.000 &  0.002\\ 
0.200 &0.250 & 0.50 & 1.00&  0.34 &   0.17&  0.64 &   0.18&  0.95 &   0.21&  0.50 &   0.27\\ 
      &      & 1.00 & 1.50&  0.19 &   0.09&  0.46 &   0.09&  0.68 &   0.14&  0.79 &   0.33\\ 
      &      & 1.50 & 2.00&  0.09 &   0.06&  0.17 &   0.06&  0.39 &   0.10&  0.14 &   0.15\\ 
      &      & 2.00 & 2.50&  0.08 &   0.07&  0.07 &   0.04&  0.31 &   0.09&  0.13 &   0.13\\ 
      &      & 2.50 & 3.00&       &       &  0.07 &   0.03&  0.12 &   0.04&  0.11 &   0.11\\ 
      &      & 3.00 & 3.50&       &       &  0.04 &   0.02&  0.08 &   0.03&  0.07 &   0.08\\ 
      &      & 3.50 & 4.00&       &       &  0.02 &   0.02&  0.06 &   0.02&  0.09 &   0.09\\ 
      &      & 4.00 & 5.00&       &       & 0.01 &  0.01  &  0.07 &   0.03&  0.05 &   0.07\\ 
      &      & 5.00 & 6.50&       &       &       &       & 0.02  &   0.01& 0.01  &  0.02\\ 
      &      & 6.50 & 8.00&       &       &       &       &       &       & 0.002 &  0.017\\

\hline
\end{tabular}
}
\end{table}
\clearpage

\begin{table}[!ht]
  \caption{\label{tab:xsec_results_Sn_pr}
    HARP results for the double-differential $p$  production
    cross-section in the laboratory system,
    $d^2\sigma^{p}/(dpd\Omega)$, for p--C interactions at 3,5,8,12~\GeVc.
    Each row refers to a
    different $(p_{\hbox{\small min}} \le p<p_{\hbox{\small max}},
    \theta_{\hbox{\small min}} \le \theta<\theta_{\hbox{\small max}})$ bin,
    where $p$ and $\theta$ are the outgoing proton momentum and polar angle, respectively.
    The central value as well as the square-root of the diagonal elements
    of the covariance matrix are given.}

\small{
\begin{tabular}{rrrr|r@{$\pm$}lr@{$\pm$}lr@{$\pm$}lr@{$\pm$}l}
\hline
$\theta_{\hbox{\small min}}$ &
$\theta_{\hbox{\small max}}$ &
$p_{\hbox{\small min}}$ &
$p_{\hbox{\small max}}$ &
\multicolumn{8}{c}{$d^2\sigma^{p}/(dpd\Omega)$}
\\
(rad) & (rad) & (\GeVc) & (\GeVc) &
\multicolumn{8}{c}{(barn/(sr \GeVc ))}
\\
  &  &  &
&\multicolumn{2}{c}{$ \bf{3 \ \GeVc}$}
&\multicolumn{2}{c}{$ \bf{5 \ \GeVc}$}
&\multicolumn{2}{c}{$ \bf{8 \ \GeVc}$}
&\multicolumn{2}{c}{$ \bf{12 \ \GeVc}$}
\\
\hline
0.050 &0.100 & 0.50 & 1.00&  0.46 &   0.16&  0.73 &   0.13&  0.94 &   0.12&  1.08 &   0.13\\ 
      &      & 1.00 & 1.50&  0.45 &   0.11&  0.73 &   0.08&  0.88 &   0.08&  0.85 &   0.07\\ 
      &      & 1.50 & 2.00&  0.35 &   0.09&  0.66 &   0.08&  0.76 &   0.07&  0.73 &   0.07\\ 
      &      & 2.00 & 2.50&  0.24 &   0.08&  0.59 &   0.08&  0.72 &   0.06&  0.59 &   0.05\\ 
      &      & 2.50 & 3.00&       &       &  0.99 &   0.10&  0.74 &   0.06&  0.72 &   0.06\\ 
      &      & 3.00 & 3.50&       &       &  1.05 &   0.09&  0.89 &   0.06&  0.68 &   0.05\\ 
      &      & 3.50 & 4.00&       &       &  1.08 &   0.11&  0.86 &   0.06&  0.71 &   0.06\\ 
      &      & 4.00 & 5.00&       &       &  1.12 &   0.13&  1.02 &   0.05&  0.81 &   0.05\\ 
      &      & 5.00 & 6.50&       &       &       &       &  0.68 &   0.04&  0.68 &   0.03\\ 
      &      & 6.50 & 8.00&       &       &       &       &       &       &  0.56 &   0.03\\ 
0.100 &0.150 & 0.50 & 1.00&  0.57 &   0.18&  0.94 &   0.15&  1.19 &   0.16&  1.28 &   0.17\\ 
      &      & 1.00 & 1.50&  0.70 &   0.15&  0.68 &   0.10&  0.89 &   0.09&  0.97 &   0.09\\ 
      &      & 1.50 & 2.00&  0.57 &   0.14&  0.67 &   0.08&  0.68 &   0.07&  0.82 &   0.08\\ 
      &      & 2.00 & 2.50&  0.64 &   0.13&  0.62 &   0.09&  0.63 &   0.07&  0.61 &   0.06\\ 
      &      & 2.50 & 3.00&       &       &  0.68 &   0.08&  0.68 &   0.06&  0.53 &   0.05\\ 
      &      & 3.00 & 3.50&       &       &  0.65 &   0.07&  0.65 &   0.06&  0.48 &   0.05\\ 
      &      & 3.50 & 4.00&       &       &  0.52 &   0.06&  0.54 &   0.04&  0.55 &   0.05\\ 
      &      & 4.00 & 5.00&       &       &  0.54 &   0.08&  0.49 &   0.04&  0.51 &   0.04\\ 
      &      & 5.00 & 6.50&       &       &       &       &  0.20 &   0.02&  0.25 &   0.02\\ 
      &      & 6.50 & 8.00&       &       &       &       &       &       & 0.140 &  0.015\\ 
0.150 &0.200 & 0.50 & 1.00&  0.57 &   0.19&  1.15 &   0.18&  1.08 &   0.15&  0.96 &   0.14\\ 
      &      & 1.00 & 1.50&  0.43 &   0.11&  0.52 &   0.07&  0.61 &   0.07&  0.63 &   0.07\\ 
      &      & 1.50 & 2.00&  0.34 &   0.11&  0.50 &   0.07&  0.54 &   0.06&  0.55 &   0.06\\ 
      &      & 2.00 & 2.50&  0.40 &   0.12&  0.39 &   0.06&  0.45 &   0.05&  0.45 &   0.05\\ 
      &      & 2.50 & 3.00&       &       &  0.36 &   0.06&  0.45 &   0.05&  0.33 &   0.04\\ 
      &      & 3.00 & 3.50&       &       &  0.24 &   0.04&  0.33 &   0.03&  0.33 &   0.04\\ 
      &      & 3.50 & 4.00&       &       &  0.16 &   0.03&  0.23 &   0.03&  0.25 &   0.03\\ 
      &      & 4.00 & 5.00&       &       &  0.18 &   0.03&  0.18 &   0.02&  0.17 &   0.02\\ 
      &      & 5.00 & 6.50&       &       &       &       & 0.065 &  0.010& 0.068 &  0.009\\ 
      &      & 6.50 & 8.00&       &       &       &       &       &       & 0.023 &  0.005\\ 
0.200 &0.250 & 0.50 & 1.00&  0.63 &   0.26&  0.72 &   0.16&  1.18 &   0.18&  0.97 &   0.16\\ 
      &      & 1.00 & 1.50&  0.17 &   0.09&  0.61 &   0.11&  0.77 &   0.11&  0.74 &   0.11\\ 
      &      & 1.50 & 2.00&  0.15 &   0.06&  0.32 &   0.07&  0.47 &   0.08&  0.40 &   0.07\\ 
      &      & 2.00 & 2.50&  0.21 &   0.08&  0.23 &   0.06&  0.45 &   0.08&  0.39 &   0.07\\ 
      &      & 2.50 & 3.00&       &       &  0.23 &   0.05&  0.29 &   0.05&  0.23 &   0.04\\ 
      &      & 3.00 & 3.50&       &       &  0.08 &   0.02&  0.23 &   0.03&  0.18 &   0.03\\ 
      &      & 3.50 & 4.00&       &       &  0.08 &   0.02&  0.15 &   0.03&  0.11 &   0.02\\ 
      &      & 4.00 & 5.00&       &       &  0.06 &   0.02&  0.12 &   0.02&  0.08 &   0.02\\ 
      &      & 5.00 & 6.50&       &       &       &       & 0.044 &  0.011& 0.037 &  0.010\\ 
      &      & 6.50 & 8.00&       &       &       &       &       &       & 0.018 &  0.005\\

\hline
\end{tabular}
}
\end{table}
\clearpage

\begin{table}[!ht]
  \caption{\label{tab:xsec_results_Ta_pim}
    HARP results for the double-differential $p$  production
    cross-section in the laboratory system,
    $d^2\sigma^{p}/(dpd\Omega)$, for $\pi^{-}$--Ta interactions at 3,5,8,12~\GeVc.
    Each row refers to a
    different $(p_{\hbox{\small min}} \le p<p_{\hbox{\small max}},
    \theta_{\hbox{\small min}} \le \theta<\theta_{\hbox{\small max}})$ bin,
    where $p$ and $\theta$ are the outgoing proton momentum and polar angle, respectively.
    The central value as well as the square-root of the diagonal elements
    of the covariance matrix are given.}

\small{
\begin{tabular}{rrrr|r@{$\pm$}lr@{$\pm$}lr@{$\pm$}lr@{$\pm$}l}
\hline
$\theta_{\hbox{\small min}}$ &
$\theta_{\hbox{\small max}}$ &
$p_{\hbox{\small min}}$ &
$p_{\hbox{\small max}}$ &
\multicolumn{8}{c}{$d^2\sigma^{p}/(dpd\Omega)$}
\\
(rad) & (rad) & (\GeVc) & (\GeVc) &
\multicolumn{8}{c}{(barn/(sr \GeVc ))}
\\
  &  &  &
&\multicolumn{2}{c}{$ \bf{3 \ \GeVc}$}
&\multicolumn{2}{c}{$ \bf{5 \ \GeVc}$}
&\multicolumn{2}{c}{$ \bf{8 \ \GeVc}$}
&\multicolumn{2}{c}{$ \bf{12 \ \GeVc}$}
\\
\hline
0.050 &0.100 & 0.50 & 1.00&  0.64 &   0.15&  0.81 &   0.12&  1.17 &   0.16&  0.91 &   0.14\\ 
      &      & 1.00 & 1.50&  0.30 &   0.08&  0.54 &   0.07&  0.84 &   0.10&  0.85 &   0.10\\ 
      &      & 1.50 & 2.00&  0.08 &   0.03&  0.40 &   0.06&  0.61 &   0.08&  0.50 &   0.08\\ 
      &      & 2.00 & 2.50& 0.01  &  0.01 &  0.26 &   0.04&  0.32 &   0.06&  0.25 &   0.07\\ 
      &      & 2.50 & 3.00&       &       &  0.09 &   0.02&  0.29 &   0.04&  0.40 &   0.06\\ 
      &      & 3.00 & 3.50&       &       & 0.04 &  0.02&  0.19 &   0.04&  0.41 &   0.05\\ 
      &      & 3.50 & 4.00&       &       & 0.02 &  0.01&  0.16 &   0.03&  0.32 &   0.04\\ 
      &      & 4.00 & 5.00&       &       & 0.01 &  0.01&  0.11 &   0.02&  0.30 &   0.03\\ 
      &      & 5.00 & 6.50&       &       &       &       & 0.029 &  0.010&  0.15 &   0.02\\ 
      &      & 6.50 & 8.00&       &       &       &       &       &       & 0.057 &  0.012\\ 
0.100 &0.150 & 0.50 & 1.00&  0.72 &   0.17&  1.16 &   0.17&  1.20 &   0.18&  1.42 &   0.21\\ 
      &      & 1.00 & 1.50&  0.53 &   0.12&  0.71 &   0.09&  0.71 &   0.09&  1.06 &   0.12\\ 
      &      & 1.50 & 2.00&  0.06 &   0.03&  0.43 &   0.06&  0.48 &   0.08&  0.53 &   0.08\\ 
      &      & 2.00 & 2.50&  0.03 &   0.02&  0.23 &   0.04&  0.40 &   0.06&  0.29 &   0.06\\ 
      &      & 2.50 & 3.00&       &       &  0.09 &   0.02&  0.24 &   0.03&  0.29 &   0.05\\ 
      &      & 3.00 & 3.50&       &       & 0.04 &    0.01&  0.16 &   0.04&  0.30 &   0.04\\ 
      &      & 3.50 & 4.00&       &       & 0.02 &    0.01&  0.12 &   0.03&  0.23 &   0.03\\ 
      &      & 4.00 & 5.00&       &       & 0.004 &  0.003&  0.11 &   0.02&  0.17 &   0.02\\ 
      &      & 5.00 & 6.50&       &       &       &       & 0.01 &  0.01 & 0.08 &  0.01\\ 
      &      & 6.50 & 8.00&       &       &       &       &       &       & 0.02 &  0.01\\ 
0.150 &0.200 & 0.50 & 1.00&  0.88 &   0.21&  0.92 &   0.15&  0.93 &   0.15&  1.06 &   0.17\\ 
      &      & 1.00 & 1.50&  0.19 &   0.06&  0.45 &   0.07&  0.51 &   0.07&  0.54 &   0.07\\ 
      &      & 1.50 & 2.00&  0.06 &   0.03&  0.31 &   0.05&  0.40 &   0.05&  0.56 &   0.08\\ 
      &      & 2.00 & 2.50&  0.07 &   0.03&  0.21 &   0.04&  0.22 &   0.05&  0.31 &   0.06\\ 
      &      & 2.50 & 3.00&       &       &  0.10 &   0.02&  0.22 &   0.04&  0.33 &   0.05\\ 
      &      & 3.00 & 3.50&       &       & 0.03 &  0.01&  0.14 &   0.03&  0.24 &   0.04\\ 
      &      & 3.50 & 4.00&       &       & 0.02 &  0.01&  0.08 &   0.02&  0.14 &   0.02\\ 
      &      & 4.00 & 5.00&       &       & 0.006 &  0.004& 0.02 &  0.01&  0.07 &   0.02\\ 
      &      & 5.00 & 6.50&       &       &       &       &       &       & 0.02 &  0.01\\ 
      &      & 6.50 & 8.00&       &       &       &       &       &       & 0.002 &  0.001\\ 
0.200 &0.250 & 0.50 & 1.00&  0.70 &   0.22&  1.10 &   0.20&  1.42 &   0.25&  1.23 &   0.22\\ 
      &      & 1.00 & 1.50&  0.53 &   0.16&  0.67 &   0.12&  1.06 &   0.17&  1.02 &   0.17\\ 
      &      & 1.50 & 2.00&  0.13 &   0.07&  0.52 &   0.09&  0.69 &   0.14&  0.61 &   0.13\\ 
      &      & 2.00 & 2.50&  0.05 &   0.05&  0.22 &   0.05&  0.36 &   0.08&  0.55 &   0.12\\ 
      &      & 2.50 & 3.00&       &       &  0.07 &   0.03&  0.26 &   0.07&  0.36 &   0.08\\ 
      &      & 3.00 & 3.50&       &       &  0.03 &   0.02&  0.04 &   0.02&  0.13 &   0.03\\ 
      &      & 3.50 & 4.00&       &       & 0.01  &  0.01 & 0.03  &  0.01 &  0.09 &   0.02\\ 
      &      & 4.00 & 5.00&       &       & 0.004 &  0.003& 0.03  &  0.01&  0.05 &   0.02\\ 
      &      & 5.00 & 6.50&       &       &       &       & 0.004 &  0.006& 0.01 &  0.02\\ 
      &      & 6.50 & 8.00&       &       &       &       &       &       & 0.002 &  0.002\\

\hline
\end{tabular}
}
\end{table}
\clearpage

\begin{table}[!ht]
  \caption{\label{tab:xsec_results_Ta_pip}
    HARP results for the double-differential $p$  production
    cross-section in the laboratory system,
    $d^2\sigma^{p}/(dpd\Omega)$, for $\pi^{+}$--Ta interactions at 3,5,8,12~\GeVc.
    Each row refers to a
    different $(p_{\hbox{\small min}} \le p<p_{\hbox{\small max}},
    \theta_{\hbox{\small min}} \le \theta<\theta_{\hbox{\small max}})$ bin,
    where $p$ and $\theta$ are the outgoing proton momentum and polar angle, respectively.
    The central value as well as the square-root of the diagonal elements
    of the covariance matrix are given.}

\small{
\begin{tabular}{rrrr|r@{$\pm$}lr@{$\pm$}lr@{$\pm$}lr@{$\pm$}l}
\hline
$\theta_{\hbox{\small min}}$ &
$\theta_{\hbox{\small max}}$ &
$p_{\hbox{\small min}}$ &
$p_{\hbox{\small max}}$ &
\multicolumn{8}{c}{$d^2\sigma^{p}/(dpd\Omega)$}
\\
(rad) & (rad) & (\GeVc) & (\GeVc) &
\multicolumn{8}{c}{(barn/(sr \GeVc ))}
\\
  &  &  &
&\multicolumn{2}{c}{$ \bf{3 \ \GeVc}$}
&\multicolumn{2}{c}{$ \bf{5 \ \GeVc}$}
&\multicolumn{2}{c}{$ \bf{8 \ \GeVc}$}
&\multicolumn{2}{c}{$ \bf{12 \ \GeVc}$}
\\
\hline
0.050 &0.100 & 0.50 & 1.00&  0.46 &   0.17&  0.98 &   0.19&  1.28 &   0.21&  1.61 &   0.63\\ 
      &      & 1.00 & 1.50&  0.21 &   0.11&  0.66 &   0.11&  0.91 &   0.14&  0.63 &   0.32\\ 
      &      & 1.50 & 2.00&       &       &  0.37 &   0.14&  0.71 &   0.11&  0.67 &   0.90\\ 
      &      & 2.00 & 2.50&       &       &       &       &  0.50 &   0.10&  0.23 &   0.54\\ 
      &      & 2.50 & 3.00&       &       &       &       &  0.15 &   0.07&  0.51 &   0.54\\ 
      &      & 3.00 & 3.50&       &       &       &       &  0.06 &   0.08&  0.62 &   1.28\\ 
      &      & 3.50 & 4.00&       &       &       &       &  0.07 &   0.04&  0.34 &   0.85\\ 
      &      & 4.00 & 5.00&       &       &       &       &       &       &  0.19 &   0.73\\ 
      &      & 5.00 & 6.50&       &       &       &       &       &       &       &       \\ 
      &      & 6.50 & 8.00&       &       &       &       &       &       &       &       \\ 
0.100 &0.150 & 0.50 & 1.00&  0.82 &   0.25&  1.11 &   0.21&  1.52 &   0.28&  1.47 &   0.62\\ 
      &      & 1.00 & 1.50&  0.39 &   0.13&  0.62 &   0.12&  0.83 &   0.13&  0.55 &   0.31\\ 
      &      & 1.50 & 2.00&  0.03 &   0.10&  0.41 &   0.11&  0.52 &   0.10&  0.60 &   0.31\\ 
      &      & 2.00 & 2.50&       &       &  0.04 &   0.07&  0.31 &   0.08&  0.43 &   0.26\\ 
      &      & 2.50 & 3.00&       &       &       &       &  0.27 &   0.07&  0.35 &   0.23\\ 
      &      & 3.00 & 3.50&       &       &       &       &  0.13 &   0.05&  0.15 &   0.16\\ 
      &      & 3.50 & 4.00&       &       &       &       &  0.13 &   0.05&  0.04 &   0.07\\ 
      &      & 4.00 & 5.00&       &       &       &       &  0.05 &   0.03&  0.14 &   0.14\\ 
      &      & 5.00 & 6.50&       &       &       &       &  0.02 &   0.14&  0.05 &   0.07\\ 
      &      & 6.50 & 8.00&       &       &       &       &       &       &  0.02 &   0.04\\ 
0.150 &0.200 & 0.50 & 1.00&  0.37 &   0.15&  1.03 &   0.20&  1.16 &   0.23&  1.39 &   0.67\\ 
      &      & 1.00 & 1.50&  0.35 &   0.11&  0.46 &   0.09&  0.75 &   0.12&  0.66 &   0.33\\ 
      &      & 1.50 & 2.00&  0.04 &   0.13&  0.38 &   0.08&  0.36 &   0.09&  0.29 &   0.22\\ 
      &      & 2.00 & 2.50&       &       &  0.20 &   0.11&  0.23 &   0.07&  0.22 &   0.20\\ 
      &      & 2.50 & 3.00&       &       &  0.03 &   0.10&  0.21 &   0.06&  0.19 &   0.18\\ 
      &      & 3.00 & 3.50&       &       &  0.01 &   0.04&  0.12 &   0.05&  0.12 &   0.15\\ 
      &      & 3.50 & 4.00&       &       &       &       &  0.07 &   0.03&  0.14 &   0.14\\ 
      &      & 4.00 & 5.00&       &       &       &       &  0.02 &   0.01&  0.06 &   0.07\\ 
      &      & 5.00 & 6.50&       &       &       &       & 0.001 &  0.001&  0.01 &  0.03\\ 
      &      & 6.50 & 8.00&       &       &       &       &       &       & 0.01 &  0.06\\ 
0.200 &0.250 & 0.50 & 1.00&  0.67 &   0.29&  0.93 &   0.27&  1.25 &   0.29&  1.19 &   0.71\\ 
      &      & 1.00 & 1.50&  0.20 &   0.11&  0.52 &   0.12&  0.69 &   0.16&  0.86 &   0.51\\ 
      &      & 1.50 & 2.00&  0.09 &   0.09&  0.19 &   0.08&  0.45 &   0.13&  0.66 &   0.53\\ 
      &      & 2.00 & 2.50&  0.15 &   0.11&  0.12 &   0.07&  0.29 &   0.10&  0.58 &   0.43\\ 
      &      & 2.50 & 3.00&       &       &  0.09 &   0.04&  0.14 &   0.05&  0.14 &   0.16\\ 
      &      & 3.00 & 3.50&       &       &  0.02 &   0.01&  0.08 &   0.03&  0.11 &   0.20\\ 
      &      & 3.50 & 4.00&       &       &  0.04 &   0.04&  0.04 &   0.02&  0.10 &  0.16\\ 
      &      & 4.00 & 5.00&       &       &  0.05 &   0.03&  0.04 &   0.02&  0.08 &   0.16\\ 
      &      & 5.00 & 6.50&       &       &       &       &  0.00 &   0.01&  0.06 &   0.16\\ 
      &      & 6.50 & 8.00&       &       &       &       &       &       &  0.02 &   0.08\\

\hline
\end{tabular}
}
\end{table}
\clearpage

\begin{table}[!ht]
  \caption{\label{tab:xsec_results_Ta_pr}
    HARP results for the double-differential $p$  production
    cross-section in the laboratory system,
    $d^2\sigma^{p}/(dpd\Omega)$, for p--Ta interactions at 3,5,8,12~\GeVc.
    Each row refers to a
    different $(p_{\hbox{\small min}} \le p<p_{\hbox{\small max}},
    \theta_{\hbox{\small min}} \le \theta<\theta_{\hbox{\small max}})$ bin,
    where $p$ and $\theta$ are the outgoing proton momentum and polar angle, respectively.
    The central value as well as the square-root of the diagonal elements
    of the covariance matrix are given.}

\small{
\begin{tabular}{rrrr|r@{$\pm$}lr@{$\pm$}lr@{$\pm$}lr@{$\pm$}l}
\hline
$\theta_{\hbox{\small min}}$ &
$\theta_{\hbox{\small max}}$ &
$p_{\hbox{\small min}}$ &
$p_{\hbox{\small max}}$ &
\multicolumn{8}{c}{$d^2\sigma^{p}/(dpd\Omega)$}
\\
(rad) & (rad) & (\GeVc) & (\GeVc) &
\multicolumn{8}{c}{(barn/(sr \GeVc ))}
\\
  &  &  &
&\multicolumn{2}{c}{$ \bf{3 \ \GeVc}$}
&\multicolumn{2}{c}{$ \bf{5 \ \GeVc}$}
&\multicolumn{2}{c}{$ \bf{8 \ \GeVc}$}
&\multicolumn{2}{c}{$ \bf{12 \ \GeVc}$}
\\
\hline
0.050 &0.100 & 0.50 & 1.00&  0.52 &   0.18&  1.12 &   0.20&  1.57 &   0.20&  1.76 &   0.24\\ 
      &      & 1.00 & 1.50&  0.45 &   0.12&  0.88 &   0.12&  1.14 &   0.11&  1.18 &   0.13\\ 
      &      & 1.50 & 2.00&  0.18 &   0.06&  0.61 &   0.11&  1.02 &   0.10&  1.16 &   0.13\\ 
      &      & 2.00 & 2.50&  0.19 &   0.07&  0.75 &   0.12&  0.86 &   0.08&  0.88 &   0.10\\ 
      &      & 2.50 & 3.00&       &       &  0.90 &   0.11&  0.98 &   0.08&  0.90 &   0.09\\ 
      &      & 3.00 & 3.50&       &       &  1.20 &   0.16&  0.91 &   0.08&  0.80 &   0.09\\ 
      &      & 3.50 & 4.00&       &       &  1.14 &   0.15&  1.08 &   0.09&  0.83 &   0.08\\ 
      &      & 4.00 & 5.00&       &       &  0.93 &   0.14&  1.20 &   0.07&  0.92 &   0.07\\ 
      &      & 5.00 & 6.50&       &       &       &       &  0.73 &   0.05&  0.87 &   0.05\\ 
      &      & 6.50 & 8.00&       &       &       &       &       &       &  0.59 &   0.05\\ 
0.100 &0.150 & 0.50 & 1.00&  0.94 &   0.26&  1.26 &   0.22&  1.75 &   0.23&  1.60 &   0.25\\ 
      &      & 1.00 & 1.50&  0.79 &   0.18&  0.82 &   0.14&  1.08 &   0.12&  1.30 &   0.15\\ 
      &      & 1.50 & 2.00&  0.69 &   0.18&  0.62 &   0.10&  0.81 &   0.10&  1.06 &   0.13\\ 
      &      & 2.00 & 2.50&  0.55 &   0.12&  0.50 &   0.10&  0.74 &   0.08&  0.80 &   0.10\\ 
      &      & 2.50 & 3.00&       &       &  0.70 &   0.10&  0.73 &   0.07&  0.62 &   0.08\\ 
      &      & 3.00 & 3.50&       &       &  0.73 &   0.11&  0.79 &   0.07&  0.63 &   0.08\\ 
      &      & 3.50 & 4.00&       &       &  0.62 &   0.11&  0.65 &   0.06&  0.61 &   0.07\\ 
      &      & 4.00 & 5.00&       &       &  0.53 &   0.09&  0.58 &   0.05&  0.63 &   0.06\\ 
      &      & 5.00 & 6.50&       &       &       &       &  0.27 &   0.03&  0.31 &   0.03\\ 
      &      & 6.50 & 8.00&       &       &       &       &       &       &  0.15 &   0.02\\ 
0.150 &0.200 & 0.50 & 1.00&  0.79 &   0.23&  1.38 &   0.25&  1.42 &   0.21&  1.70 &   0.28\\ 
      &      & 1.00 & 1.50&  0.61 &   0.15&  0.72 &   0.11&  0.91 &   0.10&  1.05 &   0.13\\ 
      &      & 1.50 & 2.00&  0.40 &   0.13&  0.65 &   0.10&  0.71 &   0.08&  0.69 &   0.10\\ 
      &      & 2.00 & 2.50&  0.37 &   0.13&  0.51 &   0.09&  0.56 &   0.07&  0.57 &   0.09\\ 
      &      & 2.50 & 3.00&       &       &  0.41 &   0.08&  0.53 &   0.06&  0.62 &   0.08\\ 
      &      & 3.00 & 3.50&       &       &  0.18 &   0.04&  0.36 &   0.04&  0.44 &   0.06\\ 
      &      & 3.50 & 4.00&       &       &  0.18 &   0.04&  0.34 &   0.04&  0.29 &   0.04\\ 
      &      & 4.00 & 5.00&       &       &  0.19 &   0.04&  0.18 &   0.03&  0.23 &   0.03\\ 
      &      & 5.00 & 6.50&       &       &       &       & 0.058 &  0.013&  0.08 &   0.02\\ 
      &      & 6.50 & 8.00&       &       &       &       &       &       & 0.034 &  0.009\\ 
0.200 &0.250 & 0.50 & 1.00&  1.37 &   0.44&  0.98 &   0.24&  1.73 &   0.27&  1.74 &   0.31\\ 
      &      & 1.00 & 1.50&  0.36 &   0.14&  0.52 &   0.12&  0.92 &   0.15&  0.96 &   0.18\\ 
      &      & 1.50 & 2.00&  0.11 &   0.05&  0.30 &   0.09&  0.39 &   0.08&  0.51 &   0.13\\ 
      &      & 2.00 & 2.50&  0.22 &   0.08&  0.20 &   0.07&  0.40 &   0.08&  0.41 &   0.11\\ 
      &      & 2.50 & 3.00&       &       &  0.28 &   0.07&  0.25 &   0.05&  0.25 &   0.06\\ 
      &      & 3.00 & 3.50&       &       &  0.10 &   0.03&  0.21 &   0.04&  0.24 &   0.05\\ 
      &      & 3.50 & 4.00&       &       &  0.10 &   0.03&  0.15 &   0.03&  0.25 &   0.05\\ 
      &      & 4.00 & 5.00&       &       &  0.12 &   0.03&  0.11 &   0.03&  0.12 &   0.03\\ 
      &      & 5.00 & 6.50&       &       &       &       & 0.047 &  0.014& 0.042 &  0.013\\ 
      &      & 6.50 & 8.00&       &       &       &       &       &       & 0.023 &  0.009\\

\hline
\end{tabular}
}
\end{table}
\clearpage

\begin{table}[!ht]
  \caption{\label{tab:xsec_results_Pb_pim}
    HARP results for the double-differential $p$  production
    cross-section in the laboratory system,
    $d^2\sigma^{p}/(dpd\Omega)$, for $\pi^{-}$--Pb interactions at 3,5,8,12~\GeVc.
    Each row refers to a
    different $(p_{\hbox{\small min}} \le p<p_{\hbox{\small max}},
    \theta_{\hbox{\small min}} \le \theta<\theta_{\hbox{\small max}})$ bin,
    where $p$ and $\theta$ are the outgoing proton momentum and polar angle, respectively.
    The central value as well as the square-root of the diagonal elements
    of the covariance matrix are given.}

\small{
\begin{tabular}{rrrr|r@{$\pm$}lr@{$\pm$}lr@{$\pm$}lr@{$\pm$}l}
\hline
$\theta_{\hbox{\small min}}$ &
$\theta_{\hbox{\small max}}$ &
$p_{\hbox{\small min}}$ &
$p_{\hbox{\small max}}$ &
\multicolumn{8}{c}{$d^2\sigma^{p}/(dpd\Omega)$}
\\
(rad) & (rad) & (\GeVc) & (\GeVc) &
\multicolumn{8}{c}{(barn/(sr \GeVc ))}
\\
  &  &  &
&\multicolumn{2}{c}{$ \bf{3 \ \GeVc}$}
&\multicolumn{2}{c}{$ \bf{5 \ \GeVc}$}
&\multicolumn{2}{c}{$ \bf{8 \ \GeVc}$}
&\multicolumn{2}{c}{$ \bf{12 \ \GeVc}$}
\\
\hline
0.050 &0.100 & 0.50 & 1.00&  0.67 &   0.16&  1.10 &   0.14&  1.22 &   0.16&  1.24 &   0.16\\ 
      &      & 1.00 & 1.50&  0.11 &   0.05&  0.54 &   0.06&  0.82 &   0.10&  0.89 &   0.09\\ 
      &      & 1.50 & 2.00&  0.10 &   0.04&  0.45 &   0.05&  0.73 &   0.09&  0.59 &   0.09\\ 
      &      & 2.00 & 2.50&  0.03 &   0.02&  0.21 &   0.03&  0.31 &   0.06&  0.13 &   0.07\\ 
      &      & 2.50 & 3.00&       &       &  0.12 &   0.02&  0.30 &   0.04&  0.35 &   0.07\\ 
      &      & 3.00 & 3.50&       &       &  0.05 &   0.02&  0.19 &   0.04&  0.32 &   0.04\\ 
      &      & 3.50 & 4.00&       &       & 0.03  &  0.01 &  0.24 &   0.04&  0.34 &   0.04\\ 
      &      & 4.00 & 5.00&       &       & 0.01 &  0.01  &  0.11 &   0.02&  0.32 &   0.03\\ 
      &      & 5.00 & 6.50&       &       &       &       &  0.03 &    0.01&  0.15 &   0.02\\ 
      &      & 6.50 & 8.00&       &       &       &       &       &       & 0.05 &  0.01\\ 
0.100 &0.150 & 0.50 & 1.00&  0.75 &   0.18&  1.30 &   0.17&  1.32 &   0.20&  1.38 &   0.19\\ 
      &      & 1.00 & 1.50&  0.34 &   0.10&  0.74 &   0.08&  0.96 &   0.11&  1.06 &   0.11\\ 
      &      & 1.50 & 2.00&  0.07 &   0.03&  0.44 &   0.06&  0.60 &   0.09&  0.53 &   0.09\\ 
      &      & 2.00 & 2.50&  0.06 &   0.03&  0.21 &   0.03&  0.36 &   0.06&  0.24 &   0.06\\ 
      &      & 2.50 & 3.00&       &       &  0.12 &   0.02&  0.31 &   0.04&  0.38 &   0.05\\ 
      &      & 3.00 & 3.50&       &       & 0.06 &  0.02&  0.17 &   0.04&  0.31 &   0.04\\ 
      &      & 3.50 & 4.00&       &       & 0.03 &  0.01&  0.15 &   0.03&  0.21 &   0.03\\ 
      &      & 4.00 & 5.00&       &       & 0.02 &  0.01&  0.08 &   0.02&  0.17 &   0.02\\ 
      &      & 5.00 & 6.50&       &       &       &       & 0.01 &  0.01& 0.08 &  0.01\\ 
      &      & 6.50 & 8.00&       &       &       &       &       &       & 0.015 &  0.004\\ 
0.150 &0.200 & 0.50 & 1.00&  0.44 &   0.14&  1.19 &   0.17&  1.37 &   0.20&  1.33 &   0.19\\ 
      &      & 1.00 & 1.50&  0.17 &   0.06&  0.47 &   0.06&  0.60 &   0.08&  0.68 &   0.08\\ 
      &      & 1.50 & 2.00&  0.03 &   0.02&  0.35 &   0.05&  0.51 &   0.07&  0.55 &   0.08\\ 
      &      & 2.00 & 2.50&  0.04 &   0.03&  0.20 &   0.04&  0.32 &   0.06&  0.37 &   0.07\\ 
      &      & 2.50 & 3.00&       &       & 0.063 &  0.015&  0.21 &   0.04&  0.30 &   0.04\\ 
      &      & 3.00 & 3.50&       &       & 0.02 &  0.01&  0.11 &   0.03&  0.23 &   0.03\\ 
      &      & 3.50 & 4.00&       &       & 0.02 &  0.01&  0.11 &   0.03&  0.17 &   0.02\\ 
      &      & 4.00 & 5.00&       &       & 0.01 &  0.01& 0.02 &  0.01&  0.10 &   0.02\\ 
      &      & 5.00 & 6.50&       &       &       &       &       &       & 0.02 &  0.01\\ 
      &      & 6.50 & 8.00&       &       &       &       &       &       & 0.003 &  0.001\\ 
0.200 &0.250 & 0.50 & 1.00&  0.55 &   0.21&  1.08 &   0.19&  1.29 &   0.23&  1.54 &   0.25\\ 
      &      & 1.00 & 1.50&  0.51 &   0.17&  0.82 &   0.12&  1.18 &   0.19&  1.10 &   0.16\\ 
      &      & 1.50 & 2.00&  0.09 &   0.06&  0.42 &   0.07&  0.74 &   0.15&  0.86 &   0.16\\ 
      &      & 2.00 & 2.50&  0.01 &   0.01&  0.34 &   0.06&  0.64 &   0.13&  0.61 &   0.12\\ 
      &      & 2.50 & 3.00&       &       &  0.08 &   0.03&  0.33 &   0.07&  0.47 &   0.08\\ 
      &      & 3.00 & 3.50&       &       & 0.03 &  0.02&  0.09 &   0.04&  0.18 &   0.04\\ 
      &      & 3.50 & 4.00&       &       & 0.007 &  0.004&  0.04 &   0.02&  0.05 &   0.02\\ 
      &      & 4.00 & 5.00&       &       & 0.003 &  0.002& 0.02 &  0.01& 0.04 &  0.01\\ 
      &      & 5.00 & 6.50&       &       &       &       & 0.001 &  0.006& 0.009 &  0.003\\ 
      &      & 6.50 & 8.00&       &       &       &       &       &       & 0.003 &  0.002\\

\hline
\end{tabular}
}
\end{table}
\clearpage

\begin{table}[!ht]
  \caption{\label{tab:xsec_results_Pb_pip}
    HARP results for the double-differential $p$  production
    cross-section in the laboratory system,
    $d^2\sigma^{p}/(dpd\Omega)$, for $\pi^{+}$--Pb interactions at 3,5,8,12~\GeVc.
    Each row refers to a
    different $(p_{\hbox{\small min}} \le p<p_{\hbox{\small max}},
    \theta_{\hbox{\small min}} \le \theta<\theta_{\hbox{\small max}})$ bin,
    where $p$ and $\theta$ are the outgoing proton momentum and polar angle, respectively.
    The central value as well as the square-root of the diagonal elements
    of the covariance matrix are given.}

\small{
\begin{tabular}{rrrr|r@{$\pm$}lr@{$\pm$}lr@{$\pm$}lr@{$\pm$}l}
\hline
$\theta_{\hbox{\small min}}$ &
$\theta_{\hbox{\small max}}$ &
$p_{\hbox{\small min}}$ &
$p_{\hbox{\small max}}$ &
\multicolumn{8}{c}{$d^2\sigma^{p}/(dpd\Omega)$}
\\
(rad) & (rad) & (\GeVc) & (\GeVc) &
\multicolumn{8}{c}{(barn/(sr \GeVc ))}
\\
  &  &  &
&\multicolumn{2}{c}{$ \bf{3 \ \GeVc}$}
&\multicolumn{2}{c}{$ \bf{5 \ \GeVc}$}
&\multicolumn{2}{c}{$ \bf{8 \ \GeVc}$}
&\multicolumn{2}{c}{$ \bf{12 \ \GeVc}$}
\\
\hline
0.050 &0.100 & 0.50 & 1.00&  0.70 &   0.23&  0.93 &   0.21&  1.47 &   0.23&  1.47 &   0.90\\ 
      &      & 1.00 & 1.50&  0.37 &   0.15&  0.71 &   0.12&  0.75 &   0.12&  1.17 &   0.66\\ 
      &      & 1.50 & 2.00&       &       &  0.31 &   0.11&  0.68 &   0.12&  0.34 &   0.52\\ 
      &      & 2.00 & 2.50&       &       &  0.03 &   0.07&  0.44 &   0.10&  0.24 &   0.42\\ 
      &      & 2.50 & 3.00&       &       &       &       &  0.21 &   0.09&  0.02 &   0.11\\ 
      &      & 3.00 & 3.50&       &       &       &       &  0.08 &   0.08&  0.15 &   0.35\\ 
      &      & 3.50 & 4.00&       &       &       &       &       &       &  0.08 &   0.19\\ 
      &      & 4.00 & 5.00&       &       &       &       &       &       &  0.17 &   0.34\\ 
      &      & 5.00 & 6.50&       &       &       &       &       &       &  1.89 &   0.30\\ 
      &      & 6.50 & 8.00&       &       &       &       &       &       &       &       \\ 
0.100 &0.150 & 0.50 & 1.00&  0.69 &   0.24&  1.23 &   0.24&  1.68 &   0.29&  0.28 &   0.37\\ 
      &      & 1.00 & 1.50&  0.30 &   0.13&  0.47 &   0.11&  0.70 &   0.12&  1.15 &   0.66\\ 
      &      & 1.50 & 2.00&  0.02 &   0.06&  0.44 &   0.12&  0.54 &   0.11&  1.22 &   0.66\\ 
      &      & 2.00 & 2.50&       &       &  0.13 &   0.09&  0.22 &   0.09&  0.45 &   0.39\\ 
      &      & 2.50 & 3.00&       &       &       &       &  0.27 &   0.15&  0.46 &   0.43\\ 
      &      & 3.00 & 3.50&       &       &       &       &  0.16 &   0.17&  0.07 &   0.19\\ 
      &      & 3.50 & 4.00&       &       &       &       &  0.10 &   0.24&  0.13 &   0.31\\ 
      &      & 4.00 & 5.00&       &       &       &       &  0.04 &   0.11&  0.11 &   0.45\\ 
      &      & 5.00 & 6.50&       &       &       &       &  0.02 &   0.11&  0.03 &   0.20\\ 
      &      & 6.50 & 8.00&       &       &       &       &       &       &  0.00 &   0.10\\ 
0.150 &0.200 & 0.50 & 1.00&  0.61 &   0.21&  1.19 &   0.23&  1.18 &   0.24&  0.23 &   0.47\\ 
      &      & 1.00 & 1.50&  0.45 &   0.13&  0.53 &   0.10&  0.65 &   0.11&  0.81 &   0.52\\ 
      &      & 1.50 & 2.00&  0.00 &   0.00&  0.34 &   0.08&  0.44 &   0.10&  0.51 &   0.47\\ 
      &      & 2.00 & 2.50&       &       &  0.07 &   0.07&  0.20 &   0.07&  0.37 &   0.39\\ 
      &      & 2.50 & 3.00&       &       &  0.01 &   0.06&  0.18 &   0.05&  0.20 &   0.37\\ 
      &      & 3.00 & 3.50&       &       &       &       &  0.07 &   0.03&  0.05 &   0.34\\ 
      &      & 3.50 & 4.00&       &       &  0.03 &   0.07&  0.05 &   0.03&  0.00 &   0.12\\ 
      &      & 4.00 & 5.00&       &       &  0.04 &   0.05&  0.02 &   0.02&  0.01 &   0.16\\ 
      &      & 5.00 & 6.50&       &       &       &       &  0.00 &   0.00&  0.01 &   0.08\\ 
      &      & 6.50 & 8.00&       &       &       &       &       &       &  0.00 &   0.13\\ 
0.200 &0.250 & 0.50 & 1.00&  0.46 &   0.27&  1.23 &   0.34&  1.30 &   0.29&  1.50 &   1.27\\ 
      &      & 1.00 & 1.50&  0.08 &   0.07&  0.61 &   0.14&  0.61 &   0.15&  1.41 &   0.96\\ 
      &      & 1.50 & 2.00&  0.15 &   0.08&  0.18 &   0.08&  0.22 &   0.08&  0.74 &   0.90\\ 
      &      & 2.00 & 2.50&  0.09 &   0.11&  0.13 &   0.08&  0.30 &   0.11&  0.66 &   1.00\\ 
      &      & 2.50 & 3.00&       &       &  0.10 &   0.06&  0.16 &   0.06&  0.07 &   0.29\\ 
      &      & 3.00 & 3.50&       &       &  0.06 &   0.05&  0.08 &   0.03&  0.18 &   0.52\\ 
      &      & 3.50 & 4.00&       &       &  0.03 &   0.03&  0.05 &   0.02&  0.16 &   0.54\\ 
      &      & 4.00 & 5.00&       &       &  0.03 &   0.02&  0.03 &   0.02&       &       \\ 
      &      & 5.00 & 6.50&       &       &       &       &  0.01 &   0.01&       &       \\ 
      &      & 6.50 & 8.00&       &       &       &       &       &       &       &      \\

\hline
\end{tabular}
}
\end{table}
\clearpage

\begin{table}[!ht]
  \caption{\label{tab:xsec_results_Pb_pr}
    HARP results for the double-differential $p$  production
    cross-section in the laboratory system,
    $d^2\sigma^{p}/(dpd\Omega)$, for p--Pb interactions at 3,5,8,12~\GeVc.
    Each row refers to a
    different $(p_{\hbox{\small min}} \le p<p_{\hbox{\small max}},
    \theta_{\hbox{\small min}} \le \theta<\theta_{\hbox{\small max}})$ bin,
    where $p$ and $\theta$ are the outgoing proton momentum and polar angle, respectively.
    The central value as well as the square-root of the diagonal elements
    of the covariance matrix are given.}

\small{
\begin{tabular}{rrrr|r@{$\pm$}lr@{$\pm$}lr@{$\pm$}lr@{$\pm$}l}
\hline
$\theta_{\hbox{\small min}}$ &
$\theta_{\hbox{\small max}}$ &
$p_{\hbox{\small min}}$ &
$p_{\hbox{\small max}}$ &
\multicolumn{8}{c}{$d^2\sigma^{p}/(dpd\Omega)$}
\\
(rad) & (rad) & (\GeVc) & (\GeVc) &
\multicolumn{8}{c}{(barn/(sr \GeVc ))}
\\
  &  &  &
&\multicolumn{2}{c}{$ \bf{3 \ \GeVc}$}
&\multicolumn{2}{c}{$ \bf{5 \ \GeVc}$}
&\multicolumn{2}{c}{$ \bf{8 \ \GeVc}$}
&\multicolumn{2}{c}{$ \bf{12 \ \GeVc}$}
\\
\hline
0.050 &0.100 & 0.50 & 1.00&  0.63 &   0.22&  1.17 &   0.23&  1.62 &   0.20&  1.92 &   0.31\\ 
      &      & 1.00 & 1.50&  0.40 &   0.12&  0.91 &   0.13&  1.21 &   0.11&  1.49 &   0.20\\ 
      &      & 1.50 & 2.00&  0.27 &   0.08&  0.77 &   0.13&  1.05 &   0.10&  1.20 &   0.17\\ 
      &      & 2.00 & 2.50&  0.08 &   0.04&  0.66 &   0.11&  0.84 &   0.08&  0.86 &   0.13\\ 
      &      & 2.50 & 3.00&       &       &  0.90 &   0.13&  0.90 &   0.09&  0.97 &   0.13\\ 
      &      & 3.00 & 3.50&       &       &  1.17 &   0.14&  1.02 &   0.09&  1.07 &   0.15\\ 
      &      & 3.50 & 4.00&       &       &  1.05 &   0.15&  1.12 &   0.08&  1.09 &   0.12\\ 
      &      & 4.00 & 5.00&       &       &  0.86 &   0.11&  1.04 &   0.07&  0.97 &   0.09\\ 
      &      & 5.00 & 6.50&       &       &       &       &  0.73 &   0.05&  0.83 &   0.07\\ 
      &      & 6.50 & 8.00&       &       &       &       &       &       &  0.63 &   0.06\\ 
0.100 &0.150 & 0.50 & 1.00&  0.59 &   0.20&  1.40 &   0.25&  1.91 &   0.25&  2.39 &   0.40\\ 
      &      & 1.00 & 1.50&  0.87 &   0.20&  0.87 &   0.15&  1.17 &   0.13&  1.56 &   0.22\\ 
      &      & 1.50 & 2.00&  0.45 &   0.15&  0.84 &   0.13&  0.92 &   0.11&  1.26 &   0.19\\ 
      &      & 2.00 & 2.50&  0.39 &   0.10&  0.77 &   0.15&  0.70 &   0.08&  0.59 &   0.11\\ 
      &      & 2.50 & 3.00&       &       &  0.97 &   0.13&  0.78 &   0.08&  0.66 &   0.11\\ 
      &      & 3.00 & 3.50&       &       &  0.87 &   0.12&  0.67 &   0.07&  0.61 &   0.10\\ 
      &      & 3.50 & 4.00&       &       &  0.66 &   0.13&  0.60 &   0.06&  0.70 &   0.10\\ 
      &      & 4.00 & 5.00&       &       &  0.59 &   0.10&  0.55 &   0.05&  0.54 &   0.07\\ 
      &      & 5.00 & 6.50&       &       &       &       &  0.29 &   0.03&  0.35 &   0.04\\ 
      &      & 6.50 & 8.00&       &       &       &       &       &       &  0.17 &   0.03\\ 
0.150 &0.200 & 0.50 & 1.00&  0.93 &   0.30&  1.52 &   0.28&  1.55 &   0.23&  1.83 &   0.34\\ 
      &      & 1.00 & 1.50&  0.52 &   0.14&  0.75 &   0.13&  0.94 &   0.10&  1.04 &   0.17\\ 
      &      & 1.50 & 2.00&  0.21 &   0.09&  0.63 &   0.11&  0.65 &   0.08&  0.79 &   0.15\\ 
      &      & 2.00 & 2.50&  0.37 &   0.13&  0.58 &   0.11&  0.50 &   0.07&  0.76 &   0.14\\ 
      &      & 2.50 & 3.00&       &       &  0.42 &   0.08&  0.43 &   0.06&  0.48 &   0.09\\ 
      &      & 3.00 & 3.50&       &       &  0.31 &   0.06&  0.30 &   0.04&  0.44 &   0.08\\ 
      &      & 3.50 & 4.00&       &       &  0.33 &   0.06&  0.22 &   0.03&  0.40 &   0.07\\ 
      &      & 4.00 & 5.00&       &       &  0.22 &   0.04&  0.23 &   0.03&  0.18 &   0.04\\ 
      &      & 5.00 & 6.50&       &       &       &       & 0.058 &  0.013&  0.13 &   0.03\\ 
      &      & 6.50 & 8.00&       &       &       &       &       &       &  0.06 &   0.02\\ 
0.200 &0.250 & 0.50 & 1.00&  0.50 &   0.30&  1.49 &   0.33&  1.56 &   0.25&  2.05 &   0.43\\ 
      &      & 1.00 & 1.50&  0.30 &   0.14&  0.84 &   0.18&  0.89 &   0.15&  0.89 &   0.22\\ 
      &      & 1.50 & 2.00&  0.21 &   0.08&  0.30 &   0.10&  0.49 &   0.10&  0.63 &   0.18\\ 
      &      & 2.00 & 2.50&  0.13 &   0.06&  0.30 &   0.09&  0.37 &   0.08&  0.41 &   0.14\\ 
      &      & 2.50 & 3.00&       &       &  0.35 &   0.09&  0.35 &   0.06&  0.31 &   0.10\\ 
      &      & 3.00 & 3.50&       &       &  0.20 &   0.06&  0.19 &   0.04&  0.17 &   0.06\\ 
      &      & 3.50 & 4.00&       &       &  0.09 &   0.03&  0.12 &   0.03&  0.11 &   0.04\\ 
      &      & 4.00 & 5.00&       &       &  0.10 &   0.03&  0.11 &   0.03&  0.13 &   0.05\\ 
      &      & 5.00 & 6.50&       &       &       &       & 0.038 &  0.015&  0.09 &   0.04\\ 
      &      & 6.50 & 8.00&       &       &       &       &       &       &  0.05 &   0.02\\

\hline
\end{tabular}
}
\end{table}
\clearpage

\end{appendix}
\end{document}